\documentclass[12pt]{article}
\usepackage{a4}
\usepackage{latexsym} 
\usepackage{amssymb}  
\usepackage{graphicx}
\usepackage{epsfig}
\usepackage[ usenames]{color}
\usepackage[rflt]{floatflt}
\begin{document}

\newcommand{\talk}[3]
{\noindent{#1}\\ \mbox{}\ \ \ {\it #2} \dotfill {\pageref{#3}}\\[-4mm]}
\newcommand{\stalk}[3]
{{#1} & {\it #2} & {\pageref{#3}}\\}
\newcommand{\snotalk}[3]
{{#1} & {\it #2} & {{#3}n.r.}\\}
\newcommand{\notalk}[3]
{\noindent{#1}\\ \mbox{}\ \ \ {\it #2} \hfill {{#3}n.r.}\\[-4mm]}
\newcounter{zyxabstract}     
\newcounter{zyxrefers}        

\newcommand{\newabstract}
{\newpage\stepcounter{zyxabstract}\setcounter{equation}{0}
\setcounter{footnote}{0}}

\newcommand{\rlabel}[1]{\label{zyx\arabic{zyxabstract}#1}}
\newcommand{\rref}[1]{\ref{zyx\arabic{zyxabstract}#1}}

\renewenvironment{thebibliography}[1] 
{\section*{References}\setcounter{zyxrefers}{0}
\begin{list}{ [\arabic{zyxrefers}]}{\usecounter{zyxrefers}}}
{\end{list}}
\newenvironment{thebibliographynotitle}[1] 
{\setcounter{zyxrefers}{0}
\begin{list}{ [\arabic{zyxrefers}]}
{\usecounter{zyxrefers}\setlength{\itemsep}{-2mm}}}
{\end{list}}

\renewcommand{\bibitem}[1]{\item\rlabel{y#1}}
\renewcommand{\cite}[1]{[\rref{y#1}]}      
\newcommand{\citetwo}[2]{[\rref{y#1},\rref{y#2}]}
\newcommand{\citethree}[3]{[\rref{y#1},\rref{y#2},\rref{y#3}]}
\newcommand{\citefour}[4]{[\rref{y#1},\rref{y#2},\rref{y#3},\rref{y#4}]}
\newcommand{\citefive}[5]
{[\rref{y#1},\rref{y#2},\rref{y#3},\rref{y#4},\rref{y#5}]}
\newcommand{\citesix}[6]
{[\rref{y#1},\rref{y#2},\rref{y#3},\rref{y#4},\rref{y#5},\rref{y#6}]}
\begin{titlepage}
\begin{flushleft}
\begin{figure}
 \includegraphics[height=2.cm]{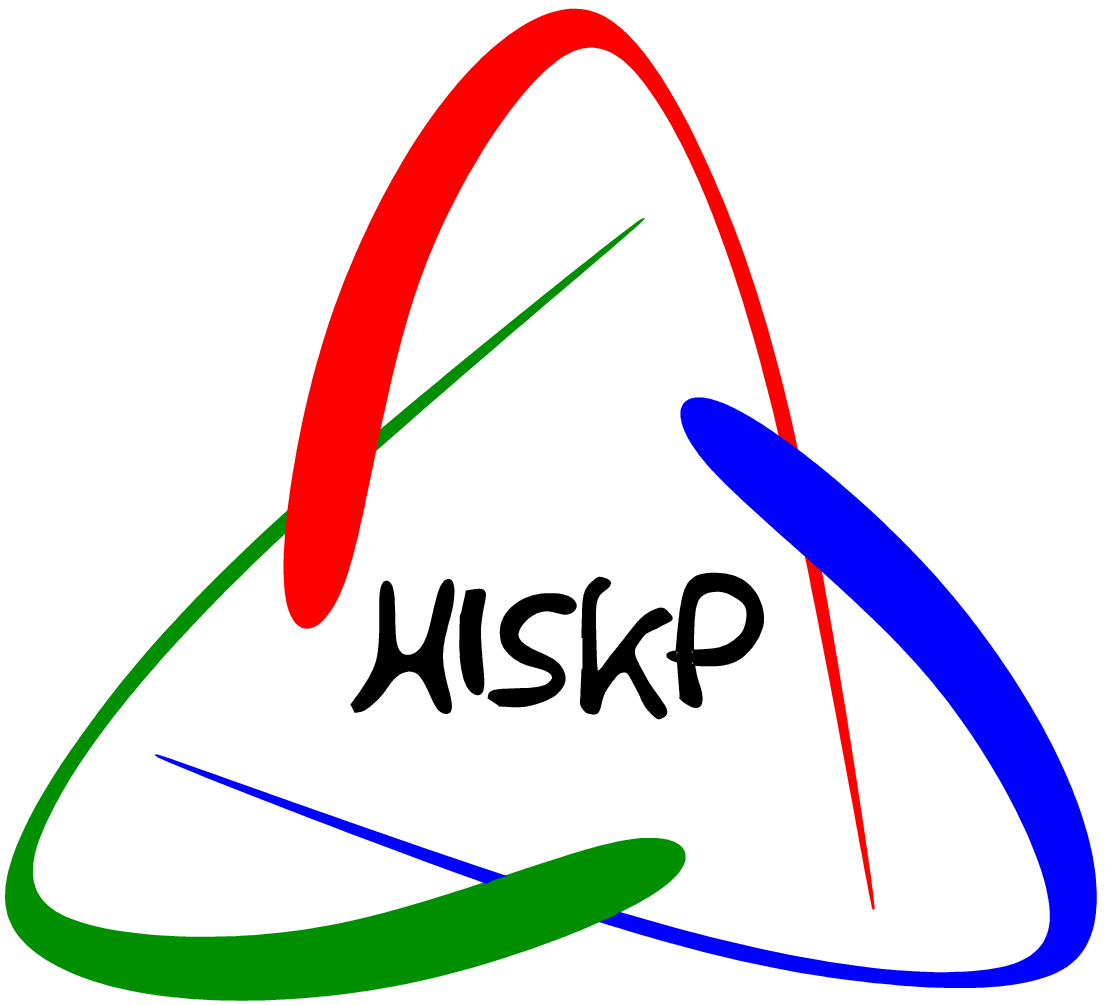} 
\end{figure}
\end{flushleft}

\vspace{-3.75cm}
\begin{flushright}
\small{
HISKP-TH 03/23
}
\end{flushright}

\vspace{1.5cm}

\begin{center}
{\large \bf Mini-Proceedings of the}\\[0.5cm]
{\large \bf Fourth International Workshop on }\\[0.5cm]
{\Large \bf  CHIRAL DYNAMICS\,:\\[0.5cm] 
             THEORY and EXPERIMENT}\\[0.5cm]
{\Large \bf  (\,CD2003\,)}\\[1cm]
Bonn, Germany\\
September 8 -- 13, 2003\\[1cm]
edited by\\[1cm]
{\bf Ulf-G. Mei\ss ner, Hans-Werner Hammer and Andreas Wirzba}\\[0.3cm]
{Helmholtz-Institut f\"ur Strahlen- und Kernphysik 
(Theorie)\\ 
Rheinische Friedrich-Wilhelms--Universit\"at Bonn\\
Nussallee 14-16, D-53115 Bonn, Germany}\\[2cm]
{\large ABSTRACT}
\end{center}
These are the proceedings of the fourth international workshop on ``Chiral
Dynamics: Theory and Experiment'' which was held at the University of Bonn,
September 8-13, 2003.  
The workshop concentrated on the various experimental and
theoretical aspects of chiral dynamics, including for the first time in this
series detailed discussions of lattice gauge theory. It consisted of an
introductory lecture, plenary talks, working group talks and working group
summaries. Included is a short contribution per talk.

\end{titlepage}

\centerline{\includegraphics*[height=8.66in]{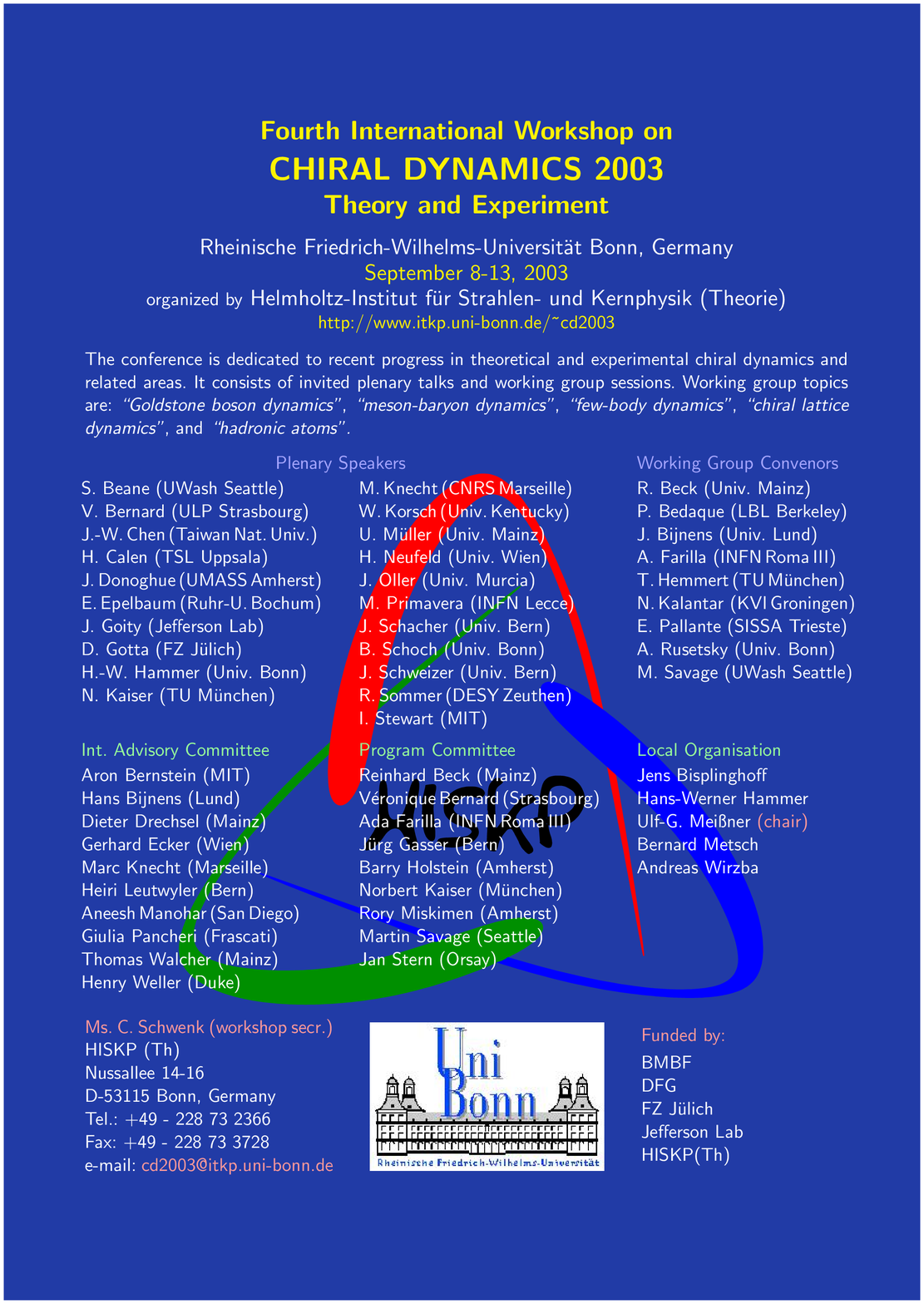}}

\thispagestyle{empty}
\newpage

\section{Introduction}

\vspace{1cm}
This volume presents the mini--proceedings of the workshop on ``Chiral
Dynamics: Theory and Experiment'' which was held at the University of Bonn,
September 8-13, 2003.  This workshop was the fourth in a series, following MIT
(1994), University of Mainz (1997) and Jefferson Laboratory (2000).  The fifth
one will be organized at TUNL-Duke in 2006.

The unique feature of this series of workshops is the approximately equal
mixture of theory and experiment. Due to the rapid developments in the
last decade, the range of topics which was discussed has considerably increased
over the years. This is reflected in the structure of the working groups,
which besides the ``classical'' topics of Goldstone boson and Goldstone
boson-baryon physics were centered around few-body systems and hadronic bound
states as well as chiral lattice dynamics.

The workshop consisted of an introductory lecture by John Donoghue, which
spurred lively discussions, and 20 plenary talks in the morning sessions. 
In the
afternoons, four working groups took place with about 90 talks in total. The
results of these working groups were summarized by the convenors in the
plenary session
on the last day of the workshop. We had about 140 participants from 20
different countries, with a large amount of younger people, showing that this
field is still growing and gaining importance. The outcome of all these talks
and the related discussions are summarized in these mini-proceedings. Although
the proceedings of the first three workshops were published in conventional
proceedings, we believe that the format chosen here better captures the actual
status of the field and allows for a quicker access to the presently available
literature. Some of the talks are available as PDF or PS file on the
conference web page under {\tt www.itkp.uni-bonn.de/$\, \tilde{} \,$cd2003/}.

Finally, we would like to take this occasion to thank the speakers, the working
group convenors, and all the participants for making this workshop a real
success.

\vspace{1.5cm}

\noindent {\large {\bf Acknowledgements}}

\vspace{0.3cm}

This workshop could not have been done without the dedicated work of many
people.  First, we would like to thank the conference secretary, 
Ms.~Claudia Schwenk together with Ms.~Barbara Mosblech and the staff of the
Helmholtz-Institut for all their efforts. Our post--docs and students Paul
B\"uttiker, Matthias Frink, Christian Haupt, Hermann Krebs, Sascha Migura,
Lucas Platter,
Udit Raha and Akaki Rusetsky performed many different tasks, which we
gratefully acknowledge. 
Special thanks  to  our co-organizers: 
Bernard Metsch, who contributed at
many different fronts, and to Jens Bispling\-hoff for help with the
organization. We are grateful to Herbert Petry for his advice.
All members of the advisory and the program committee are warmly thanked for
their help and support. Technical assistance in the Wolfgang-Paul lecture hall
was provided by Michael Kort\-mann and we are also grateful to 
Prof.~E.~Hil\-ger, the then acting director 
of the Physikalisches Institut, for support and the allocation of
some lecture halls. 
Financial support was provided by the Deutsche
Forschungsgemeinschaft, the Bundesministerium for Bil\-dung und For\-schung,
Forschungs\-zentrum J\"ulich, and Jefferson Laboratory, complemented by a
contribution from the HISKP.

\bigskip

\noindent Bonn, November 2003

\bigskip\bigskip\bigskip

\hfill Ulf-G. Mei\ss ner 

\hfill Hans-Werner Hammer

\hfill Andreas Wirzba

\newpage
\section{Participants and their email}

\begin{tabbing}
A very long namexxxxx\=a very long institutexxxxxx\=email\kill
S. Ando  \> TRIUMF \> sando@triumf.ca\\
V. Anisimovsky \> INR Moscow \> valera@al20.inr.troitsk.ru\\
R.\,A. Arndt \> George Washington Univ. \> arndtra@said.phys.vt.edu\\
D. Arndt \> Univ. of Washington \> arndt@phys.washington.edu\\
C. Aubin \> Washington Univ. \> caubin@hbar.wustl.edu\\
S.\,R. Beane \> Univ. of New Hampshire \> silas@physics.unh.edu\\
D. Be\a'cirevi\a'c \> LPT Orsay \> damir.becirevic@th.u-psud.fr\\
R. Beck \> Univ. Mainz \> rbeck@kph.uni-mainz.de\\
V. Bernard \>LPT Strasbourg\> bernard@lpt6.u-strasbg.fr\\
A.\,M. Bernstein \> MIT Cambridge\> bernstein@lns.mit.edu
% ...
\\

\end{tabbing}

\newpage
\section{Scientific program}

\vskip.5cm

\centerline{\large\bf Plenary Talks}

\noindent\mbox{}\hfill{\bf Page}

\vskip.5cm

\talk{J.\,F. Donoghue}{ Introductory Lecture: The Physics of the Chiral
Effective Field Theory}{abs:donoghue1}

\talk{J.\,L. Goity}{ $\pi^0 \to \gamma \gamma$ and the PRIMEX}{abs:goity}  

\talk{M. Primavera}{ Results from DA$\Phi$NE}{abs:primavera}  

\talk{V. Bernard}{ Spin Structure of the Nucleon --- Theory}{abs:bernard}   

\talk{B.\,H. Schoch}{ Polarized 
Total Photoabsorption Cross Sections}{abs:schoch}   

\talk{W. Korsch}{ Spin Structure of the Nucleon (at low $Q^2$)
--- Experiment}{abs:korsch}    

\talk{U. M\"uller}{ Results from the Mainz Microtron}{abs:mueller}    

\talk{H. Cal\'{e}n}{ WASA Physics -- Results and Perspectives}{abs:calen}   

\talk{H. Neufeld}{ Isospin Violation in Semileptonic Kaon Decays}
{abs:neufeld}   

\talk{M. Knecht}{ The Anomalous Magnetic Moment of the Muon}
{abs:knecht}    

\talk{J. Schweizer}{ Hadronic Atoms}{abs:schweizer}    

\talk{J. Schacher}{ Results from DIRAC}{abs:schacher}     

\talk{D. Gotta}{ PSI Results from Hadronic Atoms}{abs:gotta}    

\talk{J.\,A. Oller}{ The Chiral Unitary Approach} {abs:oller}   

\talk{E. Epelbaum}{ Improving the Convergence of 
Chiral EFT for Few--Nucleon Systems}{abs:epelbaum}     

\talk{H.-W. Hammer}{ Limit Cycle Physics}{abs:hammer}    

\talk{J.-W. Chen}{ Nuclear Electroweak Processes in Effective Field Theories}
{abs:chen}    

\mbox{}

\vspace{10mm}
\mbox{}\hfill {\bf Page\footnote{n.r.: abstract not received.}}

\vspace{7mm}
\notalk{R. Sommer}{ Lattice QCD}{ }    

\talk{S.\,R. Beane}{ Nuclear Physics and Lattice QCD}{abs:beane1}    

\talk{I.\,W. Stewart}
{ The Soft-Collinear Effective Field Theory}{abs:stewart}    

\talk{N. Kaiser}{ Chiral Dynamics of Nuclear Matter}{abs:kaiser}    

\talk{J. Bijnens, A. Farilla}{ Summary of the Working Group I: 
Goldstone Boson Dynamics}
{abs:wg1}      

\talk{R. Beck, T.\,R. Hemmert}{ Summary  of the Working Group II: 
Meson-Baryon Dynamics}{abs:wg2}     

\talk{P. Bedaque, N. Kalantar-Nayestanaki}{ Summary of the Working Group III: 
Few-Body Dynamics}{abs:wg3a}     

\talk{A. Rusetsky}{ Summary of the Working Group III: 
Hadronic Atoms}{abs:wg3b} 

\talk{E. Pallante, M. Savage}{ Summary of the  Working Group IV: 
Chiral Lattice Dynamics}{abs:wg4}     

\newpage

\centerline{\large\bf Working Group I: Goldstone Boson Dynamics}

\vskip.5cm
\centerline{Convenors: Johan Bijnens and Ada Farilla}

\mbox{}\hfill {\bf Page}

\vskip.5cm
\talk{S. Giudici}{ NA48 Results on $\chi PT$}{abs:giudici}

\talk{G. Ecker}{ $K \to 2 \pi$, $\epsilon' / \epsilon$ 
and Isospin Violation}{abs:ecker}

\talk{J. Bijnens}{ $K \to 3 \pi$ in Chiral Perturbation Theory}{abs:bijnens1}

\talk{A.\,A. Poblaguev}{ E865 Results on $K_{e3}$}{abs:poblaguev}

\talk{V. Obraztsov}{ High Statistics Study of 
the $K^- \to e \nu \pi^0$ and
 $K^- \to \pi^- \pi^0 \pi^0$ 
\\ \mbox{}\ \ \ \
Decays with ISTRA+ Setup}
{abs:obraztsov}

\talk{V. Anisimovsky}{ Measurement of the $K_{e3}$ Form Factors Using Stopped
Positive Kaons}{abs:anisimovsky}

\talk{J. Bijnens}{ $K_{l3}$ at Two Loops in ChPT}{abs:bijnens2}

\talk{E. Frle\v{z}}{ Pion Beta Decay, $V_{ud}$ and $\pi \to l \nu\gamma$}
{abs:frlez}

\talk{M.\,A. Moinester}{ A New Determination of the $\gamma \pi \to \pi\pi$ 
Anomalous Amplitude Via \\  \mbox{}\ \ \ \
$\pi^-  e \to \pi^- e\, \pi^0$ Data}{abs:moinester}

\talk{A. Gasparian}{ The $\pi^0$ Lifetime Experiment 
and Future Plans at JLab}{abs:gasparian}

\talk{M. B\"uchler}{ Renormalization Group Equations for Effective Field
Theories}{abs:buechler}

\talk{P. Dhonte}{ Scalar Form Factors to $O(p^6)$  in SU(3) Chiral
  Perturbation Theory}{abs:dhonte}

\talk{I. Scimemi}{ Hadronic Processes and Electromagnetic Corrections}
{abs:scimemi}

\talk{R. Unterdorfer}{ The Generating Functional of Chiral SU(3)}
{abs:unterdorfer}

\talk{G. Colangelo}{ Hadronic Contributions to $a_\mu$ below 
1 GeV}{abs:colangelo1}
\newpage
\mbox{}

\vspace{5mm}

\mbox{}\hfill {\bf Page}

\vspace{5mm}

\talk{F. Nguyen}{ The Measurement of the Hadronic Cross Section 
with KLOE\\  \mbox{}\ \ \ \
Via the Radiative Return}{abs:nguyen}

\talk{M. Davier}{ Hadronic Vacuum Polarization: the $\mu$ Magnetic Moment}
{abs:davier}

\talk{B. Shwartz}{ CMD-2 Results on $e^+ e^- \to$ hadrons}{abs:shwartz}

\talk{P. B\"uttiker}{ Roy--Steiner Equations for $\pi K$ 
Scattering}{abs:buettiker}

\talk{F.\,J. Yndur\'ain}{ Comments on some Chiral-Dispersive Calculations
 of $\pi\pi$ Scattering}{abs:yndurain}

\talk{G. Colangelo}{ On the Precision of the Theoretical 
Calculations for $\pi \pi$ Scattering}{abs:colangelo2}

\talk{J. Stern}{ From Two to Three Light Flavours}{abs:stern}

\talk{S. Descotes-Genon}{ Bayesian Approach to the Determination of 
$N_f$=3 Chiral Order
\\ \mbox{} \ \ \ \
Parameters}{abs:descotes}

\talk{B. Borasoy}{ Hadronic Decays of $\eta$ and $\eta'$}{abs:borasoy}

\talk{E. Oset} { $\Phi$ Radiative Decay to Two Pseudoscalars}{abs:oset2}

\talk{B. Di Micco}{ $\eta$ Decays Studies with KLOE}{abs:dimicco}

\talk{L. Roca}{ $\eta \to \pi^0 \gamma \gamma$  Decay within a Chiral
Unitary Approach}{abs:roca}


\newpage
\centerline{\large\bf Working Group II: Meson-Baryon Dynamics}

\vskip.5cm
\centerline{Convenors: Reinhard Beck and Thomas R. Hemmert}
\mbox{} \hfill {\bf Page}
\vskip.5cm

\centerline{\bf Nucleon Spin Structure / GDH}

\talk{D. Drechsel}{  GDH, Nucleon Spin Structure 
and  Chiral Dynamics: Theory}{abs:drechsel}
        
\talk{A. Braghieri}{ Helicity Dependence of Pion Photo-Production and the
GDH Sum Rule}{abs:braghieri}
        
\vskip.8cm
\centerline{\bf Real Compton Scattering / VCS}
        
\talk{H. Fonvieille}{ Virtual Compton Scattering at Low Energy and the
  Generalized 
\\ \mbox{}\ \ \ \  Polarizabilities of the Nucleon}
{abs:fonvieille}
        
\talk{R.\,P. Hildebrandt}{ Dynamical Polarizabilities from (Polarized) 
Nucleon  Compton
\\ \mbox{}\ \ \ \
Scattering}{abs:hildebrandt}
        
\talk{L. Fil'kov}{ Radiative Pion Photoproduction and Pion Polarizability}
{abs:filkov}

\vskip.8cm
        \centerline{\bf Strangeness Production/SU(3)}

\talk{K.-H. Glander}{ Strangeness Production from ELSA}{abs:glander}
        
\talk{B. Raue}{ Strangeness Production at Jefferson Lab}{abs:raue}
        
\talk{E. Oset}{ Unitarized Chiral Dynamics: SU(3) and Resonances}{abs:oset1}

\vskip.8cm
        \centerline{\bf Pion (Electro) Production and Chiral Dynamics}
        
\talk{H. Merkel}{ Threshold Pion Electroproduction: Update on Experiments}
{abs:merkel}
        
\talk{M. Kotulla}{ Two $\pi^0$ Production at Threshold}{abs:kotulla}
        
\talk{S.\,N. Yang}{ Threshold $\pi^0$ Photo- and Electro-Production}{abs:yang}

\newpage
\mbox{}

\vskip.5cm
        \centerline{\bf Hadronic Reactions / Sigma Term}
\mbox{} \hfill {\bf Page}

\talk{R.\,A. Arndt}{ Partial Wave Analysis of Pion-Nucleon
Scattering below \\
\mbox{}\ \ \ \ T$_{\rm LAB}$ = 2.1 GeV}{abs:r.arndt}
        
\talk{P. Piirola}{ Towards the Pion-Nucleon PWA}{abs:piirola}

\talk{M.\,F.\,M. Lutz}{ Meson-Baryon Scattering}{abs:lutz}
        
\talk{M. Frink}{ Order Parameters of Chiral Symmetry Breaking from
\\ \mbox{}\ \ \ \
Meson--Baryon Dynamics}{abs:frink}
        
\vskip.8cm
        \centerline{\bf Chiral Dynamics and Nucleon Resonances}
        
\talk{A.\,M. Bernstein}{ Hadron Deformation and Chiral Dynamics}{abs:bernstein}
        
\talk{V. Pascalutsa}{ Chiral Dynamics in the $\Delta$(1232) Region}
{abs:pascalutsa}

\vskip.8cm
        \centerline{\bf Recent Formal Developments}

\talk{R. Lewis}{Lattice Regularized Chiral Perturbation Theory}{abs:lewis2}
        
\talk{S. Scherer}{ Renormalization and Power Counting in 
Manifestly Lorentz-invariant
\\ \mbox{}\ \ \ \ Baryon Chiral Perturbation Theory}{abs:scherer}
        
\talk{J. Gegelia}{ Baryon ChPT with Vector Mesons}{abs:gegelia}
        
\talk{P.\,J. Ellis}
{ Baryon ChPT with Infrared Regularization: Update}{abs:ellis}
        
\talk{U.-G. Mei{\ss}ner}{ Cutoff Schemes in Chiral Perturbation Theory and
the Quark Mass \\ \mbox{}\ \ \ \ 
Expansion of the Nucleon Mass}{abs:meissner} 

\newpage
\centerline{\large\bf Working Group III:  Few-Body Dynamics and Hadronic Atoms}

\vskip.5cm
\centerline{Convenors: Paulo Bedaque, Nasser Kalantar-Nayestanaki 
and Akaki Rusetsky}
\vskip.5cm

\centerline{\bf Few-Body Dynamics I}
\mbox{}\hfill {\bf Page}

\talk{U. van Kolck}{ Charge Symmetry Breaking in Pion Production}{abs:vanKolck}
        
\talk{E.\,J. Stephenson}{ Observation of the Charge Symmetry Breaking
$d$+$d$ $\to$ $^4$He+$\pi^0$
\\ \mbox{}\ \ \ \
 Reaction near Threshold}{abs:stephenson}
        
\talk{A. Nogga}{ Probing Chiral Interactions in Light Nuclei}{abs:nogga}
        
\talk{F. Myhrer}{ Solar Neutrino Reactions and Effective Field Theory}
{abs:myhrer}

\talk{S. Ando}{ Solar-Neutrino Reactions on Deuteron in 
EFT* and\\ \mbox{}\ \ \ \  Radiative Corrections of Neutron Beta Decay}
{abs:ando}
        
\talk{M.\,C. Birse}{ EFT's for the Inverse-Square Potential and Three-Body 
Problem}{abs:birse}

\vskip1.2cm
        \centerline{\bf Hadronic Atoms}

\talk{M. Cargnelli}{ DEAR - Kaonic Hydrogen: First Results}{abs:cargnelli}
        
\talk{J. Gasser}{ Comments on $\bar K N$ Scattering}{abs:gasser}
        
\talk{A.\,N. Ivanov}{ On Pionic and Kaonic Hydrogen}{abs:ivanov}
        
\talk{P. Zemp}{ Deser-type Formula for 
Pionic Hydrogen}{abs:zemp}

\talk{L. Nemenov}{ Atoms Consisting of $\pi^+\pi^-$ and $\pi K$ 
Mesons}{abs:nemenov}
        
\talk{L. Girlanda}{ Deeply Bound Pionic Atoms: Optical Potential at $O(p^5)$ 
in ChPT}{abs:girlanda}

\newpage
\mbox{}

\vskip1.2cm
        \centerline{\bf Few-Body Dynamics II}
\mbox{}\hfill {\bf Page}

\talk{H.\,W. Grie{\ss}hammer}{ All-Order Low-Energy Expansion in the 
3-Body System}{abs:griesshammer}

\talk{T. Black}{ New Results in the Three and Four Nucleon Systems} 
{abs:black}
        
\talk{H. Paetz gen. Schieck}{ $N d$ Scattering at Low Energies}{abs:schieck}
        
\talk{W. Tornow}{Nucleon-Deuteron Scattering at Low Energies}{abs:tornow}

\talk{S. Kistryn}{ Three-Nucleon System Dynamics Studied in the
$d p$-Breakup
\\ \mbox{} \ \ \ \
 at 130 MeV}{abs:kistryn}
        
\talk{J.\,G. Messchendorp}{ Few-Body Studies at KVI}{abs:messchendorp}

\talk{H. Sakai}{ $d p$ Elastic Scattering Experiments}{abs:sakai}

\vskip1.2cm
        \centerline{\bf Few-Body Dynamics III}

\talk{J.\,F. Donoghue}
{ Nuclear Binding Energies and Quark Masses}{abs:donoghue2}
        
\talk{F. Rathmann}{Proton-Proton Scattering Experiments}{abs:rathmann}
        
\talk{C. Hanhart}{ Systematic Approaches to Meson Production in $N N$ and 
$d d$ Collisions}{abs:hanhart}
        
\talk{D. Cabrera}{ Vector Meson Properties in Nuclear Matter}{abs:cabrera}

\talk{A. Schwenk}{ A Renormalization Group Method for Ground State 
Properties\\ \mbox{}\ \ \ \  of Finite Nuclei}{abs:schwenk}
        
\talk{L. Platter}{ The Four-Body System in EFT}{abs:platter}
        
\talk{H. Krebs}{ Electroproduction of Neutral Pions 
Off the Deuteron Near Threshold} {abs:krebs}

\newpage
\centerline{\large\bf Working Group IV: Chiral Lattice Dynamics}

\vskip0.5cm
\centerline{Convenors: Elisabetta Pallante and Martin Savage}
\vskip0.5cm

\centerline{\bf  Formal Lattice Issues}
\mbox{}\hfill {\bf Page}

\talk{I. Montvay}{ Partially Quenched Chiral Perturbation Theory 
and Numerical
\\ \mbox{}\ \ \ \  Simulations}{abs:montvay}
        
\talk{C. Aubin}{ Pion and Kaon Properties in Staggered Chiral
Perturbation Theory}{abs:aubin}
        
\notalk{R. Mawhinney}{ Recent Developments in Domain Wall Fermions}{ }
        
\talk{S. Sint}{ Twisted Mass QCD and Lattice Approaches to the 
$\Delta I = 1/2$ Rule}{abs:sint}
        
\talk{C. Hoelbling}{ Overlap Phenomenology}{abs:hoelbing}
 
\vskip1.2cm 
\centerline{\bf Mesons Properties from the Lattice}

\talk{N. Shoresh}{ Lattice QCD and Chiral Effective Lagrangians}{abs:shoresh}
        
\talk{D. Be\'cirevi\'c}{ Chiral Perturbation Theory, Heavy-light Mesons and 
Lattice QCD}{abs:becirevic}
        
\talk{S. D\"urr}{ The Pion Mass in a Finite Volume}{abs:duerr}
        
\talk{G. Villadoro}{ The Lattice Determination of $K\to \pi\pi$ 
in the $I=0$ Channel}{abs:villadoro}
        
\talk{T. Chiarappa}{ Investigation of the $ \epsilon$ Regime of ChPT: 
Meson Correlation Functions}{abs:chiarappa}
        
\talk{S. Peris}{ $\epsilon'/ \epsilon$ and the $\Delta $I=1/2 Rule in the 
$1/N_c$ Expansion}
{abs:peris}
 
\newpage
\mbox{}
 
\vskip1.2cm
\centerline{\bf Nucleon Properties from the Lattice}
\mbox{}\hfill {\bf Page}

\talk{M. G\"ockeler}{ Nucleon Form Factors from the QCDSF Collaboration}
{abs:goeckeler}
        
\talk{S.\,R. Beane}{ N and NN Properties in PQQCD}{abs:beane2}
        
\talk{D. Arndt}{ Partially Quenched Chiral Perturbation Theory 
for the Baryon and
\\ \mbox{}\ \ \ \ 
Meson Charge Radii}{abs:d.arndt}
        
\talk{R. Lewis}{ Strange Matrix Elements in the Nucleon}{abs:lewis1}
        
\talk{M. Procura}{ Nucleon Mass and Nucleon Axial Charge 
from Lattice QCD}{abs:procura}
 
\vskip1.2cm
\centerline{\bf Future Directions and Focus}

\talk{M. Golterman}{ What should be addressed in the next five 
years}{abs:golterman}


\newabstract 
\label{abs:donoghue1}

\begin{center}
{\large\bf The Physics of the Chiral Effective Field Theory}\\[0.5cm]
John F. Donoghue\\[0.3cm]
Dept. of Physics, University of Massachusetts,\\
Amherst, MA 01003, USA\\[0.3cm]
\end{center}

In this talk I looked at the effective field theory aspects of chiral
perturbation theory. We have known for a long time about the physics
that determines the low energy constants that appear in the chiral
lagrangian - these are very often dominated by the effects of the
nearest resonances. However, effective field theory goes beyond the
lagrangian by including the quantum effects that come from loop
corrections. The physics of chiral loops is less understood and the
talk focussed on this aspect. Some of the conclusions are perhaps
controversial, but I presented the experimental evidence that supports
them. A copy of the full talk will be available on my web site,
http://www.physics.umass.edu/jdonoghue/ , for the moderate future.

Effective field theory is a procedure for extracting the low energy
predictions of a theory. An effective field theory will contain an
energy scale that defines the separation of low energy (where the
effective field theory is valid) from high energy (where a more
complete theory is needed). Loop diagrams always involve the
integration over all momentum, including momenta that occur beyond
this separations scale. These high energy components of loops in the
effective field theory will not be correct because the effective field
theory is no longer valid/accurate at these high energies. However,
since this inaccuracy arises from short distance physics, it can be
corrected for by an adjustment of the local terms in the chiral
lagrangian.  Nevertheless, it is interesting to explore where this
separation occurs in loop processes and to see when the loops are
accurate.

I first explored loops as a function of energy. In very many low
energy processes, loops are relatively unimportant - the main
ingredients to the processes are the low energy constants of the
chiral lagrangian. Indeed, in the classic meson sector, the only
situation where loops are important is when the two intermediate pions
can be in an S-wave. The clearest probe of this physics that I know of
is the reaction $\gamma \gamma \to \pi^0 \pi^0 $, which to order $E^4$
occurs purely through loops.  In this case, the two photon initial
state provides a variable energy to probe the loop diagram.  Here we
have the one-loop and two-loop chiral analysis, plus data, as well as
a dispersive analysis based on the Omnes function. (References can be
found in \cite{two}.) We see clearly that the one-loop analysis fails
at $\sqrt{s}= 600$~MeV. The reason for this is clear - loop diagrams
are made of tree level vertices and the lowest order $\pi\pi$ S-wave
scattering amplitude fails at this energy. A two loop calculation or
the use of the Omnes function can improve results as one approaches
this energy, but still 600~MeV appears as the scale where the
effective field theory breaks down in S-wave pion loops.

I then explored loops as a function of mass. For this exercise, the
dispersive analysis of the $V-A$ vacuum polarization functions is
useful because the Weinberg and DMO sum rules allow one to probe the
chiral logarithms using data (A portion of this analysis has been
published in \cite{bpaper}.). The point here is that chiral logarithms
arise from the threshold behavior of the dispersive spectral
functions, yet in the data one can also see the onset of the high
energy regime (which occurs at the rho mass in this channel) beyond
which the effective theory is no longer effective. The region between
the threshold and the onset of the high energy regime is the region
where the effective theory can be applied. We can probe this region as
a function of the meson mass by using the chiral prediction for the
spectral function in this region. It is easy to see that kaon loops
fall outside the region of validity of the effective field
theory. Their threshold starts at $\sqrt{s} = 1$~GeV, which is well
beyond the separation scale. This and other evidence points to the
conclusion that kaon loops are not valid parts of the chiral effective
field theory. Moreover, one can compare the chiral parameterization
for the loop process (in terms of $m^2$ and $m^2 \ln m^2$) and the low
energy part of the dispersive result. Doing this one learns that the
chiral representation is excellent near the pion mass, but fails near
meson masses of $m= 250$~MeV. This is lower than most of us would
initially expect. However part of the difference is just a factor of
two which arises from the fact that the threshold occurs at $\sqrt{s}
= 2 m$. Thus for this mass, the kinematic threshold suppression is not
yet overcome before passing out of the low energy regime.

Finally I discussed regularization schemes, in particular the
difference between scale-invariant schemes (variants of dimensional
regularization) and scale-dependent schemes (such as various cutoff
methods). In the region where the effective theory is valid, there
cannot be a difference due to this scheme dependence. However, scale
dependent schemes have the ability to probe how much of the loop
physics arises from short distances - this is a useful
diagnostic. Various examples of this procedure reinforce the
conclusion that kaon loops lie outside the range of validity of the
effective field theory.

My conclusions were: Pion loops are only important in the $\pi\pi$ S
wave; The transition from low-energy to high-energy happens around E =
600 MeV and m = 250 MeV; Pion loops are firmly in EFT region; If you
are doing kaon loops, you are not doing effective field theory; Nature
uses a cutoff regularization.

\newabstract 
\label{abs:goity}

\begin{center}
{\large\bf {\boldmath${{\pi^0\to\gamma\gamma}}$} and the PRIMEX}\\[0.5cm]
Jose L. Goity \\[0.3cm]
Department of Physics, Hampton University,\\
Hampton, VA 23668, USA, and\\
Jefferson Lab, Newport News, VA 23606, USA.\\[0.3cm]
\end{center}

In the chiral limit, the amplitude of the decay $\pi^0\to\gamma\gamma$
is determined by the anomaly induced on the neutral iso-triplet axial
current by the EM field. The amplitude is thus given in terms of the
only available quantities in that limit, namely, the fine structure
and the pion-decay constants. The prediction for the $\pi^0$ width
using that decay amplitude is in good agreement with the current
experimental determination, which has a rather poor precision of about
8\%. The deviations from the chiral limit prediction can be
theoretically pinned down with rather good precision, as several
analyses have shown \citethree{BM}{GBH}{AM}. The chief correction is induced
by state mixing: due to isospin breaking the $\pi^0$ is not a pure
isospin state, but it has an admixture with the two iso-singlet
states, identified in the chiral limit with the $\eta$ and $\eta'$
mesons. The mixing with the first is driven by the ratio
$(m_u-m_d)/(m_s-\hat{m})$, and that with the second by the ratio $N_c
(m_u-m_d)/\Lambda_\chi$. It turns out that these two mixing effects
give each an enhancement of about 1\% to the $\pi^0$ amplitude. Such
an enhancement had been pointed out long ago \cite{Kitazawa}, and the
more recent analyses have put it on solid ground, with calculations of
strong interaction \cite{GBH} and electromagnetic \cite{AM}
corrections at NLO in the chiral and the fine structure constant
respectively.

The NLO strong interaction corrections were analyzed in the framework
of ChPT combined with the $1/N_c$ expansion \cite{KL} (to allow for a
consistent inclusion of the $\eta'$). In this approach the mixing
corrections mentioned above are LO.  The NLO corrections to the
$\pi^0$ amplitude stem from order $p^2$ and order $1/N_c$ corrections
to the decay constants and order $p^4$, $p^2/N_c$ and $1/N_c^2$ to the
masses of the $\pi^0$, $\eta$ and $\eta'$, as well as a correction
from the order $p^6$ Lagrangian of unnatural parity. These corrections
are dominated by effects proportional to $m_s/\Lambda_\chi$ and
$1/N_c$, which are expected to correct the enhancement that results
from the LO result by an amount of the order of 30\%. The actual
analysis, based on a global fit involving as inputs the meson masses
and decay constants and the $\eta$ and $\eta'$ partial widths to two
photons, shows that the NLO correction to the $\pi^0$ amplitude is
within that expectation. The strong interaction effects are then
determined to give an increase of the $\pi^0\to\gamma\gamma$ width
(with respect to the result obtained using the chiral limit amplitude)
of $4.7\pm 0.8\%$. The chief uncertainty resides here in the input
required for the ratio $R=(m_s-\hat{m})/(m_u-m_d)$. The uncertainty in
the EM piece of the $K^+-K^0$ mass difference leads to the dominant
uncertainty in the determination of $R$.  The other sub leading
corrections to the $\pi^0$ amplitude are EM. It was shown that at
order $\alpha^2$ the amplitude is only affected via EM corrections to
the masses and decay constants. The analysis entails the estimate of a
few low energy constants, and the result is that the EM effects tend
to reduce the $\pi^0$ width by roughly 0.03 eV (or 0.4 \%).

In summary, the theoretical prediction that results from the analyses
just described is
\[\Gamma_{\pi^0\to\gamma\gamma}=8.11\pm 0.07 \;{\rm eV}.\]

The upcoming Jefferson Lab experiment PRIMEX \cite{PRIMEX} will
determine the $\pi^0$ width by means of the Primakoff effect. The
coherent photo-production of the $\pi^0$ off the Coulomb field of a
nucleus ( $^{12}C$, $^{120}Sn$ and $^{208}Pb$) is dominated at small
angles by that effect. The Primakoff cross section gives direct access
to the $\pi^0\to\gamma\gamma$ width because it is proportional to
it. The measurement of this cross section to high accuracy is
therefore the goal of this experiment.  The PRIMEX represents a new
generation of such experiments (three previous experiments of this
kind have a very important weight in the present average value of the
width). It has a tagged bremmstrahlung photon beam with uncertainties
of less than 1\% in photon flux and $10^{-3}$ in energy, which
represents a big improvement over previous experiments. Also, a high
resolution hybrid $Pb$ glass and crystal calorimeter provides for a
precise determination of the $\pi^0$ production angle (uncertainty of
the order of $0.01^o$). With this the maping the Primakoff peak is
possible and in this way the determination of other production
mechanisms can be also determined, such as the coherent production via
strong interactions whose interference with the Primakoff effect gives
the dominating deviation to be taken into account.  With this setup,
the PRIMEX aims at a determination of the $\pi^0$ width at the 1.4\%
level of precision. This experimental accuracy implies that the
enhancement effect of over 4\% predicted by theory can be put to the
test and in this way uncover experimentally chiral symmetry breaking
effects in the $\pi^0$ lifetime.  The PRIMEX is scheduled to run in
early to mid 2004.

\newabstract 
\label{abs:primavera}

\begin{center}
{\large\bf Results from DA$\Phi$NE}\\[0.5cm]
{\bf Margherita Primavera}$^1$ on behalf 
of the {\bf KLOE Collaboration}\cite{kloecol}\\[0.3cm]
$^1$INFN Lecce, 
Via Arnesano, 73100 Lecce, Italy\\[0.3cm]
\end{center}

DA$\Phi$NE\cite{dafne}, the Frascati $\phi$ factory, is an $e^+e^-$
accelerator operating at $W\approx m_\phi\approx 1.02\ GeV/c^2$. The
design luminosity is $\approx 5\cdot 10^{32}\ cm^{-2}s^{-1}$.  One of
the two machine interaction regions is occupied by the KLOE experiment
and the other one is shared in time between the two experiments DEAR
and FINUDA.

After the 2002 data taking campaign, in which data of kaonic nitrogen
($\sim 17\ pb^{-1}$) and kaonic hydrogen ($\sim 60\ pb^{-1}$) has been
collected, DEAR \cite{dear} has measured the three transitions
$7\rightarrow 6, 6\rightarrow 5, 5\rightarrow 4$ in kaonic nitrogen
and performed a feasibility test of a precision measurement of $K^-$
mass, and has obtained a preliminary measurement of the shift
$\varepsilon$ and the width $\Gamma$ of the $K_\alpha$ line of kaonic
hydrogen.

FINUDA \cite{finuda} physics program is concentrated on $\Lambda$
hypernuclei spectroscopy and decays, especially the non mesonic
decays, which are only possible in a nucleus. The detector has been
installed during this year and is now ready to collect data.

KLOE \cite{kloe} is a general purpose detector, optimized to study CP
simmetry violation in the neutral kaon system. It consists of a large
drift chamber \cite{camera} surrounded by a lead/scintillating fiber
electromagnetic calorimeter \cite{calor}.  The detector is immersed in
a 0.52 T solenoidal magnetic field, provided by a superconducting
coil.  The momentum resolution of the drift chamber is $\sigma_p/p\leq
0.4$\% on tracks with $45^\circ < \theta < 135^\circ$, while
calorimeter energy and time resolution are, respectively, $\sigma_E/E
= 5.7\%/\sqrt{E(GeV)}$ and $\sigma_t = 54\ ps/\sqrt{E(GeV)}\oplus 50 \
ps$. The whole data set collected by KLOE during 2001-2002 amounts to
about 450 $pb^{-1}$. Much of this data set has been analized and has
yielded improved results on $K_S$ and radiative decays, as well as
studies concerning a wide range of topics in kaon and hadronic
physics.

{\bf\em Kaon physics at KLOE.}  At DA$\Phi$NE, a tagging tecnique can
be applied in order to signal the presence of $K_S$($K_L$) or
$K^+$($K^-$) in an event with a $K_L$($K_S$) or $K^-$($K^+$) in the
opposite hemisphere. KLOE has measured the ratio
$\Gamma(K_S\rightarrow\pi^+\pi^-(\gamma))/
\Gamma(K_S\rightarrow\pi^0\pi^0)=2.236\pm
0.003\pm 0.015$ \cite{kshpipi}, from which the difference of $\pi\pi$
phase shifts in $K\rightarrow\pi\pi$ transitions with I=0 and 2 has
been determined: $\chi_0-\chi_2= (48\pm 3)^\circ$, thus improving the
agreement with present predictions. KLOE has improved its previous
measurement of BR($K_S\rightarrow\pi^\pm e^\mp\nu$)\cite{kshsemil}
obtaining $(6.81\pm0.12\pm 0.10)\cdot 10^{-4}$ and the first-ever
measurement of the $K_S$ semileptonic asymmetry $A_S=(19\pm 17\pm
6)\cdot 10^{-3}$. For the ratio $\Gamma (K_L\rightarrow\gamma\gamma)/
\Gamma(K_L\rightarrow 3\pi^0)$ KLOE obtains the value $
(2.793\pm0.022\pm 0.024)\cdot 10^{-3}$\cite{klgg} and is currently
evaluating the BR of the decay channels of $K_L$ to charged
particles. On 187 pb$^{-1}$ of 2001-2002 data, KLOE measures
$BR(K^\pm\rightarrow\pi^\pm\pi^0\pi^0) = 1.781\pm 0.013\pm 0.016\%$
\cite{taup}. KLOE should be able to measure all $K_{l3}$ BR's, thus
improving the experimental error on the CKM matrix element $\mid
V_{us}\mid$.

{\bf\em $\phi$ decays and continuum physics at KLOE.} KLOE has
recently published \cite{phirhopi} an analysis of the decay
$\phi\rightarrow\pi^+\pi^-\pi^0$, mainly proceeding through $\rho\pi$
final states.  The fit to the Dalitz plot provides values for the
$\rho$ mass and width: $m_\rho= 775.8\pm 0.5\pm 0.3\ MeV$ and
$\Gamma_\rho=143.9\pm 1.3\pm 1.1\ MeV$. From the ratio of BR's for the
decays $\phi\rightarrow\eta'\gamma$ and $\phi\rightarrow\eta\gamma$
KLOE has obtained a value of the pseudoscalar mixing angle, $\varphi_P
= (41.8_{-1.6}^{+1.9})^\circ$ and a limit on the gluonium content of
the $\eta'$ \cite{etaetap}. KLOE has analized, using 17 $pb^{-1}$ of
2000 data, \citetwo {f0}{a0} the decays $\phi\rightarrow\pi^0\pi^0\gamma$
and $\phi\rightarrow\eta\pi^0\gamma$, where the dominant contributions
come from production and decay of the scalar mesons $f_0$ and $a_0$,
respectively. Finally, KLOE is concluding the measurement of
$\sigma(e^+e^-\rightarrow\pi^+\pi^-)$ for $0.3 < s_\pi < 1\ GeV^2$
\cite{sigmaadr} by using initial state radiation and a MC generator to
relate $\sigma(e^+e^-\rightarrow\pi^+\pi^-\gamma)$ to
$\sigma(e^+e^-\rightarrow\pi^+\pi^-)$. The KLOE data provide
$a_\mu^{hadr}\times10^{10}(0.37<s_\pi<0.95\ GeV^2)= 374.1\pm 1.1\pm
5.2 \pm 2.6_{th}$ plus a $0-2\%$ correction for loss of ISR+FSR events
in the simulation.

\newabstract 
\label{abs:bernard}

\begin{center}
{\large\bf Spin Structure of the Nucleon  --- Theory}\\
[0.5cm]
{\bf V. Bernard}$^1$, T.R. Hemmert$^2$ and Ulf-G. Mei\ss ner $^3$
\\[0.2cm]
$^1$Laboratoire de Physique Th\'eorique, F67084 Strasbourg, France\\
[0.1cm]
$^2$Technische Universit{\"a}t M{\"u}nchen, D-85747 Garching, Germany\\
[0.1cm]
$^3$Universit\"at Bonn, D-53115 Bonn,
Germany.\\[0.1cm]
\end{center}

Understanding the spin structure of the nucleon is a central topic of 
present nuclear and particle physics interest. In the low energy region
it allows to test QCD chiral dynamics with processes involving two 
photons. It is also of particular interest
to obtain an understanding of how in QCD the
transition from the non-perturbative to the perturbative regime takes place,
guided by the precise experimental mapping of spin-dependent observables
from low momentum transfer to the multi-GeV region, as it is one of the main
thrusts of the research carried out at Jefferson Laboratory (see W. Korsch, 
these mini-proceedings and \citetwo{CLAS}{E94010}). 
The theoretical investigation of the nucleon's spin structure in the 
non-perturbative regime of QCD can be done using a dispersive analysis
(see D. Drechsel,  these mini-proceedings). We are utilizing
chiral perturbation theory (CHPT) which is based on the spontaneous and 
explicit chiral symmetry breaking QCD is supposed to undergo. 
However, baryon CHPT is complicated by the fact that the nucleon mass $m$
does not vanish in the chiral limit and thus introduces a new mass scale apart
from the one set by the quark masses. Therefore, any power of the quark masses
can be generated by chiral loops, spoiling the one-to-one
correspondence between the loop expansion and the one in the small parameter
$q$ denoting the pion mass or a small external momenta.  
One method to overcome this is  
heavy baryon chiral perturbation theory, (HBCHPT)
where the nucleon  mass is
transformed from the propagator into a string of vertices with
increasing powers of $1/m$. 
However, this method has the disadvantage
 that certain types of diagrams are at odds with strictures from analyticity.
In a fully relativistic treatment,
such constraints from analyticity are automatically fulfilled. It was recently
argued \cite{ET} that relativistic one--loop integrals can be separated
into ``soft'' and ``hard'' parts. While for the former the power counting
as in HBCHPT applies, the contributions from the latter can be absorbed in
certain LECs. The underlying method  is called
``infrared regularization'' \cite{BL}. It was shown that in this scheme one 
actually resums 
graphs with n kinetic insertions on the nucleon line which in the case of
HBCHPT would be $1/m^n$ suppressed. This of course leads to better convergence
of the chiral serie and should allow to increase the range of applicability 
of the theory. However this range is limited since one 
introduces in the calculation of the loop functions an {\it infinite} Feynman
parameter which might result in divergences.
In the case of forward double virtual scattering (V$^2$CS)
which is the process
under investigation here, they appear at $Q^2= 1 {\rm GeV}^2$ 
in principle well beyond the range of validity of CHPT. However dealing with
higher derivatives of the loop functions the influence of the singularity
is felt at smaller values of  $Q^2$, restricting our range to $Q^2_{\rm max}
=0.2 \cdots 0.4 {\rm GeV}^2$.  We have performed an ${\cal O}(q^4)$  parameter 
free calculation of the V$^2$CS 
spin structure functions in the infrared regularization scheme
\cite{BHM}.
These are intimately connected to the ones probed in inelastic 
electroproduction experiments because of unitarity. The $\Delta(1232)$ being 
well known to play a significant role in the spin sector of the
nucleon, we have calculated relativistic Born graphs
to get an estimate of its contribution 
to the spin structure functions.

\vspace{-0.3cm}

\begin{figure} [h]
 \centering
 \begin{minipage}[c]{.45\textwidth}
{A typical result for  the case of the proton is depicted in the figure
which shows the so called generalized Gerasimov-Drell-Hearn sum rule.
The zero crossing around 0.25 GeV$^2$
nicely reproduces the data \cite{CLAS}. The inclusion 
of the $\Delta$ and the vector mesons improves the comparison between theory
and experiment as confirmed by the very preliminary data obtained by the CLAS 
collaboration. However the calculation of the  $\Delta$  is very conservative
and thus leads to large theoretical uncertainties.}
 \end{minipage}
 \hfill
 \begin{minipage}[c]{.50\textwidth}
   \centering
   \includegraphics [scale=.34]{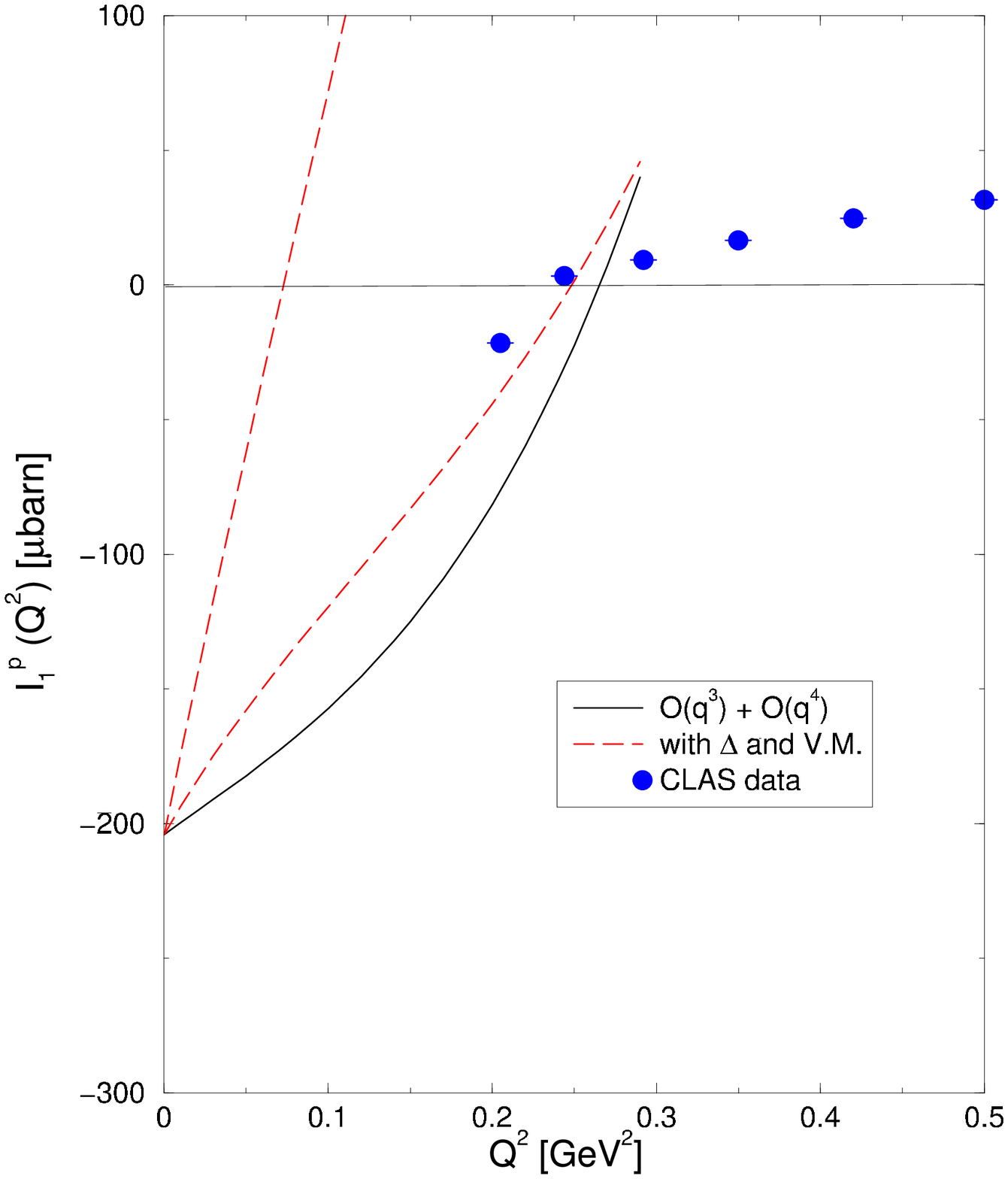}
 \end{minipage}
\end{figure}


\vspace{-0.45cm}

\noindent A fully systematic
treatment is underway, a consistent extension of the 
Lorentz-invariant method to spin-3/2 fields having been formulated \cite{BHM1}.
We have performed a chiral expansion 
of the forward and longitudinal-transverse spin 
polarizabilities and 
found bad convergence in the case of the forward one. The reason rests
entirely on the slow convergence of the pole (Born) graphs. A resummation
of these is needed, work in this direction is in progress.
An understanding of the spin structure of the nucleon in the non perturbative
regime of QCD and of the transition to the perturbative one is clearly
underway.

\newabstract 
\label{abs:schoch}

\begin{center}
{\large\bf Polarized Total Photoabsorption Cross Sections}\\[0.5cm]
Berthold H. Schoch\\[0.3cm]
Physikalisches Institut, Universit\"at Bonn,\\
Nussallee 12, D53115 Bonn, Germany\\[0.3cm]

\end{center}

\vspace{1pt}
\vspace{1pt}Fundamental static properties of the nucleons are connected via
sum rules with the total photon absorption cross sections of the nucleons.
For the case of the unpolarized cross section $\sigma _{T}$ \ the sum of the
electric $\alpha (E_{\gamma })$ and magnetic $\beta (E_{\gamma })$
polarizabilities can be expressed by Baldins sum rule:
\begin{center}
\vspace{1pt}

$\ \ \ \ \ \ \ \ \ \ \ \ \ \ \ \ \alpha +\beta =\frac{1}{2\pi ^{2}}%
\int_{E_{0}}^{\infty }\frac{\sigma _{T}(E_{\gamma }^{^{\prime }})}{E_{\gamma
}^{^{\prime 2}}}dE_{\gamma }^{^{\prime }}$ $\hspace{1in}(1)$
\end{center}
\vspace{1pt} The energy dependence of the energy weighted total absorption
cross section exhibits already the most relevant nucleon excitations for
these static properties.Contributions from non resonant single pion
production and the first and second resonance region dominate the integral.
For the energies above the second resonance region the total \ absorption
cross section can be described by a Regge parametrization which reflects the
dominance of \ diffractive processes. The contributions to the integral from
these processes are small due to the inverse quadratic energy weighting of
the cross section.

By using a circulary polarized photon beam and a longitudinal -- in beam
direction -- polarized target, helicity 3/2 ($\sigma _{3/2}$) and helicity 1/2
($\sigma _{1/2}$) cross sections can be measured.  The knowledge of these
cross sections provides again insight into the origin of static responses for
low energy photon scattering. The GDH sum rule connects the anomalous magnetic
moment, $\kappa _{N}$ ,\ with the difference of the helicity cross sections:
\vspace{1pt}

\begin{center}
$\frac{\pi \cdot e^{2}\cdot \kappa _{N}^{2}}{2m^{2}}=\int_{E_{threshold}}^{%
\infty }\frac{\sigma _{3/2}(E_{\gamma }^{^{\prime }})-\sigma
_{1/2}(E_{\gamma }^{^{\prime }})}{E_{\gamma }^{^{\prime }}}dE_{\gamma
}^{^{\prime }}=I_{GDH}\hspace{1in}(2)$

\vspace{1pt}
\end{center}
\vspace{1pt}

Due to the only inverse weighting with the gamma energy possible contributions
from above the second resonance region of the nucleon may contribute. The
value of $\kappa _{N}$ is well known. However, it is of great interest where
the strength in the excitation spectrum resides in building up $\kappa _{N}.
$The helicity cross sections have been measured at MAMI $(200\leq E_{\gamma
}/MeV\leq 800)$ \cite{Ahrens} and ELSA $(680\leq E_{\gamma }/MeV\leq 2900)$
\cite{Zeitler}.  The cross section, $\sigma _{3/2}(E_{\gamma }^{^{\prime
  }})-\sigma _{1/2}(E_{\gamma }^{^{\prime }})$, differs fom the total
unpolarized cross section ($\sigma _{3/2}(E_{\gamma }^{^{\prime }})+\sigma
_{1/2}(E_{\gamma }^{^{\prime }}))/2$ by an almost complete suppression of the
diffractive processes dominating the total unpolarized cross section for
$E_{\gamma }\succ 1200MeV$. Thus the nucleon resonances stand up more clearly,
up to a visible fourth resonance region. This behaviour of the cross sections
could have been expected by realizing that the anomalous magnetic moment is a
C ( charge conjugation) = $-1$ quantity whereas the polarizabilities transform
like C=1 \cite{Bass}.  This makes a difference concerning the coupling of the
photon to the constituent quarks and the sea quark/meson cloud, respectively.
For the GDH  sum rule the following contributions have been extracted from
the experiments: MAMI$(200\leq E_{\gamma }/MeV\leq 800),$ $I_{GDH}=226\pm 5\pm
12\mu b,$ ELSA $(800\leq E_{\gamma }/MeV\leq 2800),$ $I_{GDH}=28.0\pm
2.1(overall).$ A multipole analysis \cite{Tiator} yields for the energy range,
$E_{\pi -threshold}^{\gamma }\leq E_{\gamma }\leq 200MeV,$ $I_{GDH}$
$=-27.5\mu b.$ Based on Regge extrapolations and fits to deep inelastic
electron scattering the asymtotic high energy contribution has been estimated
to be $I_{GDH}=-15\mu b $\cite{Bianchi}. The sum of these contributions adds
up to $I_{GDH}=211.5\mu b.$ The overall uncertainty amounts to $ \pm 6\%.$
Thus, $I_{GDH}$ agrees quite well with the expected value of $I_{GDH}^{\exp
  ected}=205\mu b.$ Of great interest are the measurements on the neutron.
Measurements on the deuteron have been carried out on ELSA and MAMI up to
$E_{\gamma }=1900MeV$. It remains to be seen, how reliable information on the
neutron can be extracted from these data.  With the knowledge of $\sigma
_{3/2}-\sigma _{1/2}$ the forward spin polarizability $\gamma _{0}^{p}$ can be
extracted via the sum rule:

\vspace{1pt}

\begin{center}
$\gamma _{0}^{p}=-\frac{1}{4\pi ^{2}}\int_{E_{threshold}}^{\infty }\frac{%
\sigma _{3/2}(E_{\gamma }^{^{\prime }})-\sigma _{1/2}(E_{\gamma }^{^{\prime
}})}{E_{\gamma }^{^{\prime 3}}}dE_{\gamma }^{^{\prime }}\hspace{1in}(3)$
\end{center}
The different contributions yield:\newline 
 MAMI$(200\leq E_{\gamma }/MeV\leq
800),
\gamma _{0}^{p} = -\left[ 1.87\pm 0.08(stat)\pm 0.1(syst)\right] \cdot
10^{-4}fm^{4}$, ELSA $ (800\leq E_{\gamma }/MeV\leq 2800),$ $\gamma
_{0}^{p}=-\left[ -2.7\pm 0.2(overall)\right] \cdot 10^{-6}fm^{4}.$ A
multipole analysis \cite{Tiator} yields for the energy
range, $E_{\pi -threshold}^{\gamma }\leq E_{\gamma }\leq 200MeV,$ $\gamma
_{0}^{p}=\left[ 0.9\right] \cdot 10^{-4}fm^{4}.$ The sum amounts to : $\gamma
_{0}^{p}=-\left[ 1\pm 0.1\right] \cdot 10^{-6}fm^{4}.$ This value poses a
challenge for theory. Various recent  calculations are at variance 
with this result \cite{Drechsel}.

\newabstract 
\label{abs:korsch}

\begin{center}
{\large\bf Spin Structure of the Nucleon 
(at low $\mathbf {Q^2}$) --- Experiment}\\[0.5cm]
Wolfgang Korsch$^1$ \\[0.3cm]
$^1$ Dept. of Physics and Astronomy, University of Kentucky, \\
177 Chemistry - Physics Bldg.,\\
Lexington, KY 40506, USA\\ [0.3cm]
\end{center}

Recent progress in the development of polarized few-nucleon
targets and electron beams allows us to perform high precision
measurements on the $Q^2$ evolution of fundamental sum rules 
related to the spin of the nucleon. During the last 
15 years the main focus of nucleon spin structure studies
has been the decomposition of the nucleon spin in terms of 
quarks, gluons, and their orbital angular momenta. However, already in the 
1960's Drell, Hearn, and Gerasimov derived a 
very fundamental sum rule which related the
difference  of the total 
photon absorption cross sections, $\sigma_{1/2}-\sigma_{3/2}$, 
to the square of the anomalous magnetic moments of
the nucleon~cite{gdh}. Here $\sigma_{3/2}$ and $\sigma_{1/2}$ 
refer to the cross sections with the nucleon spin aligned  parallel or
anti-parallel to the spin of the photon. 
This result can be extended using the low energy expansion of the forward
scattering amplitudes in the long wavelength limit of the photon. 
The first two orders of this expansion 
for the transverse-transverse and longitudinal-transverse interference 
amplitudes are given in Eqs. (\ref{eqn1}) and (\ref{eqn2}): 
\begin{eqnarray}%
\label{eqn1}
 {\mathit Re~} {\tilde g}_{TT}(\nu, Q^2)  &=& \frac{2 \alpha}{M^2}
 I_{1}(Q^2) \nu + \gamma_0(Q^2) \nu^3 + {\mathcal {O}}(\nu^5) 
\\
\label{eqn2}
 {\mathit Re~} {\tilde g}_{LT}(\nu, Q^2) 
&=& \frac{2 \alpha}{M^2} Q I_{3}(Q^2) 
+ Q \delta_{LT}(Q^2) \nu^2 + {\mathcal {O}}(\nu^4)
\end{eqnarray}%
 
In the $Q^2 \to 0$ limit $I_1$ reduces to the GDH sum rule ($\propto
\kappa_N^2$). $\gamma_0(0)$ describes the forward spin polarizability of the
nucleon and $I_3(0)$ is proportional to $e_N \kappa_N$ ($e_N$ being the charge
of the nucleon). The quantity $\delta_{LT}(0)$ is the longitudinal-transverse
polarizability.  By means of dispersion relations and the optical theorem the
above amplitudes can be related to spin dependent structure functions and
therefore cross sections~\cite{drechsel}. The $Q^2$ evolution of these static
properties of the nucleon is of fundamental theoretical interest, since the
low $Q^2$ behavior can be calculated within the framework of chiral
perturbation theory (Ch.P.T.)  ~\cite{veronique}.

The first experiments using a polarized electron beam and polarized targets at
Jefferson Lab focused on the $Q^2$ evolutions of the GDH integral for the
proton, deuteron, and the neutron ($^3$He). CEBAF routinely delivers a
polarized beam with polarization values of 75\% or higher and energies up to
5.7 GeV. In Hall B the large acceptance spectrometer, CLAS, was used in
combination with dynamically longitudinally polarized NH3 and ND3 targets to
evaluate these integrals from $\pi$ production threshold to invariant masses,
$W$, up to $\approx$ 3 GeV. Longitudinal asymmetries were measured in
inclusive electron $-$ proton (deuteron) scattering. Fits to world data and
models were used for the unpolarized structure functions and transverse spin
contributions in order to extract the spin structure functions $g_1^{p,d}(x,
Q^2)$. The first moments of $g_1^{p,d}(x, Q^2)$ were extracted in a $Q^2$
range from about 0.1 GeV$^2$ to about 3.5 GeV$^2$~\cite{fatemi}.  It was found
that both, $g_1^{p}$ and $g_1^{d}$ depend strongly on $Q^2$.  $g_1^{p}$
changes sign at a $Q^2$ value around 0.3 GeV$^2$ whereas the sign change for
$g_1^{d}$ appears to occur between 0.5 GeV$^2$ and 0.6 GeV$^2$.  The zero
crossings of the structure functions are an indication of the transition where
non-partonic degrees of freedom such as resonances become dominant.
Therefore, resonance contributions seem to be slightly less important for the
proton.

The high precision data are in nice agreement with chiral
perturbation theory calculations for $Q^2$ values less than 0.2 GeV$^2$. 

Hall A used a dense polarized $^3$He gas target to measure the generalized GDH
integral~\cite{amarian1}, the generalized spin polarizability, and the
longitudinal$-$transverse polarizability for the neutron. The spin dependent
cross sections were measured directly by polarizing the target longitudinally
and transversely to the direction of the incident beam.  Nuclear corrections
were applied according to Ref.~\cite{degliAtti}. Results in a $Q^2$ range from
0.10 GeV$^2$ to 0.90 GeV$^2$ were presented~\cite{amarian}. The GDH integral
for the neutron exhibits a strong $Q^2$ dependence as well.  The MAID
model~\cite{drechsel1} underestimates the GDH integral at low values of $Q^2$
but describes both, $\gamma_0(Q^2)$ and $\delta_{LT}(Q^2)$, quite well.
Ch.P.T. calculations are in good agreement with the data for the GDH integral
and $\gamma_0(Q^2)$ for $Q^2 < 0.2$ GeV$^2$~\cite{veronique}.

The combination of the Hall A and Hall B data will allow us to evaluate the
difference of the first moments $\Gamma_1^p(Q^2) - \Gamma_1^n(Q^2)$.  In the
high $Q^2$ regime this difference is known as the Bjorken sum rule.  It has
practically no sensitivity to sea quarks in the asymptotic limit and is also
less sensitive to the $\Delta$ excitation in the low $Q^2$ regime. Therefore,
the comparison of this difference can serve as a crucial test for theoretical
calculations.

\vspace*{-0.5cm} l

\newabstract 
\label{abs:mueller}

\begin{center}
{\large\bf Results from the Mainz Microtron}\\[0.5cm]
{\bf Ulrich M\"uller}$^1$\\
representing the A1, A2, and A4 collaborations\\[0.3cm]
$^1$Institut f\"ur Kernphysik, Universit\"at Mainz,\\
Johann-Joachim-Becher-Weg 45, 55099 Mainz, Germany\\[0.3cm]
\end{center}

\noindent Some highlights of recent experimental results
at MAMI-B~\cite{Herminghaus} are presented.

The neutron electric form factor $G_{E,n}$
has been measured at momentum transfers of $Q^2 = 0.30$, $0.59$, and
$0.79\,(\mathrm{GeV}/c)^2$ in a $\mathrm{D}(\vec e,e'\vec n)p$
experiment of the A1 collaboration, using a spin precession method
and neutron recoil polarimetry.
Preliminary values of
$G_{E,n} = 
 0.0552 \pm 0.0061\,\hbox{(stat)} \pm 0.0055\,\hbox{(syst)}$,
$0.0469 \pm 0.0073\,\hbox{(stat)} \pm 0.0054\,\hbox{(syst)}$, and
$0.047  \pm 0.009\,\hbox{(stat)} {+0.001 \atop -0.002}\,\hbox{(syst)}$,
have been obtained, respectively~\cite{gen}.

An extensive programme of charged pion electroproduction off the
proton has been performed by the A1 collaboration. The reaction
$\mathrm{H}(e,e'\pi^+)n$ has been measured at a constant invariant
mass of $W = 1125\,\mathrm{MeV}$ and four-momentum transfers of\
$Q^2 = 0.058$, $0.117$, $0.195$, and $0.273\,(\mathrm{GeV}/c)^2$.
For each $Q^2$ value, measurements at different values of~$\epsilon$,
the virtual photon polarisation parameter, allow for a Rosenbluth
separation of longitudinal and transverse cross sections.
Analysis of the data taken in 2000 and 2002 is in progress~\cite{axial}.

Threshold production of neutral pions has been measured in an
$\mathrm{H}(e,e'p)\pi^0$ experiment by the A1 collaboration at a
momentum transfer of $Q^2 = 0.05\,(\mathrm{GeV}/c)^2$ and
$\epsilon = 0.933$.
Cross sections of the structure functions $\sigma_0$,
$\sigma_\mathrm{LT}$, $\sigma_\mathrm{TT}$, and $\sigma_\mathrm{LT'}$
and the helicity asymmetry $A_\mathrm{LT'}$ have been
determined~\cite{pi0}.
The helicity-independent cross sections are described by the
MAID\,2000 model~\cite{maid2000} within their errors.
The magnitude of the transverse-transverse part is significantly
overestimated by \mbox{ChPTh}~\cite{chpth}.

The A2 collaboration used the DAPHNE detector to measure the helicity
dependent exclusive cross sections for the absorption of polarised
photons by longitudinally polarised nucleons.
The cross section integral of the Gerasimov-Drell-Hearn sum rule 
from $200$ to $800\,\mathrm{MeV}$ was calculated from the data as
$I^\mathrm{GDH}_p = (-226 \pm 5\,\hbox{(stat)} \pm
12\,\hbox{(syst)})\,\mathrm{\mu b}$~\cite{gdh}.
After a correction for the missing region below $200\,\mathrm{MeV}$
and by including ELSA data above $800\,\mathrm{MeV}$~\cite{elsa},
a value of
$I^\mathrm{GDH}_p = (-227.5 \pm 5\,\hbox{(stat)} \pm
12\,\hbox{(syst)} \pm 3\,\hbox{(theo)})\,\mathrm{\mu b}$
is obtained.

The A2/TAPS collaboration determined the E2 amplitude of the
$N \to \Delta$ transition in an $\mathrm{H}(\vec\gamma,\pi^0)p$
experiment using linear polarised photons in the energy range from
$200$ to $790\,\mathrm{MeV}$.
At the position of the $\Delta$ resonance,
$E_\gamma = 340\,\mathrm{MeV}$, a preliminary value of
$R_{EM} = (-2.40 \pm 0.16 \pm 0.24)\cdot 10^{-2}$
is obtained~\cite{e2m1}. This is in agreement with a previous
A2/DAPHNE measurement~\cite{e2m1daphne}.

The reaction $\mathrm{H}(\gamma,p\pi^0\gamma')$ has been measured by
the A2 collaboration with the TAPS calorimeter for energies between
$\sqrt{s} = 1221$ and $1331\,\mathrm{MeV}$.
Cross sections differential in angle and energy have been determined
for all particles in the final state in three bins of the excitation
energy.
A value for the magnetic dipole moment of the $\Delta^+(1232)$
resonance of
$2.7{+1.0 \atop -1.3}\,\hbox{(stat)} \pm 1.5\,\hbox{(syst)}
\pm 3\,\hbox{(theo)}$ nuclear magnetons has been
extracted for the first time~\cite{mudelta}.

Parity violating electron scattering by the A4 collaboration using
counting techniques gives access to the strange vector form factors of
the nucleon.
A~preliminary value of
$A_\mathrm{phys} = (-5.6 \pm 0.6 \pm 0.2)\cdot 10^{-6}$
corresponding to a strange quark contribution of
$A_\mathrm{phys}-A_0 = (0.9 \pm 0.7)\cdot 10^{-6}$
has been extracted at a beam energy of $854.3\,\mathrm{MeV}$,
based on $4.6\cdot 10^{12}$ events~\cite{maas}.
Analysis of data taken at $570.1\,\mathrm{MeV}$ is in progress.

\newabstract 
\label{abs:calen}
\begin{center}
{\large\bf WASA Physics - Results and Perspectives}\\[0.5cm]
Hans Cal\'en$^1$ \\ for the CELSIUS/WASA collaboration\\[0.3cm]
$^1$The Svedberg Laboratory, Uppsala University,\\
P.O. Box 533, SE-751 21 Uppsala, Sweden\\[0.3cm]
\end{center}

The CELSIUS accelerator and storage ring at the The Svedberg Laboratory in
Uppsala is used for hadron physics research in the GeV region.
The programme includes production and decay of $\pi$ and $\eta$ mesons in 
light ion collisions at the WASA 4$\pi$ detector facility \cite{CW}.

The CELSIUS ring provides protons of kinetic energies up
to 1360 MeV.  For proton energies up to 500 MeV, electron cooling can be used, 
and this gives a beam size around 1 mm 
and a relative momentum spread down to the $10^{-5}$ region.

The recently developed pellet-target system provides small spheres of frozen
hydrogen as internal targets. This allows high luminosity and high detection
coverage for meson decay products like photons, electrons and charged pions.
The target system is now operating close to design performance with hydrogen
pellets having a diameter of about 25 micron, a speed of 80 m/s and occuring
at a rate of up to 15000 pellets/s.  The pellets are guided 2.5 m in a thin
tube to the CELSIUS beam and then further to a dump. The pellet beam diameter
at the interaction region is 2-3 mm.

Deuterium pellet generation has been tested successfully at different
occasions and now a useful deuterium pellet beam will be developed using a
separate test station that is being prepared.

The pellet target is integrated in the WASA 4$\pi$ detector facility.  WASA
consists of a forward part for measurements of charged target-recoil particles
and scattered projectiles, and a central part designed for measurements of the
meson decay products. The forward part consists of eleven planes of plastic
scintillators and of proportional counter drift tubes.  The central part
consists of an electromagnetic calorimeter of CsI(Na) crystals surrounding a
superconducting solenoid.  Inside of the solenoid are placed a cylindrical
drift chamber and a barrel of plastic scintillators.

Initially, production of $\eta$s and multiple $\pi$s in pp collisions have
been measured. The two-pion production studies started in the mid nineties at
the CELSIUS cluster-jet target \cite{PW2pi} have been repeated and extended to
higher energies exploiting the much larger acceptance at WASA, in particular
for the 2$\pi^0$ channel.  Data from threshold up to excess energies, Q,
around 100 MeV are well described by model calculations \cite{Val} involving
heavy meson exchange and excitation of the $N^{\ast}(1440)$ Roper resonance.
At the highest CELSIUS energies, with Q above 200 MeV, a preliminary analysis
shows that data rather follow flat phase space than simple expectations for a
dominant mechanism based on $N^{\ast}$ or $\Delta\Delta$ excitations.  At the
highest energies, data on $\eta \rightarrow \gamma\gamma$ and $\eta
\rightarrow 3\pi^0$ are regularly collected for control and development of
detector performance.  These data are also being analysed for detailed studies
of $\eta$ production mechanism. The $\eta \rightarrow 3\pi^0$ channel turned
out to be cleaner than expected for the pp reaction. The $pp \rightarrow pp
3\pi^0$ prompt production has not been measured in this energy region.  By
assumimg flat phase space distributions we obtain an upper limit of 1
$\mu$barn for the total cross section at 1360 MeV.
 
The experiment is being developed for studies of $\eta$ charged-particle
decays with a final goal to measure rare decays.  Deuterium pellets as a
target will allow very clean tagging of $\eta$s in the $pd
\rightarrow^3$He$\eta$ reaction at threshold. It will also allow studies of
meson production in pd, dd and other reactions, one of special interest being
the quasi-free pn reaction.

\newabstract  
\label{abs:neufeld}

\begin{center}
{\large\bf Isospin Violation in Semileptonic Kaon Decays}\\[0.5cm] 
{Helmut Neufeld}\\[0.3cm]
Institut f\"ur Theoretische Physik der Universit\"at Wien,\\
Boltzmanngasse 5, A-1090 Wien, Austria.\\[0.3cm]
\end{center}

Current and future experiments on the semileptonic decays of kaons and 
pions are already sensitive to the following subleading effects: 

\begin{itemize}
\item
strong contributions of order $p^6$ 
\item 
isospin-violation of order $(m_d - m_u) p^2, e^2 p^2$ generated by the mass 
difference of the light quarks and  by electromagnetism
\end{itemize}

All these contributions are of comparable size and have to be included in 
an up-to-data analysis. The appropriate theoretical framework is provided 
by chiral perturbation theory with virtual photons and leptons 
\cite{KNRT00}. This effective quantum field theory 
describes the interactions of the pseudoscalar octet, the photon and the 
light leptons at low energies and allows a comparison of experimental data 
with the predictions of the standard model. So far, this machinery has 
been applied to the following problems:

\begin{itemize}
\item
$K_{\ell 3}$ form factors including electromagnetic contributions 
to order $e^2 p^2$ \cite{Kl3}
\item
numerics of $K^+_{e3}$ \cite{Kl3} and $K^0_{e3}$ to ${\cal{O}} (p^6, 
(m_d-m_u)p^2, e^2 p^2)$ \cite{CKM}
\item
pionic beta decay \cite{pibeta}
\end{itemize}

The extraction of the CKM matrix element $|V_{us}|$ from $K^+_{e3}$ decay 
data may serve as an illustration. The relevant 
formula is given by \cite{Kl3}

\begin{equation} \label{extr}
|V_{us}|=\frac{16 \pi^{3/2} \Gamma(K^+_{e3(\gamma)})^{1/2}}
              {G_F M_{K^+}^{5/2} S_{\rm EW}^{1/2} |f_+^{K^+ \pi^0}(0)| 
I(\lambda_+)^{1/2}}.
\end{equation}
Theory has to provide the form factor at zero momentum transfer 
including all contributions up to ${\cal{O}} (p^6, (m_d-m_u)p^2, e^2 
p^2)$ \cite{CKM}: 
\begin{equation} \label{formfac}
              f_+^{K^+ \pi^0}(0) = 1.002 \pm 0.010.
\end{equation}
The error is largely dominated 
by the uncertainty of 
the ${\cal{O}} (p^6)$ contribution \cite{p6}. The application of 
Eq. (\ref{extr}) 
requires a consistent treatment of rleal photon emission in 
$\Gamma(K^+_{e3(\gamma)})$ (experimental input) and in the phase space 
integral $I(\lambda_+)$. A possible choice is to accept all pion and 
positron energies in the whole $K^+_{e3}$ Dalitz plot without any 
further cut on the photon energy. In this case, radiative corrections 
reduce $I(\lambda_+)$ by $1.27 \%$.

Recently, a new high statistics measurement of the $K^+_{e3}$ branching 
ratio has been performed by the E865 Collaboration at Brookhaven
\cite{E865}. Their analysis of more than 70,000 $K^+_{e3}$ events yielded a 
branching ratio which was about $2.3 \sigma$ larger than the current  
value of the Particle Data Group (PDG 2002). If the E865 result can be 
confirmed, the CKM unitarity puzzle would 
be resolved. Inserting the E865 rate 
in Eq. (\ref{extr}) gives \cite{CKM}
\begin{equation}
|V_{us}| = 0.2238 \pm 0.0033
\end{equation}
being in good agreement with the unitarity of the first row of the CKM 
matrix:
\begin{equation}
|V_{ud}|^2 + |V_{us}|^2 + |V_{ub}|^2 - 1 = -0.0024 \pm 0.0032.
\end{equation} 
It should be noted, however, that the E865 result does not only differ from 
older $K^+_{e3}$ data but is also inconsistent with the present $K^0_{e3}$ 
rate given by PDG 2002 (based again on very old and imprecise data).
We propose \cite{CKM} a rather powerful consistency check of present and 
future high-precision $K^+_{e3}$ and $K^0_{e3}$ data by considering the 
ratio 
\begin{equation}
r_{+0} = f_+^{K^+ \pi^0}(0) \Big/ f_+^{K^0 \pi^-} (0).
\end{equation}
This quantity is  largely insensitive to the dominating theoretical 
uncertainties, in particular the contributions of order $p^6$. The 
theoretical prediction \cite{CKM} 
\begin{equation}
r_{+0}^{\rm th} = 1.022 \pm 0.002 - 16 \pi \alpha X_1
\end{equation}
depends only on the (unknown) electromagnetic low-energy coupling $X_1$ 
\cite{KNRT00}. 
Already simple dimensional analysis ($|X_1| \le 1/(4 \pi)^2$) confines 
$r_{+ 0}$ to a rather narrow band leading to a  stringent test for 
the observable quantity
\begin{equation}
r_{+ 0}^{\exp} = \Bigg( \frac{2 \Gamma(K^+_{e3(\gamma)}) M_{K^0}^5 I_{K^0}}
{\Gamma(K^0_{e3(\gamma)}) M_{K^+}^5 I_{K^+}} \Bigg)^{1/2}.
\end{equation}

\newabstract 
\label{abs:knecht}

\begin{center}
{\large\bf The Anomalous Magnetic Moment of the Muon
\footnote{Work supported in part by TMR, EC contract No.
HPRN-CT-2002-00311 (EURIDICE).} }\\[0.5cm]
{ Marc Knecht}\\[0.3cm]
Centre de Physique Th\'eorique,\\
CNRS Luminy, Case 907, F-13288 Marseille Cedex 9, France
\end{center}

Over the last few years, the experimental situation concerning the
measurement of the anomalous magnetic moment of the muon has
witnessed tremendous improvements. 
The present world average 
\begin{equation}
a_{\mu^+}^{\mbox{\scriptsize{exp}}}\,=\,11\,659\,203(8)\ [0.7\,{\mbox{ppm}}],
\end{equation}
is dominated by the latest result from the Brookhaven E821 experiment
\cite{Bennett:2002jb}.
This situation has triggered a lot of theoretical activity
devoted to the evaluation of the muon g-2 in the standard model,
as well as to the contributions of possible degrees of freedom
beyond the standard model. Within the standard model, the
contributions to $a_{\ell}$, $\ell = e^\pm, \mu^\pm, \tau^\pm$,
are conveniently decomposed as $
a_{\ell}^{\mbox{\scriptsize{SM}}}
=a_{\ell}^{\mbox{\scriptsize{QED}}}
+
a_{\ell}^{\mbox{\scriptsize{had}}}
+
a_{\ell}^{\mbox{\scriptsize{weak}}}
$.
The first contribution arises from loops containing 
virtual leptons and photons only. Including
also quark and gluon loops induces the hadronic
part $a_{\mu}^{\mbox{\scriptsize{had}}}$. Finally,
the remaining degrees of freedom of the standard model,
neutrinos, electroweak gauge bosons, are responsible
for the last piece, $a_{\mu}^{\mbox{\scriptsize{weak}}}$.
%

The contributions to $a_{\ell}^{\mbox{\scriptsize{QED}}}$
have been computed analytically up to three loops
(for references, and for an introduction to
the subject, see \cite{schladming03}). The four loop
contribution has been evaluated numerically, while the
five loop contribution in the muon case is an estimate
only. The results
\begin{eqnarray}
&&\!\!\!\!\!\!\!\!\!\!
a_{e}^{\mbox{\scriptsize{QED}}}
 =
0.5 \textstyle{\left(\frac{\alpha}{\pi}\right)}
- 0.328\,478\,444\,00
\textstyle{\left(\frac{\alpha}{\pi}\right)^2}
+1.181\,234\,017
\textstyle{\left(\frac{\alpha}{\pi}\right)^3}
-1.750\,2(38\,4)
\textstyle{\left(\frac{\alpha}{\pi}\right)^4}
\nonumber\\
&&\!\!\!\!\!\!\!\!\!\!
a_{\mu}^{\mbox{\scriptsize{QED}}}
 =
0.5 \textstyle{\left(\frac{\alpha}{\pi}\right)}
+ 0.765\,857\,399(45)
\textstyle{\left(\frac{\alpha}{\pi}\right)^2}
+24.050\,509\,5(2\,3)
\textstyle{\left(\frac{\alpha}{\pi}\right)^3}
\nonumber\\
&&\qquad 
 +125.08(41)
\textstyle{\left(\frac{\alpha}{\pi}\right)^4}
+930(170)
\textstyle{\left(\frac{\alpha}{\pi}\right)^5}
\end{eqnarray}
include the recent reevaluation of the four loop
calculation \cite{KinoshitaNio03}, which affects
$a_{e}^{\mbox{\scriptsize{QED}}}$, but only
marginally
$a_{\mu}^{\mbox{\scriptsize{QED}}}$.
Using the comparison between the calculation of 
$a_{e}^{\mbox{\scriptsize{SM}}}$ and the
experimental value 
$a_{e^-}^{\mbox{\scriptsize{exp}}}=11\,596\,521\,88.3(4.2)$ 
$[3.7\,{\mbox{ppb}}]$ in order to obtain a
determination of the fine structure constant
$\alpha$, then yields $a_{\mu}^{\mbox{\scriptsize{QED}}}
= 11\,658\,470.35(28) \times 10^{-10}$.
%

Present limitations in the standard model 
evaluation of $a_{\mu}$ come from the 
presence of hadronic contributions which
involve the low energy region. These can be
decomposed into leading and next-to-leading
hadronic vacuum polarization, denoted
as $a_{\mu}^{\mbox{\scriptsize{hvp;1}}}$
and $a_{\mu}^{\mbox{\scriptsize{hvp;2}}}$, 
respectively, and hadronic light-by-light scattering, 
$a_{\mu}^{\mbox{\scriptsize{h L$\times$L}}}$.
The evaluation of the latter relies on specific
models, and the present situation is summarized
by $a_{\mu}^{\mbox{\scriptsize{h L$\times$L}}}
=+8(4)\times 10^{-10}$, where the error reflects
a conservative estimate of the model dependence.
In the case of $a_{\mu}^{\mbox{\scriptsize{hvp;1}}}$,
a data based evaluation is possible, using measurements
of the cross section for $e^+ e^- \to {\mbox{hadrons}}$
and/or of the hadronic decays of the $\tau$. A new analysis
\cite{CMD-2new} of the CMD-2 data for $e^+ e^- \to \pi^+ \pi^-$, 
which give the most important contribution, has
resulted in several recent reevaluations, 
$a_{\mu}^{\mbox{\scriptsize{hvp;1}}}(e^+e^-)=696.3(7.2)\times 10^{-10}$
\cite{davieretal03},
$691.8(6.1)\times 10^{-10}$
\cite{teubneretal03}, and
$694.8(8.6)\times 10^{-10}$
\cite{jegerlehner03}. While these values are quite
compatible, their average gives
$a_{\mu}^{\mbox{\scriptsize{hvp;1}}}(e^+e^-)=694.3(7.5)\times 10^{-10}$,
they still show a noticeable and puzzling discrepancy with
the evaluation obtained using the $\tau$ data,
$a_{\mu}^{\mbox{\scriptsize{hvp;1}}}(\tau)=711.0(5.8)\times 10^{-10}$
\cite{davieretal03}. Adding the remaining pieces
(see references in \cite{schladming03}),
$a_{\mu}^{\mbox{\scriptsize{hvp;2}}}=-10.0(0.6)\times 10^{-10}$,
$a_{\mu}^{\mbox{\scriptsize{weak}}}=+15.4(0.3)\times 10^{-10}$,
one arrives at the standard model value (the errors
are associated to vacuum polarization, hadronic light-by-light 
scattering, and to QED and weak contributions, respectively)
\begin{equation}
a_{\mu}^{\mbox{\scriptsize{SM}}}(e^+e^-)=11\,159\,178.1\pm
7.5\pm 4.0 \pm 0.4
\end{equation}
if only $e^+e^-$ data are used, or
\begin{equation}
a_{\mu}^{\mbox{\scriptsize{SM}}}(\tau)=11\,159\,194.8\pm
5.9\pm 4.0 \pm 0.4
\end{equation}
if the $\tau$ data are used at low energies. 
The deviation
with respect to the experimental value (1)
amounts to 2.2$\sigma$ or 0.8$\sigma$,
respectively. Excluding further problems
with the data analyses, the only possibility
to explain the difference between the $e^+e^-$ and 
the $\tau$ based evaluations seems to be unaccounted 
for isospin violating corrections.
The last paper of \cite{jegerlehner03}
presents an interesting discussion
in this respect. Future data from 
KLOE or BABAR will also be helpful.

\newabstract 
\label{abs:schweizer}

\begin{center}
{\large\bf Hadronic Atoms}\\[0.5cm]
Julia Schweizer\\[0.3cm]
Institute for Theoretical Physics, University of Bern,\\
Sidlerstr. 5, CH-3012 Bern, Switzerland\\[0.3cm]
\end{center}
Nearly fifty years ago, Deser et al. \cite{Deser:1954vq} derived the formula
for the width of pionic hydrogen at leading order in isospin symmetry
breaking. Similar formulas also hold for pionium and the $\pi^- K^+$ atom,
which decay predominantly into $2\pi^l0$ and $\pi^0 K^0$, respectively. These
Deser-type relations allow to extract the scattering lengths from measurements
of the decay width and the strong energy shift. The DIRAC collaboration at
CERN \cite{Adeva:1994xz} aims to measure the pionium lifetime to $10\%$
accuracy which allows to determine the S-wave $\pi\pi$ scattering lengths
difference $|a_0^0-a_0^2|$ at $5\%$ precision. The experimental result can
then be confronted with the very precise theory prediction
$a_0^0-a_0^2=0.265\pm0.004$ \cite{Colangelo:2000jc}. Particularly interesting
is that one may determine in this manner the nature of the SU(2)$\times$ SU(2)
chiral symmetry breaking in QCD by experiment. In a new experiment on pionic
hydrogen at PSI \cite{PSI}, the pionic hydrogen collaboration plans to measure
the strong energy shift of the ground state at $0.2\%$ and the decay width at
$1\%$ accuracy. In principle, the $\pi N$ scattering lengths can then be
extracted from data on pionic hydrogen alone. Finally, the DEAR collaboration
\cite{Bianco:1998wb} will measure the strong energy shift and decay width of
the 1$s$ state in kaonic hydrogen and kaonic deuterium.

In order to determine the scattering lengths from these precision
measurements, the theoretical expressions for the decay width and the strong
energy shift must be known to an accuracy that matches the experimental
precision. The non-relativistic effective Lagrangian framework has proven to
be the most efficient method to investigate bound-state characteristics
\cite{NRLagr}. In particular, this technique allows one to evaluate the higher
order corrections to the Deser-type formulae. In the following, we present the
result for the decay width of the $\pi^- K^+$ atom at next-to-leading order in
isospin symmetry breaking. The non-relativistic $\pi K$ Lagrangian provides a
systematic expansion in powers of the isospin breaking parameter $\delta$,
where both the fine-structure constant $\alpha$ and the quark mass difference
$m_u-m_d$ count as order $\delta$. The result for the width of the ground
state of the $\pi^- K^+$ atom at order $\delta^{9/2}$ yields,
\begin{equation}
  \Gamma_{\pi^0 K^0} =
  8\alpha^3\mu_+^2p^*\mathcal{A}^2\left(1+K\right), 
\quad\mathcal{A} = -\frac{1}{8\sqrt{2}\pi}\frac{1}{M_{\pi^+}+M_{K^+}}{\rm
  Re}\,A^{00;\pm}_{\rm thr}+{\it o}(\delta),
\label{Gammapi0K0}
\end{equation}
with 
\begin{eqnarray}
p^* &=& \left[\frac{M_{\pi^+}
  \Delta_K+M_{K^+}\Delta_\pi}{M_{\pi^+}
+M_{K^+}}-\alpha^2\mu_+^2+\frac{(\Delta_K-\Delta_\pi)^2}
{4(M_{\pi^+}+M_{K^+})^2}\right]^{\frac{1}{2}},\nonumber\\
K &=& \frac{M_{\pi^+}
  \Delta_K+M_{K^+}\Delta_\pi}{M_{\pi^+}
+M_{K^+}}{a^+_0}^2-4\alpha\mu_+\left[{\rm
  ln}\alpha-1\right](a^+_0+a^-_0)+{\it o}(\delta).
\end{eqnarray}
Here $\mu_+$ denotes the charged reduced mass and the meson mass differences
read $\Delta_\pi = M_{\pi^+}^2-M_{\pi^0}^2$, $\Delta_K = M_{K^+}^2-M_{K^0}^2$.
The quantity ${\rm Re}\,A^{00;\pm}_{\rm thr}$ is determined as follows. One
evaluates the relativistic $\pi^- K^+ \rightarrow \pi^0 K^0$ amplitude at
order $\delta$ near threshold, see Refs.~\citetwo{Kubis:2001ij}{Nehme:2001} and
removes the infrared-divergent Coulomb phase. The real part of this matrix
element contains a singularity $\sim 1/|\bf{p}|$ at threshold ($\bf{p}$
denotes the center of mass momentum of the charged pion and kaon). The
constant term in the threshold expansion corresponds to ${\rm
  Re}\,A^{00;\pm}_{\rm thr}$. Further, the normalization of $\mathcal{A}$ is
chosen such that in the isospin symmetry limit it coincides with the isospin
odd scattering length $a_0^-$. The isospin even and odd $\pi K$ scattering
lengths\footnote{We use the same notation as in Ref. \cite{Kubis:2001ij}.}
$a_0^+$ and $a_0^-$ are defined in QCD at $m_u = m_d$ and $M_\pi \doteq
M_{\pi^+}$, $M_K \doteq M_{K^+}$.

\newabstract 
\label{abs:schacher}

\begin{center}
{\large\bf Results from DIRAC}\\[0.5cm]
J\"urg Schacher$^1$\\
for the DIRAC Collaboration\\[0.3cm]
$^1$Lab for High Energy Physics, Bern University,\\
Sidlerstrasse 5, CH-3012 Bern, Switzerland\\[0.3cm]
\end{center}

The $\pi^+\pi^-$ atom or $A_{2\pi}$ is a hydrogen-like atom 
consisting of $\pi^+$ and $\pi^-$ mesons. 
This atom decays predominantly by strong interaction into
$\pi^0\pi^0$. Hence the $A_{2\pi}$ lifetime is in good approximation 
inversely proportional to the squared difference between
the S-wave $\pi\pi$ scattering lengths for isospin 0 and 2,
$|a_0 -a_2|$. This value is predicted in chiral perturbation theory
(ChPT). An experimental determination of the $A_{2\pi}$ atom lifetime provides
a possibility to check predictions ofl ChPT in a model-independent way.

Using ChPT at next-to-leading order in isospin breaking 
the $A_{2\pi}$ lifetime is predicted to be~\cite{LJUB01}:
$
\tau = (2.9 \pm 0.1) \cdot 10^{-15} s.
$
A lifetime measurement with 10\% accuracy allows
to determine the above $\pi\pi$ scattering length difference at 5\%.

The $\pi^+\pi^-$ atoms are produced by Coulomb interaction in the final state
of $\pi^+\pi^-$ pairs generated in proton--target 
interactions~\citetwo{NEME85}{AFAN93}.
After production these atoms travel through the target and
some of them are broken up due to their interaction with matter: ``atomic
pairs'' are produced, characterized by small pair c.m. relative momenta
$Q < 3$~MeV/$c$. These pairs will be detected in the DIRAC setup. Other atoms
annihilate into $\pi^0\pi^0$. The amount of broken up atoms $n_A$ depends
on the lifetime which defines the decay rate. Therefore, the breakup
probability is a function of the $A_{2\pi}$ lifetime.

In addition, the proton--target interaction produces $\pi^+\pi^-$ pairs 
with Coulomb (``Coulomb pairs'') and without  
Coulomb final state interaction. The latter category includes 
pion pairs with one pion from the decay of long-lived 
resonances (``non-Coulomb pairs'') as well as two pions from 
different interactions (``accidental pairs''). ``Coulomb'' and 
``non-Coulomb pairs'' together are the so-called ``free pairs''.
The total number of produced atoms ($N_A$) is proportional 
to the number of ``Coulomb pairs'' ($N^C$) with low relative momenta 
($N_A=K \cdot N^C$). The coefficient $K$ is precisely calculable. 

DIRAC aims to measure the $A_{2\pi}$ breakup probability: 
$P_{\rm br}(\tau)$ is defined as ratio of the observed number $n_A$
of ``atomic pairs'' to the number $N_A$ of produced $\pi^+\pi^-$ atoms
calculated from the measured number of ``Coulomb pairs''.

The purpose of the DIRAC setup at the CERN PS  
is to record $\pi^+\pi^-$ pairs with small relative momenta~\cite{ADEV03}. 
The 24~GeV proton beam hits a thin target (typically 100 $\mu$m 
thick Ni foil). Emerging $\pi^+\pi^-$ pairs travel  
in vacuum through the upstream spectrometer part with 
coordinate and ionisation detectors, 
before they are split by the 2.3~Tm bending magnet into the ``positive''    
and ``negative'' arm. Both arms are equipped with high precision drift 
chambers, time of flight detectors, cherenkov, preshower and muon counters. 
The relative time resolution between the two arms is around 180~ps.

The momentum reconstruction makes use of the drift chamber information of the
two arms as well as of the measured hits in the upstream coordinate detectors.
The relative momentum resolution for $Q_X$ and $Q_Y$ is estimated to be
0.4~MeV/c, whereas the $Q_L$ resolution is 0.6~MeV/c. A system of fast trigger
processors selects low $Q$ events.

In the analysis the signal extraction of ``atomic pairs'' from $A_{2\pi}$ 
ionisation takes advantage of the specific signature of such events, 
namely time-correlation and very small relative momenta. After rejection 
of electrons and muons and subtraction of ``accidental pairs'', 
the reduced sample of time-correlated low $Q$ events consists of 
``atomic'' and ``free pairs'' only. The number of ``atomic pairs'' 
is found directly from the excess of $\pi^+\pi^-$ pairs with 
very low relative momenta ($Q < 3$~MeV/$c$ or $|Q_L| < 1.5$~MeV/$c$)
relative to the expected number of ``free pairs'' simulated using 
``accidental pairs''. Such a $|Q_L|$-distribution of $\pi^+\pi^-$ pairs 
from nickel data in 2001 is shown (Fig. left).
A preliminary analysis of this data sample leads to a breakup probability of 
$P_{\rm br}(\tau) =0.46$ with a statistical error of 9\%. There are
also systematic errors due to the accuracy in describing multiple scattering 
in the target, membranes and detector planes and also in describing  
detector responses. Exploiting the dependence of the breakup probability 
on lifetime (Fig. right) a preliminary estimate for
the $\pi^+\pi^-$ atom lifetime yields 
$
\tau = [3.1^{+0.9}_{-0.7} (stat) \pm 1 (syst)] \cdot 10^{-15} s.
$

Up to now DIRAC has extracted more than 12000 ``atomic pairs'' and hence 
will be able to improve substantially the lifetime measurement. 
\begin{center}
\begin{tabular}{c c c}
\includegraphics[width=6.5cm]{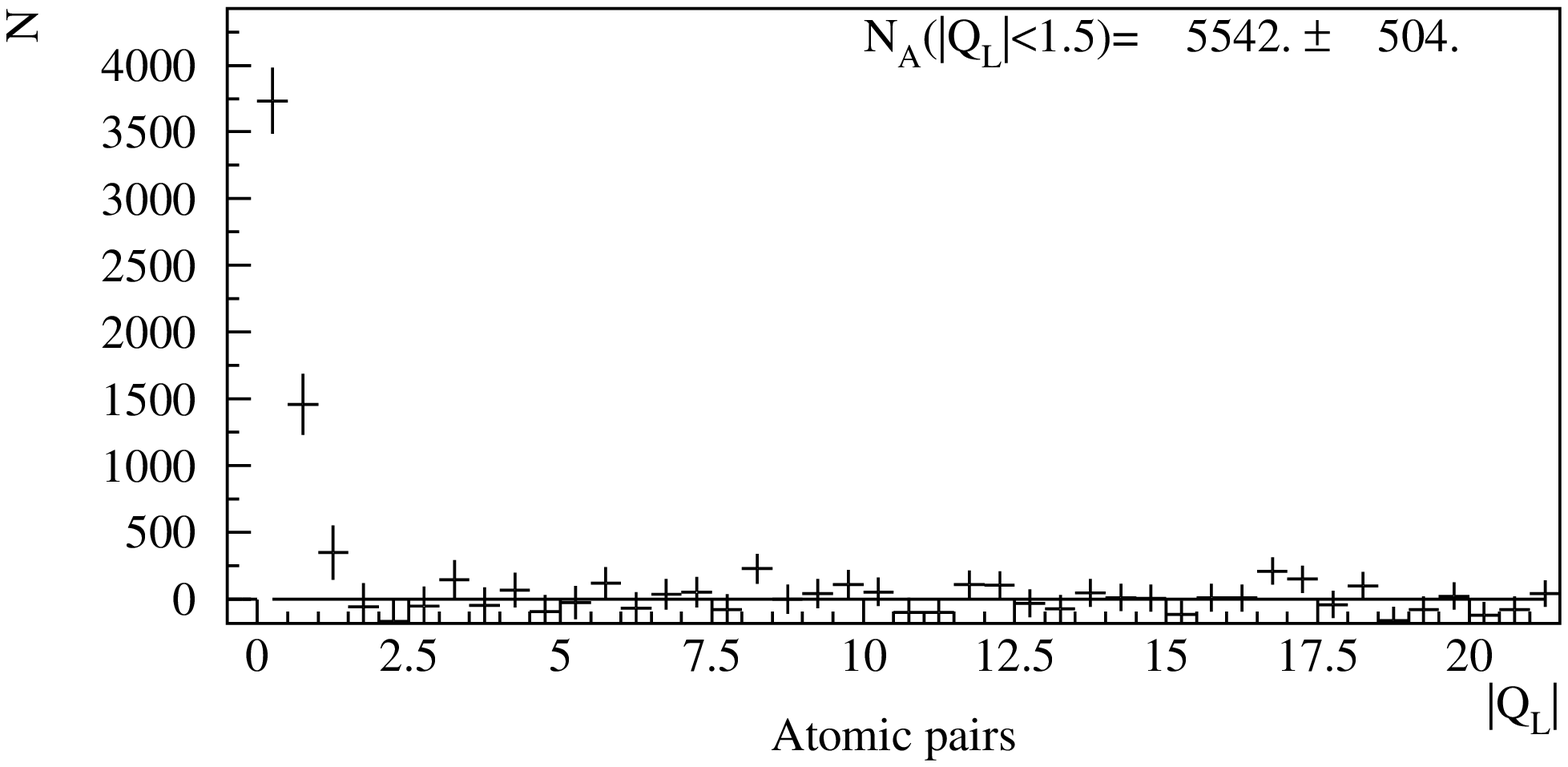}
&\hspace*{.3cm}&
\includegraphics[width=6.5cm]{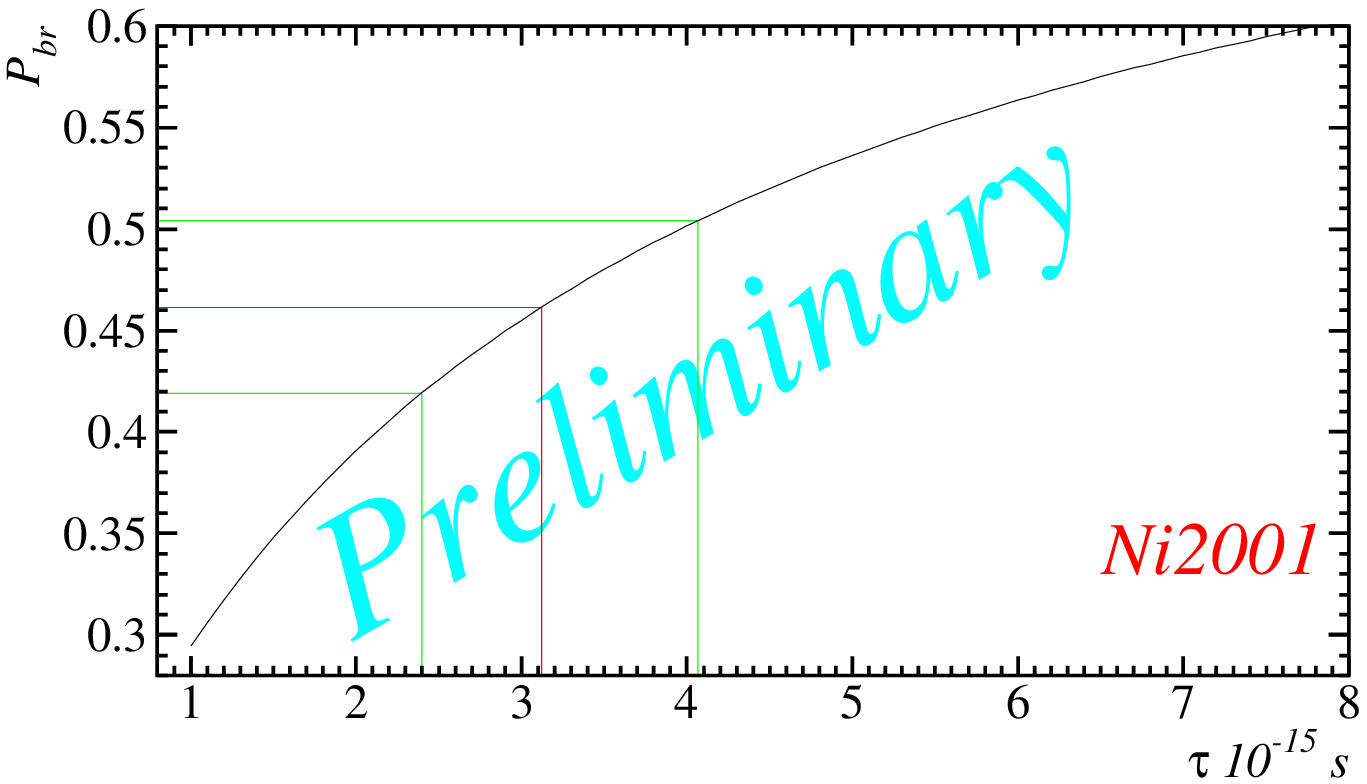}
\end{tabular}
\end{center}

\newabstract 
\label{abs:gotta}

\begin{center}
{\large\bf PSI Results from Hadronic Atoms}\\[0.5cm]
Detlev Gotta\\[0.3cm]
Forschungszentrum J\"ulich, D-52425 J\"ulich\\[0.3cm]
\end{center}

In pionic hydrogen ($\pi$H) the hadronic interaction manifests itself by a
change of the energies and of the natural line width of K X--rays as compared
to a purely electromagnetically bound atomic system.  Measurement of
ground--state transitions allows the determination of the $\pi$N s--wave
interaction, which is described by the isoscalar and isovector scattering
lengths $a^{+}$ and $a^{-}$\,\citethree{Lyu00}{Gas03}{Zem03}. 
In addition, from the
line broadening $\Gamma_{1s}$, which depends only on $a^{-}$, the $\pi$N
coupling constant can be extracted by the Goldhaber--Miyazawa--Ohme sum
rule\,\cite{Gol55}.

To improve on the accuracy achieved by previous measurements\,\cite{Sch01}, a
thorough study of the atomic cascade is essential. For that purpose, a first
series of measurements has been completed by the new pionic--hydrogen
experiment at the Paul--Scherrer--Institut (PSI R--98.01\,\cite{R98.01}),
using the new cyclotron trap, a cryogenic target, a Bragg spectrometer
equipped with spherically bent silicon and quartz crystals and a large--area
CCD array. Data analysis is in progress.

Except radiative decay, de--excitation during the atomic cascade is due to
collisions of the $\pi$H system with other atoms of the target, which leads to
density effects because of different collision probabilities. In order to
identify radiative de--excitation of the $\pi$H atom -- when bound into
complex molecules formed during collisions $\pi^{-} p+H_{2}\rightarrow
[(pp\pi^{-})p]ee$\,\cite{Jon99} -- the energy of the $\pi$H(3p-1s) transition
was measured at various target densities corresponding to a pressure range
from 3.5\,bar to liquid. X--ray transitions from molecular states should then
show up as low--energy satellites with density dependent intensity. The new
data do not show any density effect and, consequently, the measured line shift
$\epsilon_{1s}$ can be attributed exclusively to the strong interaction.  The
value of $\epsilon_{1s} =7.120\pm 0.013$\,eV was found to be in good agreement
with the result of the previous experiment. Precision was improved by more
than a factor of two\,\citetwo{Ana03}{Hen03}.

At present, the accuracy for $\Gamma_{1s}$ (7\%) is limited by the not
sufficiently well known correction for the Doppler broadening of the X--ray
lines\,\cite{Sch01}.  The broadening is caused by conversion of de--excitation
energy into kinetic energy during collisions (Coulomb
de--excitation)\,\cite{Bad01}. For that reason the precisely measured
1s--level shift in pionic deuterium was used together with the shift of
hydrogen in the determination of the $\pi$N scattering lengths\,\cite{Eri02}.
This procedure, however, requires a sophisticated treatment of the 3--body
system $\pi$D. In addition, up to now it cannot be excluded that the radiative
decay channel after molecule formation is strongly enhanced in deuterium
compared to hydrogen.

To study the influence of Coulomb de--excitation on the cascade, the three
transitions $\pi$H(2p-1s) (2.4~keV), $\pi$H(3p-1s) (2.9~keV) and $\pi$H(4p-1s)
(3.0~keV) were studied at a target density equivalent to 10~bar.  An increase
of the line width was found for the (2p-1s) line compared to the (3p-1s)
transition, which is attributed to the higher energy release available for the
acceleration of the $\pi$H system. This result is corroborated by a reduced
line width of the (4p-1s) transitionl\,\citetwo{Ana03}{Hen03}. 
The response of the
crystal spectrometer was obtained here from the $\pi^{12}$C(5g-4f) line
(3.0~keV), which is negligibly narrow compared to the experimental resolution.
From the (4p-1s) line width, a safe upper limit for the 1s--level broadening
of $\Gamma_{1s} < 850$\,meV is extracted, which is smaller than the result of
\,\cite{Sch01} but still consistent within the errors.

For further improvement, Coulomb de--excitation must be studied in more
detail. For this reason a second series of measurements is foreseen starting
with a high statistics study of ground--state transitions from muonic hydrogen
($\mu$H), where no strong--interaction effects occur. Beforehand, the crystal
resolution function, which has to be known with better accuracy than available
from a $\pi$C spectrum, will be determined with X--rays emitted from
helium--like argon, ionised by means of an electron--cyclotron resonance ion
trap (ECRIT) presently set up at PSI\,\cite{Bir03}. With that, studies of the
Bragg crystals can be performed with the neccessary statistics within a
reasonable time scale. The $\mu$H experiment will then be followed by a
remeasurement of $\pi$H with high statistics.

With the detailed knowledge of the crystal response together with a newly
developed cascade code\,\cite{Jen02}, which includes the velocity development
during the atomic cascade, a sufficiently accurate correction for the Doppler
broadening in pionic hydrogen should be achievable to extract $\Gamma_{1s}$ at
the level of about 1\%.

\newabstract 
\label{abs:oller}

\begin{center}
{\large\bf The Chiral Unitary Approach}\\[0.3cm]
Jos\'e A. Oller\\[0.15cm]
Departamento de F\'{\i}sica, Universidad de Murcia, E-30071 Murcia,
 Spain.\\[0.15cm]
\end{center}

\vspace{-1mm}
The effective field theory of low energy QCD is Chiral Perturbation Theory
(CHPT).  It is a perturbative expansion in powers of the external fourmomenta
of the pseudo-Goldstone bosons (the canonical ones being the pions) over a
typical hadronic scale, $M_\rho\simeq 4\pi f_\pi\simeq 1$ GeV.  Nevertheless,
even at low energies there are several examples where such perturbative
procedure cannot be applied or the convergence is very slow. The most
important ones are the S-waves corresponding to low energy isosclar pion-pion
interactions, $\bar{K}N$ scattering close to threshold or nucleon-nucleon
interactions. In all the three cases, its non-perturbative nature originates
from large unitarity corrections that, indeed, give rise to the appearance of
poles in the physical amplitudes: the broad $\sigma$ resonance with a mass
around 450 MeV, the $\Lambda(1405)$ in $\bar{K}N$, the deuteron in $^3 S_1$
and an antibound state in $^1 S_0$ just below threshold in $NN$.  Such large
unitarity corrections are due to numerical enhancements in $\pi\pi$ scattering
\cite{nd} and to the large kaon and nucleon masses in $\bar{K}N$ and $NN$, see
ref.\cite{wein2}. In the last case, there is also a problem associated with
the unnatural values of the S-wave $NN$ scattering lengths, much larger than
the inverse of the pion mass.
On the other hand, CHPT can only be used for rather low energies, although the
chiral constraints are beyond the pure perturbative regime since they
constitute Ward identities that must be satisfied by the Green functions at
any energy. It should be then interesting to dispose of parameterizations
valid for higher energies that could apply the chiral constraints in a large
energy window, e.g. from threshold up to around 2 GeV in meson-meson, where
one could try to match with perturbative+OPE of QCD.  When going to higher
energies, apart from the right hand cut (unitarity corrections) one should
also face the appearance of preexisting resonances (that is, resonances that
are not due to the meson-meson self interactions), e.g. the $\rho(770)$ in
P-wave $\pi\pi$ scattering, the $\Delta(1232)$ in $\pi N$ P-waves, etc. Of
course, these resonances also contribute to other partial waves but not
through their direct exchanges in the s-channel but because of their exchanges
in the crossing ones.  These facts can be included systematically through the
so called Chiral Unitary Approach (UCHPT) \citethree{nn}{kn}{nd}. The general
idea is to match {\it algebraically} the general expression of a partial wave
amplitude, or the one of a production process (e.g. a form factor), that takes
into account the right hand cut to all orders, with the perturbative
expression from CHPT (or alike chiral effective field theories like Heavy
Baryon CHPT, Kaplan-Savage-Wise counting, etc). The simplest way to proceed is
by performing a dispersion relation for the inverse of the $T$ matrix, since
unitarity requires above threshold that
$\displaystyle{\hbox{Im}T^{-1}_{ij}=-\rho_i \delta_{ij}}$ with $\rho_i=
\theta(s-s_{th})\,q/8\pi\sqrt{s}$~. Then, performing a dispersion integral
along this cut one has,
{\small\begin{equation}
\label{master}
T^{-1}_{ij}=R^{-1}_{ij}
+\delta_{ij}\underbrace{\left(g(s_0)_i-\frac{s-s_0}{\pi}\int_{s_{th}}^\infty 
\frac{\rho(s')}{(s'-s-i0^+)(s'-s_0)} 
ds'\right)}_{g(s)_i:~ \hbox{single unitarity bubble}}~,
\end{equation}} 

\vspace{-3mm}
\noindent thus $T=\left[1/R+ g\right]^{-1}$~, 
where $g(s_0)_i$ is a subtraction constant and $1/R$ is the remnant not fixed
by the right hand cut.  The next step is to fix the unknown input $R$, or
interaction kernel, by matching algebraically with the effective field theory
calculation of $T$. E.g. if we know $T=T_2+T_4+{\cal O}(p^6)$ for meson-meson
scattering then taking $g$ naturally as order $p^0$ one has,
$R_2=T_2~,~R_4=T_4-T_2 g T_2$ , and so on. In this way, the interaction kernel
$R$ is fixed order by order with increasing precision, with the advantage that
eq.(\ref{master}) can be also applied when the interactions between the
hadrons are not perturbative.  The perturbation is now performed on $R$,
instead of directly on $T$, because the latter is affected by the large
unitarity corrections. In addition, since many of the effects that limit a
straightforward application of CHPT to higher energies are overcome with this
procedure, because unitarity is fulfilled to all orders and explicit
resonances can also be included as part of $R$, then this scheme is also
applicable to increase the energy range where chiral Lagrangians can be
applied. An analogous expression to eq.(\ref{master}) can also be derived for
production mechanisms \cite{scalar}
Many applications of the UCHPT can be found in the literature in meson-meson,
meson-baryon \cite{kn2} and baryon-baryon \cite{nn} scattering.  Its
repercussion in the strong interacting S-waves for all the previous scattering
processes has been particularly important and clarifying, providing a reliable
and systematic way of procedure for these particularly problematic waves, out
of the range of standard vector meson like models for scalars, naive quark
models and straightforward applications of large $N_c$ QCD insights.
Furthermore, it has also provided the necessary final(initial) state
corrections to production mechanisms, e.g. $\phi$ decays, $\gamma\gamma$
fusion into two mesons, production of exotic resonances, $J/\Psi$ decays,
electromagnetic interactions of meson and baryons \cite{varios}.


\newabstract 
\label{abs:epelbaum}

\begin{center}
{\large\bf Improving the Convergence of 
Chiral EFT for}\\[0.1cm]
{\large\bf Few--Nucleon Systems }\\[0.5cm]
{\bf Evgeny Epelbaum}$^1$, Walter Gl\"ockle$^2$ 
and Ulf-G. Mei\ss ner$^{3,4}$\\[0.3cm]
$^1$Jefferson Laboratory, \\ Newport News, VA 23606, USA\\[0.3cm]
$^2$Institut f\"ur Theoretische Physik II, Ruhr--Universitaet Bochum, \\
D-44780 Bochum, Germany\\[0.3cm]
$^3$Universit\"at Bonn, 
Helmholtz--Institut f\"ur Strahlen-- und Kernphysik (Theorie), \\
Nu{\ss}allee 14 - 16, D-53115 Bonn, Germany\\[0.3cm]
$^4$Forschungszentrum J\"ulich, Institut f\"ur Kernphysik (Theorie), \\
D-52425 J\"ulich, Germany\\[0.3cm]
\end{center}

In the past decade chiral EFT has been applied by several groups to study the
properties of few--nucleon systems based on the original idea of Weinberg
\cite{Weinb}, see \cite{Beda02} for a recent review. It is of utmost
importance for all these applications to verify the convergence of the
low--momentum expansion. As pointed out by Kaiser et al. \cite{KBW},
nucleon--nucleon peripheral scattering provides an excellent ground to test
the convergence of the chiral expansion. Indeed, the high angular momentum
scattering states are dominated by the long--range part of the nuclear force,
where chiral dynamics is important.  For example, no short--range contact
interactions (with, in general, unknown coupling constants) contribute to D--
and higher partial waves at next--next--to--leading order (NNLO), which allows
for parameter--free calculations. In addition, the interaction between two
nucleons becomes weak due to the centrifugal barrier, so that the S--matrix
can be obtained using perturbative methods.  It was found in \cite{KBW} that
the subleading two--pion exchange (TPE) contribution, which arises at NNLO and
depends on the low--energy constants (LECs) $c_{1,3,4}$ shows unphysically
strong attraction already at intermediate distances, if the values for these
LECs are taken from pion--nucleon scattering. This results in strong
disagreement between theoretical predictions and the data for $E_{\rm lab} >
50$ MeV ($E_{\rm lab} > 150$ MeV) for D-- (F--) waves and might therefore
indicate problems with the convergence of the chiral expansion, see also
\cite{ep02} for more discussion.

In \cite{EGMnew} we have investigated the origin of the unphysical attraction
of chiral TPE and proposed a new regularization scheme for calculating pion
loop integrals, which allows to improve the convergence and is closely related
to the long--distance regularization method applied in baryon SU(3) chiral
perturbation theory \cite{don98_2}.  The proposed scheme makes use of the
spectral function representation of the TPE potential
\begin{equation}
\label{spectrepel}
V  (q) = \frac{2}{\pi} \int_{2 M_\pi}^\infty d \mu \, \mu
\, \frac{\rho (\mu)}{\mu^2 + q^2},
\end{equation}
where the spectral function $\rho (\mu )$ can be obtained via $\rho (\mu ) =
\Im \left[ V (0^+ - i \mu ) \right]$.  In \cite{EGMnew} we have demonstrated
that the short--range high--$\mu$ components of the TPE spectral function
calculated using dimensional regularization (DR) or infinite momentum cut--off
regularization start to dominate the potential already at intermediate
distances.  For example, the dominant contributions to the isoscalar central
part at distances $r \sim M_\pi^{-1}$ arise from the region $\mu \sim 600$
MeV. Chiral EFT predictions for the spectral function can hardly be expected
to converge for such and larger values of $\mu$. These spurious short--range
components picked up in DR or equivalent regularization schemes have been
shown to be responsible for the unphysical behavior of the subleading chiral
TPE potential.

Following the lines of \cite{don98_2}, the convergence of the chiral expansion
can be greatly improved by keeping only truly long--distance portion of the
pion loop integrals. This has been achieved introducing the finite cut--off
$\Lambda$ in the spectral function representation eq.~(\ref{spectrepel}).  The
results for nuclear forces obtained using the new regularization scheme are
equivalent to the DR ones modulo an infinite set of higher--order contact
interactions (this holds true if one uses a sharp cut--off). Both
regularization schemes thus lead to exactly the same result for observables
provided one keeps terms in all orders in the EFT expansion.

The new regularization scheme has been successfully applied to the
two--nucleon scattering problem at NNLO \citetwo{EGMnew}{EGMnew2}.  
Extensions to
N$^3$LO \cite{EGMnew3} and to systems with more than two nucleons are
underway.

\newabstract 
\label{abs:hammer}

\begin{center}
{\large\bf Limit Cycle Physics}\\[0.5cm]
{H.-W. Hammer}$^1$ \\[0.3cm]
$^1$Universit{\"a}t Bonn, Helmholtz-Institut f{\"u}r 
Strahlen- und Kernphysik (Theorie),\\ Nussallee 14-16,
53115 Bonn, Germany\\[0.3cm]
\end{center}

The development of the renormalization group (RG)
has had a profound effect on many branches of physics.
Its successes range from explaining the universality
of critical phenomena in condensed matter physics
to the nonperturbative formulation of quantum field theories
that describe elementary particles.
The RG can be reduced to a set of
differential equations that define a flow in the space of
coupling constants.
Scale-invariant behavior at long distances, as in critical phenomena,
can be explained by RG flow to an infrared fixed point.
Scale-invariant behavior at short distances, as in
asymptotically-free field theories,
can be explained by RG flow to an ultraviolet fixed point.
However, a fixed point is only the simplest topological feature
that can be exhibited by a RG flow.
Another possibility is a limit cycle,
which is a 1-dimensional orbit that is closed under the RG flow.
The possibility of RG flow to a limit cycle was proposed by
Wilson in 1971 \cite{Wilson:1970ag}.
However, few physical applications of RG limit cycles have emerged.

In 1970, Efimov discovered a remarkable effect
in the 3-body sector for nonrelativistic particles
with short-ranged S-wave 2-body interactions \cite{Efi71}.
The strength of the interaction is governed by the scattering
length $a$. Efimov showed that if $|a|$ is much larger than
the range $r_0$ of the interaction, there are shallow 3-body bound states
whose number increases logarithmically with $|a|/r_0$.
In the limit $a \to \pm \infty$,
there are infinitely many shallow 3-body bound states
with an accumulation point at the 3-body scattering threshold.
The ratio of the binding energies of successive states
rapidly approaches the universal constant
$\lambda_0^2 \approx (22.7)^2$. 

The Efimov effect was revisited by Bedaque, Hammer,
and van Kolck within the framework of effective field theory (EFT)
\cite{Bedaque:1998kgkm}.
The nonperturbative solution of the EFT in the 3-body sector
can be obtained by solving integral equations numerically.
These integral equations have unique solutions only
in the presence of an ultraviolet cutoff $\Lambda$.
The resulting predictions for 3-body observables,
although finite, depend on the cutoff and are
periodic functions of $\ln(\Lambda)$.
In Ref.~\cite{Bedaque:1998kgkm}, it was shown that the EFT
could be fully renormalized to remove the residual dependence
on $\Lambda$ in the 3-body sector by adding a 3-body interaction term
$g_3(\Lambda)(\psi^* \psi)^3$ to the effective Lagrangian
\cite{Bedaque:1998kgkm}.
The dependence of 3-body observables on the cutoff
decreases like $1/\Lambda^2$ if
$g_3(\Lambda) \propto H(\Lambda)/\Lambda^2$,
where $H(\Lambda)=\cos[s_0 \ln(\Lambda/\Lambda_*) + \arctan(s_0)]/
\cos[s_0 \ln(\Lambda/\Lambda_*) - \arctan(s_0)]$ with 
$s_0\approx 1.006$.
With this renormalization, 3-body observables have well-defined
limits as $\Lambda \to \infty$, but they depend on
the parameter $\Lambda_*$ introduced by dimensional transmutation.
Since $H(\Lambda)$ is a periodic function of $\ln(\Lambda)$,
the renormalization of the field theory involves
a limit cycle.
This contact EFT has also been applied to the 3-nucleon problem
\cite{Bedaque:1998kgkm}. The success of this program suggests
that physical QCD is close to an infrared limit cycle. 

We conjecture that QCD can be tuned
to the critical RG trajectory for an infrared limit cycle
by adjusting the up and down quark masses $m_u$ and $m_d$
\cite{Braaten:2003eu}.
Our argument is based on recent work in which an EFT
with explicit pions was used to extrapolate nuclear forces
to the chiral limit of QCD \citetwo{Beane:2002xf}{Epelbaum:2002gb}.
In this limit, the masses $m_u$ and $m_d$ of the up and down quarks
are zero and the pion is an exactly massless Goldstone boson associated
with spontaneous breaking of the chiral symmetry of QCD.
According to these chiral extrapolations,
the small binding energy 2.2 MeV of the deuteron is a fortuitous consequence
of the physical values of $m_u$ and $m_d$. 
When extrapolated farther from the chiral limit to slightly larger
pion masses, the deuteron's binding energy decreases to zero
and then it becomes unbound. In this region, the inverses of both 
the spin-singlet and spin-triplet scattering lengths $1/a_s$ and
$1/a_t$ are close to zero \citetwo{Beane:2002xf}{Epelbaum:2002gb}. 

In the next-to-leading order chiral extrapolation of
Ref.~\cite{Epelbaum:2002gb}, only quark mass dependent
operators proportional to $m_u + m_d$ enter.
The extrapolation in $m_\pi$ can be interpreted as an
extrapolation in the sum $m_u + m_d$,
with $m_u - m_d$ held fixed. Changing  $m_u - m_d$
changes the degree of isospin-symmetry breaking.
Since $a_s$ and $a_t$ correspond to different
isospin channels, they respond differently
to changes in $m_u - m_d$.  Therefore it may be possible
by tuning both $m_u$ and $m_d$
to make $1/a_t$ and $1/a_s$ vanish simultaneously.
This point corresponds to a critical RG trajectory for an infrared
limit cycle. At this critical point,
the triton has infinitely-many increasingly-shallow excited states
with an accumulation point at the 3-nucleon threshold.
The ratio of the binding energies of successively shallower states
rapidly approaches a constant $\lambda_0^2$ close to 515. This 
opens the exciting possibility of observing an RG limit cycle 
in QCD using a combination of lattice QCD and chiral EFT. 

This talk is based on work done in collaboration with P.F. Bedaque, 
Eric Braaten, and U. van Kolck.

\newabstract 
\label{abs:chen}

\begin{center}
{\large \textbf{Nuclear Electroweak Processes in Effective Field Theories}}\\%
[0.5cm]
Jiunn-Wei Chen\\[0.3cm]
Department of Physics, National Taiwan University, Taipei 10617, Taiwan\\[%
0.3cm]
\end{center}

In this talk, I review some selected topics in nuclear electroweak processes
using effective field theories (EFT), including applications in
astrophysics, nucleon property extractions from nuclear systems and some
applications in parton physics.

In nuclear astrophysics, EFT was applied to several very low energy
processes such that pions can be treated as heavy particles and integrated
out. In Big Bang Nucleosynthesis (BBN), the $np\rightarrow d\gamma $ cross
section is an important input contributing the largest uncertainty to the
primordial abundance rate of $^{7}$Li. Using EFT, the theoretical
uncertainty is reduced to 1\% \cite{npdgamma} in the energy range relevant
to BBN. The EFT predictions also agree with a very recent measurement of the
inverse process in the same energy region \cite{Tornow}.

The same theory is also applied to neutrino-deuteron breakup processes which
are important inputs for the measurement of solar neutrino flux at Sudbury
Neutrino Observatory (SNO). However, there was no precise, direct
measurement for those cross sections and two state of the art potential
model calculations differ by 5-10\%. Using EFT, this difference was
successfully described by a parameter $L_{1,A}$, which is an axial two-body
current encoding the unknown short distance physics \cite{BCK}. Since $%
L_{1,A}$ is the \emph{only} unknown parameter that appears in all the low
energy weak interaction deuteron breakup process at the
next-to-next-to-leading order, any weak deuteron breakup measurement will
constrain $L_{1,A}$. By now, several independent constraints on $L_{1,A}$
have been obtained with consistent results (see \cite{CHR} for a summary).

Another recent development is the model independent calculations on $%
NN\rightarrow NNX$ in supernova (SN) 1987a \cite{HPR}, where $X$ is some
undetected light particle. From the missing energy upper bound of SN1987a,
several improved constraints have been obtained with $X$ taken as graviton
and dilaton---to probe the size of extra dimensions; as axion, saxion and
light neutralinos, and as $\nu \bar{\nu}$ for neutron star cooling 
\cite{NNBrem}.

In the extractions of the nucleon polarizabilities from deuteron Compton
scattering, ref.\cite{GPBS} used EFT with pion integrated out to obtain the
isoscalar electric polarizability $\alpha _{N}=7.2\pm 2.1\pm 1.6$ and
magnetic polarizability $\beta _{N}=6.9\mp 2.1\mp 1.6,$ in units of $%
10^{-4}fm^{3}$. This is consistent with $\alpha _{N}=9.0\pm
1.5_{-0.8}^{+3.6} $ and $\beta _{N}=1.7\pm 1.5_{-0.6}^{+1.4}$ found in ref. 
\cite{BMMPvK} using EFT with pions plus some higher energy data. However,
higher precision Compton scattering data are clearly needed to better
determine $\alpha _{N}$ and $\beta _{N}$.

Recently chiral perturbation theory has been applied to study parton physics
through computations of hadronic twist-2 matrix elements \cite{ASCJ}. The
computed quark mass dependence of twist-2 matrix elements are useful to
guide the chiral extrapolations of lattice data. The method can also be
applied to generalized parton distributions, SU(3) flavor structure, large $%
N_{c}$ behavior and nuclear modifications (EMC effects) \cite{CJBJ}. A
recent study \cite{CS} in deeply virtual Compton scattering (DVCS) indicates
that $\gamma ^{\ast }N\rightarrow \gamma N\pi $ involves twist-2 matrix
elements which cannot be determined from $\gamma ^{\ast }N\rightarrow \gamma
N$ alone, contrary to previous claims \cite{GMV}.


\newabstract 
\label{abs:beane1}

\def\si{{}^1\kern-.14em S_0}
\def\siii{{}^3\kern-.14em S_1}
\def\diii{{}^3\kern-.14em D_1}
\begin{center}
{\large\bf Nuclear Physics and Lattice QCD}\\[0.5cm]
Silas R.~Beane$^{1,2}$\\[0.3cm]
$^1$Department of Physics, 
University of New Hampshire, Durham, NH 03824\\[0.3cm]
$^2$Jefferson Laboratory, 12000 Jefferson Avenue, 
Newport News, VA 23606\\[0.3cm]
\end{center}

\noindent Impressive progress is currently being made in computing
properties and interactions of the low-lying hadrons using lattice
QCD.  However, cost limitations will, for the foreseeable future,
necessitate the use of quark masses, $M_q$, that are significantly
larger than those of nature, lattice spacings, $a$, that are not
significantly smaller than the physical scale of interest, and lattice
sizes, $L$, that are not significantly larger than the physical scale
of interest.  Extrapolations in the quark masses, lattice spacing and
lattice volume are therefore required. The hierarchy of mass scales
is: $L^{-1}\ll M_q \ll \Lambda_\chi \ll a^{-1}\ \ $.  The appropriate
EFT for incorporating the light quark masses, the finite lattice
spacing and the lattice size into hadronic observables is $\chi$-PT,
which provides systematic expansions in the small parameters
$e^{-{m_\pi{L}}},\quad {1/{{\ L}{\Lambda_\chi}}}, \quad
{{p}/\Lambda_\chi},\quad {{M_q }/\Lambda_\chi}$ and ${\
a}{\Lambda_\chi}\ $.  The lattice introduces other unphysical scales
as well.  Lattice QCD quarks will increasingly be artificially
separated into non-degenerate ``sea'' and ``valence'' quarks, a
procedure known as partial-quenching. Continuum EFT methods are being
developed which properly account for
finite-$a$~\citetwo{Sharpe:1998xm}{RSa}, finite-$L$~\citetwo{GandLV}{durr} and
partial-quenching effects~\cite{Pqqcd1} and which will allow for a
systematic extrapolation from unphysical simulations of hadron
properties to nature.  Recently, partially-quenched $\chi$-PT
(PQ$\chi$-PT) has been formulated for baryons~\citetwo{CSn}{BSn} with
finite lattice spacing corrections~\cite{BeSafina}, and many
quantities have been computed. (See Ref.~\cite{Arndt3} and references therein.)

Progress in few-nucleon effective field theory has reached the
point where one may meaningfully ask how nuclear energy levels change
as we vary the quark masses in the QCD Lagrangian.  A practical
motivation for understanding the quark mass dependence of nuclear
physics is that future lattice QCD results will require extrapolation.
Fortunately, the quark-mass dependence of few-nucleon systems can be
studied in effective field theory~\citethree{BBSvK}{BSmq}{EMmq}, and
amusingly, with existing results, the successful predictions of
Big-Bang nucleosynthesis (BBN) can be used to constrain the time
dependence of the Higgs vev and $\Lambda_{QCD}$, and thereby search for
physics beyond the standard model~\cite{yoo}\cite{gail}.  There has
been one attempt to compute nucleon-nucleon (NN) scattering parameters
in lattice QCD; Ref.~\cite{fuku} computes the $\si$ and $\siii$
scattering lengths in quenched QCD (QQCD) using L\"uscher's
method~\cite{Luscher}.  The NN potential is modified in an unphysical
way in QQCD and PQQCD~\cite{BSpot}, however it is straightforward to
develop the partially-quenched EFT which matches to a
partially-quenched lattice simulation~\cite{BSnn}.  A second approach
to the NN system on the lattice is to study the simplified problem of
two interacting heavy-light particles~\cite{JLabMIT}.  It has
been suggested that lattice QCD simulations of the potential between
hadrons containing a heavy quark will provide insight into the nature
of the intermediate-range force between two
nucleons~\cite{JLabMIT}. While the NN potential is not itself an
observable, one may instead consider heavy systems.  The
$\Lambda_Q\Lambda_Q$ interaction is an ideal system to
investigate~\cite{ABSl}.  Since the $\Lambda_Q$ is an isosinglet,
there is no OPE, and the leading large-distance behavior is governed
by two-pion exchange (TPE), which is physics analogous to the
intermediate-range attraction in the NN potential.

\newabstract 
\label{abs:stewart}

\newcommand{\bn}{\bar n}
\newcommand{\bnP}{\bar {\cal P}}
\newcommand{\nn}{\nonumber} 
\newcommand{\mcdot}{\!\cdot\!}


\begin{center}
{\large\bf The Soft-Collinear Effective Field Theory}\\[0.3cm]
{ Iain W. Stewart}\\[0.2cm]
Massachusetts Institute for Technology, Cambridge, MA 02139, USA\\[0.3cm]
\end{center}

The soft-collinear effective theory~\cite{SCET} (SCET) provides a formalism for
systematically investigating processes with both energetic and soft hadrons
based solely on the underlying structure of QCD.  Most effective theories that
we are familiar with are designed to separate the physics for hard $p_h^\mu\sim
Q$ and soft $p_s^\mu$ momenta ($Q^2 \gg p_s^2$).  Examples include chiral
perturbation theory (ChPT), few-nucleon effective theory, heavy quark effective
theory, or the high density effective theory. In SCET we incorporate an
additional possibility, namely energetic hadrons where the constituents have
momenta $p_c^\mu$ nearly collinear to a light-like direction $n^\mu$.  Both the
energetic hadron and its collinear constituents have $\bn\cdot p_c \sim Q$,
where we have made use of light-cone coordinates
$(p_c^+,p_c^-,p_c^\perp)=(n\mcdot p_c,\bn\mcdot p_c,p_c^\perp)$.  However, the
collinear constituents still have small offshellness $p_c^2\sim p_s^2$.  The
process of disentangling the interactions of hard-collinear-soft particles is
known as factorization, and is simplified by the SCET framework. I compare some
of the basic features of SCET with ChPT in Table~\ref{tab:compare}.


Much like any effective theory the basic ingredients of SCET are its field
content, power counting, and symmetries. The Lagrangian and operators,
\begin{eqnarray}
  {\cal L}= {\cal L}^{(0)} + {\cal L}^{(1)} + \ldots \,,\qquad
  {\cal O}= {\cal O}^{(0)} + {\cal O}^{(1)} + \ldots\,,
\end{eqnarray}
are organized in a series where only ${\cal L}^{(0)}$ and ${\cal O}^{(0)}$ are
relevant at LO, an additional ${\cal L}^{(1)}$ or ${\cal O}^{(1)}$ is needed
at NLO, etc. The expansion parameter will be $\lambda=\sqrt{\Lambda_{\rm
    QCD}/Q}$ or $\eta=\Lambda_{\rm QCD}/Q$ depending on whether the collinear
fields describe an energetic jet of hadrons or an individual energetic hadron.
The effective theory with an expansion in $\lambda$ is called ${\rm SCET}_{\rm
  I}$, while the one with an expansion in $\eta$ is called ${\rm SCET}_{\rm
  II}$. For a review of the field content, power counting and symmetries
see~\cite{rev1}, for more detail see~\cite{detail}.  A useful advantage of
using the symmetries of SCET rather than full QCD is that transformations like
$P,C,T$, baryon number, isospin, etc., can be implemented separately for the
soft and collinear sectors.

In its simplest form SCET is applied to hadrons built of light quarks, for
detailed examples see~\cite{bfprs}.  Progress has been made for processes such
as exclusive hadronic form factors and deeply virtual Compton
scattering~\cite{bfprs}, as well as deep inelastic scattering and jet
production~\cite{bwm}.  Many new results have also been obtained for
$B$-decays, where we combine SCET with heavy quark effective theory.  Here the
large mass of the $B$ meson provides plenty of energy for light hadrons in the
final state (cf.~\cite{rev2} for a short review and further references to the
literature).  Interesting results are also obtained by combining SCET with
non-relativistic QCD, an effective field theory for heavy $Q\bar Q$
systems~\cite{scet_nrqcd}.  Very recently progress has been made on combining
SCET with ChPT~\cite{scet_chpt} to describe light quark mass dynamics in
energetic hadrons with an effective field theory. MIT-CTP 3444.

\begin{table}[t!]
\vspace{-.2cm}
\caption{A comparison of two effective field theories, chiral perturbation
  theory (ChPT) and the soft-collinear effective theory (SCET). 
  For notation see~\cite{SCET}.
  \label{tab:compare}}
\vspace{-.2cm}
\begin{center}
\begin{tabular}{c|c}
 {\color{blue} ChPT} & {\color{red} SCET} \\ \hline\hline
Soft hadrons, pions, nucleons &
Soft \& Energetic hadrons, quarks, gluons \\[3pt] \hline
 Nonlinear fields: ${\color{red}\Sigma} =\xi^2$ &
  Nonlinear fields: {\color{red} W}, {\color{OliveGreen} S},
 {\color{blue} Y} (Wilson Lines)\\
 ${\color{red}\Sigma} = \exp(2iM/f)$ 
  & ${\color{red} W}= P \exp( -g/\bnP\, \bn\mcdot A_n)$ 
   \\[3pt] \hline
 Chiral transformations: &  Gauge transformations: \\
 $\Sigma\to {\color{red} L} \Sigma {\color{blue} R^\dagger}$ &
    $W\to {\color{red} U_c} W$, $W\to {\color{blue} U_u} 
  W {\color{blue} U_u^\dagger}$ \\[3pt] \hline
  Labels on fields: $N_{\color{OliveGreen} v}$ & 
  Labels on fields: $h^{(b)}_{\color{OliveGreen} v}$, 
  $\xi_{\color{red} n}$, $A^\mu_{\color{red} n}$ \\[3pt] \hline
Calculable matrix elements &
Calculable coefficients \\[3pt] \hline
 Perturbative loops  &
  Non-perturbative \& perturbative loops \\[3pt] \hline
 Unknown coefficients: {\color{OliveGreen} $L_i$,  $\ldots$}
 & Unknown matrix elements: 
   {\color{OliveGreen} $\phi_\pi(x)$, $f_{i/p}(\xi)$, $\ldots$}\\ \hline
\phantom{x} \\[-10pt]
 Power counting: {\Large  {\color{red} $\frac{E}{\Lambda}$},
  {\color{blue} $\frac{m_{\pi}}{\Lambda}$} } &
   Power counting: {\Large {\color{red} $\frac{\Lambda}{E}$}, 
  {\color{blue} $\frac{m_\rho}{E}$} } \\
\end{tabular}
\end{center}
\vspace{-.2cm}
\end{table}

\newabstract 
\label{abs:kaiser}

\begin{center}
{\large\bf Chiral Dynamics of Nuclear Matter}\\[0.5cm]
S. Fritsch$^{1}$, {\bf N. Kaiser}$^1$ and W. Weise$^{1,2}$\\[0.3cm]
$^1$Physik-Dept. T39, TU M\"unchen, D-85747 Garching, Germany\\[0.3cm]
$^2$ECT$^{*}$, Villa Tambosi, I-38050 Villazzano (Trento), Italy.\\[0.3cm]
\end{center}
In recent years a novel approach to the nuclear matter many-body problem based
on effective field theory (in particular chiral perturbation theory) has
emerged \citetwo{lutz}{pap1}. 
The key element there is a separation of long- and 
short-distance dynamics and an ordering scheme in powers of small momenta. At 
nuclear matter saturation density, $\rho_0 \simeq 0.17\,{\rm fm}^{-3}$, the 
Fermi-momentum $k_{f0}$ and the pion mass $m_\pi$ are comparable scales
($k_{f0}\simeq 2m_\pi$), and therefore pions must be included as explicit 
degrees of freedom in the description of the nuclear many-body dynamics. In
ref.\cite{pap1} we have calculated the equation of state of isospin-symmetric 
nuclear matter in the three-loop approximation of chiral perturbation theory. 
The contributions to the energy per particle $\bar E(k_f)$ as they arise from 
$1\pi$- and $2\pi$-exchange (closed vacuum) diagrams are ordered in powers of 
the Fermi-momentum $k_f$ as
\begin{equation}\bar E(k_f) = \sum_{\nu = 2}^5 k_f^\nu {\cal F}_\nu(k_f/m_\pi)
\,. \end{equation}
Note that each expansion coefficient ${\cal F}_\nu(k_f/m_\pi)$ for $\nu \geq 3$
is itself a function of the ratio of the two small scales, $k_f$ and $m_\pi$,
inherent to this calculation. It has been demonstrated that the empirical
saturation point, $\bar E(k_{f0})\simeq -15.3 $\,MeV, $\rho_0 \simeq
0.17\,$fm$^{-3}$, and the nuclear matter compressibility $K = k^2_{f0}\bar
E''(k_{f0})\simeq 255\,$MeV can well reproduced at order ${\cal O}(k_f^5)$ in
the small momentum expansion with just one single momentum cut-off of $\Lambda
=7f_\pi \simeq 0.65$\,GeV which parametrizes the necessary short-range
NN-dynamics. The underlying saturation mechanism is surprisingly simple (in the
chiral limit $m_\pi=0$), namely the proper combination of an attractive
$k_f^3$-term and a repulsive $k_f^4$-term. In calculation, the attractive
$k_f^3$-term comes mainly from iterated $1\pi$-exchange (regulated by a
momentum cut-off $\Lambda$) whereas the stabilizing $k_f^4$-term is induced by
Pauli-blocking in the medium. 

In the same framework, we calculate the density-dependent asymmetry energy 
$A(k_f)$ and obtain at nuclear matter saturation density $\rho_0$ the value 
$A(k_{f0})\simeq 34\,$MeV. This prediction is in good agreement with the 
empirical value of the asymmetry energy $A(k_{f0})= (33\pm 4)\,$MeV. The
equation of state of pure  neutron matter is also in fair agreement with
sophisticated many-body calculations and a resummation result \cite{steele} 
but only for low neutron densities $\rho_n <0.25\,$fm$^{-3}$. The mere fact
that pure neutron matter comes out unbound is already non-trivial.  

Furthermore, we evaluate the momentum and density dependent complex 
single-particle potential $U(p,k_f)+i\,W(p,k_f)$ \cite{pap2} (i.e. the average
nuclear mean field). Chiral $1\pi$- and $2\pi$-exchange give rise to a 
potential depth for a nucleon at the bottom of the Fermi sea of $U(0,k_{f0})
= -53.2\,$MeV. This value is in good agreement with the depth of the empirical 
optical  model potential and the nuclear shell model potential. The imaginary 
single-particle potential $W(p,k_f)$ is generated entirely by iterated 
$1\pi$-exchange. The half-width of a nucleon-hole state at the bottom of the 
Fermi sea comes out as $W(0,k_{f0})= 29.7\,$MeV. This number is comparable 
to results of many-body calculations based on the Paris-potential or effective
Gogny forces. The basic theorems of Hugenholtz-Van-Hove and Luttinger are 
satisfied in our perturbative calculation. 

Next, we extend this calculation of nuclear matter to finite temperatures 
\cite{pap3}. The free energy per particle $\bar F(\rho,T)$ is constructed from
the interaction kernels by convoluting them with Fermi-Dirac 
distributions. The calculated pressure isotherms $P= \rho^2\,\partial\bar F/
\partial \rho$ display the familiar first-order liquid-gas phase transition of
isospin-symmetric nuclear matter. The predicted critical point, $T_c \simeq
25.5\,$MeV and $\rho_c \simeq 0.09\,$fm$^{-3}$, lies however somewhat too high
in temperature. This feature originates from the too strong momentum 
dependence of the real single-particle potential $U(p,k_f)$ near the 
Fermi-surface $p=k_f$. We consider also pure neutron matter at $T>0$ and find 
fair agreement with sophisticated many-body
calculations for low neutron densities $\rho_n <0.2\,$fm$^{-3}$.  

Finally, chiral $1\pi$- and $2\pi$-exchange determine a nuclear energy density 
functional of the form \cite{pap4}:
\begin{eqnarray} {\cal E}[\rho,\tau,\vec J\,] &=& \rho\bar E(k_f)+\bigg[\tau-{3
\over 5} \rho  k_f^2\bigg] {1\over 2\widetilde M^*(\rho)} \nonumber \\ && + 
(\vec \nabla \rho)^2F_\nabla(k_f)+  \vec \nabla \rho \cdot\vec J\, F_{so}(k_f)
+ \vec J\,^2 \, F_J(k_f) \,, \end{eqnarray}
with $\rho(\vec r\,) =2k_f^3(\vec r\,)/3\pi^2$  the local nucleon density, 
$\tau(\vec r\,)$ the local kinetic energy density and $\vec J(\vec r\,)$ the 
local spin-orbit density. This functional (with explicitly density dependent 
strength functions $F_{\nabla,so,J}(k_f)$) may open a novel approach to nuclear
structure calculations which truly start from the long-range pion-induced 
NN-interaction.

\newabstract 
\label{abs:wg1}

\begin{center}
{\large\bf Summary of the Working Group I:}\\[0.1cm]
{\large\bf Goldstone Boson Dynamics}\\[0.5cm] 
{\bf Johan Bijnens}$^1$ and {\bf Ada Farilla}$^2$\\[0.3cm]
$^1$Dep. Theor. Phys. 2, Lund University,\\
S\"olvegatan 14A, S22362 Lund, Sweden\\[0.3cm]
$^2$Dipartimento di Fisica dell'Universit\`a ``Roma Tre'' e Sezione INFN,\\
Via della Vasca Navale 84, I-00146 Roma, Italy.
\\[0.3cm]
\end{center}

The Goldstone Boson working group had a rather extensive program. We had
planned 29 presentations of which 13 were experimental and 16 theoretical. 
Two of the
experimental presentations were canceled just before the meeting.
The presentations were divided in a series of rather overlapping sessions
in particular Kaons, $V_{us}$-$V_{ud}$, the Anomaly,
Chiral Perturbation Theory (ChPT),
hadronic
contributions to the muon $g-2$, Roy Equations, the transition from 2 to
3 flavours as well as eta and phi physics. Each section contained experimental
and theoretical talks. For more details and references we refer to the
speakers own summaries.

{\bf Kaons:} Here we had two talks on nonleptonic Kaon decays
in the framework of ChPT. G.~Ecker reported on ongoing work in isospin breaking
effects in $K\to2\pi$ and CP-violation in that system. J.~Bijnens discussed
the calculation of $K\to3\pi$ decays in the isospin limit and first
results on local isospin breaking. S.~Giudici described the new NA48 rare
decay data with an emphasis on those testing ChPT predictions.

{\bf\boldmath$V_{us}$-$V_{ud}$:} This topic was covered in the plenary talk by
Neufeld. Here we concentrated on a series of experimental talks
regarding both $K_{\ell3}$ decays and pion beta decay. 
The new results on $K_{\ell3}$ were from ISTRA+ (V.~Obrazstov)
 and KEP-PS E246 (V.~Anisimovsky) both 
presenting new measurements of the slopes. The most surprising new result
presented was the new BNL E865 measurement of the branching ratio for
$K^+_{e3}$ which differs by several sigma from the PDG average and if confirmed
would solve the unitarity problem in this sector. E.~Frlez showed the
contribution from the PIBETA experiment to this puzzle, the branching
ratio of pionic beta decay leading to a value of $V_{ud}$ compatible with the
PDG one.
J.~Bijnens presented ChPT results at two loops for $K_{\ell3}$ in the isospin
limit with two main conclusions. The curvature of the $f_+(t)$ form factor
might be important and the needed $p^6$ low energy constants for $f_+(0)$
can be obtained from the slope and curvature of $f_0(t)$.

{\bf Anomaly:}
The anomaly plays an important role in field theory.
Its role in the decay $\pi^0\to\gamma\gamma$ was discussed in the plenary
talk by Goity. In the WG M.~Moinester described uncertainties on the
result for $F_{3\pi}(0)$ resulting from the choice of different form factors
in the experimental subtraction towards a possible future COMPASS measurement.
The $\pi^0$ lifetime experiments and the present status of the JLab experiment
PRIMEX were discussed by A.~Gasparian. 

{\bf ChPT issues:} M.~B\"uchler presented work on the renormalization group
in ChPT at all orders. It was shown how the leading logarithms can in
principle be obtained by calculating only one loop diagrams.
The scalar form factor of pions and kaons were
presented at two loops in ChPT by P.~Dhonte, several large corrections
were found and a comparison with dispersive results was made.
I.~Scimemi discussed in detail the problem of defining precisely what are
electromagnetic corrections in an effective theory while R.~Unterdorfer
described how the generating functional method of calculating processes
has been extended to the case of three propagators.

{\bf\boldmath$(g-2)_\mu$ hadronic contributions:} A major part of the WG was
the session devoted to hadronic corrections. G.~Colangelo presented the
constraints from the $\pi\pi$ phase shift analysis on the $\rho$ contribution.
The updated CMD-2 results on $e^+e^-\to$hadrons at low energy were
discussed by B.~Shwartz while M.~Davier gave a general overview of all
the hadronic input used and their final results. He also compared
the tau decay spectral functions with those from $e^+e^-$. Finally, F.~Nguyen
showed the new KLOE analysis for $e^+e^-\to$hadrons using the radiative return
method.

{\bf Roy Equations:} Using dispersion relations and crossing leads to a series
of integro-differential equations for amplitudes generically referred to
as Roy Equations. P.~B\"uttiker presented a reanalysis of this system of
equations in the $\pi K$ sector. The Roy equations for the $\pi\pi$ system
were reanalyzed in great detail by Colangelo et al. (CGL) The precision claimed
in this calculation has been challenged by Yndurain and Pelaez. This
criticism was presented by F.~Yndurain and the reply to these by G.~Colangelo.
The original precision claimed is very high and thus needs to be checked
independently. The situation at present is that the CGL analysis seems to
have survived this challenge but the debate is still ongoing.

{\bf From two to three Flavours:} J.~Stern and S.~Descotes both discussed
aspects of the difference between ChPT for two and three flavours concentrating
on the issue of whether disconnected $\bar{s}s$ loops have a large effect.
They tested this question using a Bayesian approach in dealing with the input
as well as proposed a possible solution to this problem by not
rewriting the quark masses at higher order in terms of the meson masses.

{\bf\boldmath$\eta$, $\eta^\prime$ and $\phi$ decays:} 
On the theory side three talks presented results here. B.~Borasoy treated
mainly $\eta^\prime$ decays using a Bethe-Salpeter based resummation
scheme and found good agreement with many decays. 
E.~Oset and
L.~Roca discussed several decays using the chiral unitary approach
presented by L.~Oller in his plenary talk. E.~Oset talked about
$\phi\to\pi^0\pi^0\gamma, \pi^0\eta\gamma$ and was very happy with the
new KLOE data which agreed much better with his predictions. L.~Roca showed
how this method describes $\eta\to\pi^0\gamma\gamma$ and the medium dependence
of the $\sigma$ mass. The KLOE eta decay results and the status of many
ongoing analysises was presented by B.~Di~Micco.


\newabstract 
\label{abs:wg2}

\begin{center}
{\large\bf Summary of Working Group II: Meson-Baryon Dynamics}\\[0.5cm]
{\bf R. Beck$^1$ and T.R. Hemmert$^2$}\\[0.3cm]
$^1$Institut f\"ur Kernphysik, Universit\"at Mainz,
Becherweg 45, D-55099 Mainz \\[0.3cm]
$^2$ Theoretische Physik T39, TU M\"unchen, James-Frank-Str., 
D-85747 Garching \\[0.3cm] 
\end{center}

With 22 presentations from 10 different countries (of which 14 had a 
theoretical and 8 an experimental focus) this working group on chiral 
dynamics in the single nucleon sector covered a wide range of topics, 
attesting to the fact that 9 years after the first Chiral Dynamics 
conference this continues to be a very active area of research, 
driven by the close interaction between theory and experiment.

For individual summaries of these 22 contributions we refer to the 
following pages of these proceedings. Here we only want to highlight 
a few developments which we expect generate plenty of interesting results 
in the next few years and are likely to concern this working group at 
the time of next Chiral Dynamics conference in 2006:

\begin{itemize}
\item SU(3) meson-baryon dynamics: 
New $K^+ \Lambda$ and $K^+\Sigma$ data have been shown in this working group 
for the differential cross section and the hyperon polarization. In the future 
precise polarization observables (polarized photons and recoil polarization) 
will become available to separate the different contributions (background, 
nucleon resonances, ...).
Encouraging theoretical progress has been made by different groups utilizing 
various forms of non-perturbative/unitarization methods. As we expect new 
precision data in Kaon-photoproduction near threshold in the next few years, 
it will be a challenge for theory to find out whether there exists a (small?) 
region of applicability for purely perturbative SU(3) BChPT and at which 
energy such a framework could be matched onto the existing 
unitarized/non-perturbative theories of SU(3) meson-baryon dynamics.

\item $\Delta$(1232) resonance region: 
Over the last few years double polarization experiments have been performed 
to test the GDH sum rule and the forward spin polarizability. Very precise 
data were shown for the proton and for the future the experimental 
focus will be on the neutron. New data on 2$\pi^0$ photoproduction off the     
proton were presented, which allowed for the first time a test of ChPT in the 
2$\pi^0$ threshold region. In addition a preliminary value for the pion 
polarizability has been extracted from radiative pion-photoproduction 
experiment.
Compton scattering and 
pion-photoproduction have to be addressed simultaneously in upcoming 
theoretical calculations to obtain results in ChEFT which are competitive 
with existing models. Two different ChEFT frameworks 
(SSE and $\delta$-expansion) exist and should both be further 
explored/developed. The non-perturbative treatment of the width of the 
resonance in this region does not pose a problem for either framework. 
Improved calculations of these two scattering processes up into the 
resonance region will serve as the basis for better theoretical predictions 
of nucleon dynamical polarizabilities, $N\Delta$-transition form factors etc.  

\item Pion threshold region: 
The large discrepancy between ChPT and pion-electroproduction data reported 
at the Chiral Dynamics workshop 2000 at JLab seems to be resolved. A new 
measurement has been performed at $Q^2<0.1$ GeV$^2$, which gives significant 
higher cross sections and a better agreement with ChPT. 
Existing non-relativistic calculations of pion photo- and electroproduction 
near threshold are being extended to ${\cal O}(p^4)$ in relativistic BChPT. 
From a theoretical point of view this should shed new light on the still 
puzzling experimental situation in pion-electroproduction for 
$Q^2<0.1$ GeV$^2$. From the comparison between relativistic and 
non-relativistic calculations one also hopes to learn more about the 
(radius of) convergence of BChPT.

\item ${\cal O}(p^4)$ BChPT: The role of the large "pole-contributions" 
identified in forward virtual Compton scattering needs further study. 
The hope is that a resummation leads to a better-behaved perturbative 
expansion. A satisfying solution of this problem will be mandatory for 
any further progress in ${\cal O}(p^4)$ ChEFT calculations pertaining 
to the low-energy spin structure of the nucleon, VCS on the proton, etc.

\item Regularization Schemes: A host of schemes and ideas pertaining 
to regularization in ChEFT has been presented in the working group. 
We expect that the "effectiveness" in the sense of "simplification of 
large calculations at high chiral orders" will decide which ideas/proposals 
are taken up by the community. "Cutoff methods" are likely to play a 
prominent role in the future as they can serve as a microscope to detect 
problems of convergence in a chiral series. 
\end{itemize} 

New experimental capabilities, polarized beams, polarized targets and full 
acceptance instruments at the different laboratories will open the 
possibilities to isolate small components of the cross section via measurements
of interference terms. We hope this will result in exciting new data for the 
topics of Chiral Dynamics.

Finally, we would like to thank all the participants of this working group 
for their active participation and all speakers for the high quality of their 
presentations resulted in a stimulating and productive working group.
We also thank the organizer of the workshop, Ulf Mei{\ss}ner, as well as the 
hospitality of the Bonn university. We hope to see all of you again at 
Duke University in 1000 days!

\newabstract 
\label{abs:wg3a}

\begin{center} 
{\large\bf Summary of Working Group III: \\[0.1cm] 
Few-Body Dynamics}\\
[0.5cm] Paulo Bedaque$^1$ and Nasser Kalantar-Nayestanaki$^2$\\
[0.3cm] $^1$LBL, \\
 One Cyclotron Road, Berkeley, California, U.S.A. \\ 
$^2$KVI, \\ Zernikelaan 25, Groningen, The Netherlands.\\
[0.3cm] 
\end{center}   
In the theory part of the working group we had the presentation of 12 talks:
\begin{enumerate}       
\item U. van Kolck, Charge Symmetry Breaking in Pion Production
\item A. Nogga, Probing chiral interactions in light nuclei
\item F. Myhrer, EFT and low energy neutrino deuteron reactions
\item S. Ando, Solar-neutrino reactions on deuteron in EFT* and radiative
  corrections of neutron beta decay
\item M. Birse, EFT's for the inverse-square potential and three-body problem
\item H. Griesshammer, All orders power counting for the three-nucleon system
\item J. Donoghue, Nuclear Binding Energies and Quark Masses
\item C. Hanhart, Effective field theory treatment for meson production in NN
  collisions
\item D. Cabrera, Vector meson properties in nuclear matter
\item A. Schwenk, A renormalization group method for ground state properties
  of finite nuclei
\item L. Platter, The 4-body system in effective field theory
\item H. Krebs, Electroproduction of neutral pions off deuteron near the
  threshold
\end{enumerate}  
The talks aimed at complementing the topics discussed in the plenary sessions.
Griesshammer, Birse and Platter talked about the effective theory obtained by
integrating out pions and, in particular, the subtle issues arising in three
or more nucleon systems. Birse presented a configuration space analysis of the
renormalization in the thre-body problem that agrees with the standard power
counting.  Griesshammer showed how to extend this power counting to arbitrary
orders in the low energy expansion and exemplify it with NNLO calculations of
neutron-deuteron phase shifts. van Kolck and Hanhart discussed charge symmetry
breaking and how the recent and future experiments can help to determine the
low energy constants involved. Myhrer, Ando and Krebs used effective field
theories and related methods to make prediction on innelastic processes
involving weak current (Myhrer, Ando) and pion production (Krebs). Myhrer
advocated that well established models of exchange currents make a prediction
for the two-nucleon weak currents with very small uncertanties. Nogga showed
the first many-nucleon calculations (through the no-core shell model method)
using a potential derived from chiral perturbation theory while Schwenk
discussed a renormalization group inspired method to deal with many-body
problems. Donoghue presented a way of estimating two-nucleon low energy
constants using dispersion relations. Cabrera used a chiral model to discuss
vector meson properties in a medium.

For the experimental work, there were a total of 8 experimental talks in 
the working group. They were given by:  
\begin{enumerate}       
\item T. Black, Precision scattering length measurements        
\item St. Kistryn, pd break-up scattering experiments   
\item J. Messchendorp, Few-body studies at KVI  
\item F. Rathmann, pp scattering experiments     
\item H. Sakai, pd elastic scattering experiments       
\item H. Schieck, Nd scattering at low energies         
\item E. Stephenson, CSB experiment at IUCF     
\item W. Tornow, Nd scattering at low energies 
\end{enumerate}    

A summary of all these talks can be found in these proceedings.  
The bulk of the presentations concentrated on studying the three-body 
systems  with the aim of understanding possible effects of the three-nucleon
 forces. It  is clear from these measurements that calculations relying 
only on the two-body  forces fail to predict almost all the physical 
observables at all energies and  the need to add a three-nucleon force 
is present. However, more work needs to  be done to formulate three-body 
forces which can explain the bulk of the  data measured with relatively 
high precision. In order to understand the  three-body systems fully, 
more spin observables should be measured but  also the break-up channel
 in nucleon-deuteron scattering seems to be very  promising due to its 
much richer phase space compared to the elastic channel.  Although there
 is a large data-base for the two-body system, attention should be paid 
to some spin observables. Bremsstrahlung results in the two-body  sector
 also lack a good theoretical understanding of the process.  
Measurements at low energies are promising since they could deal 
with more  fundamental quantities which are needed in all the theories. 
More experimental  efforts are underway. The relatively high-precision 
results from IUCF show  very clearly the charge-symmetry breaking. More
 theoretical work is being  performed to describe this experimental 
observation.  


\newabstract  
\label{abs:wg3b}

\begin{center}
{\large\bf Summary of Working Group III: 
Hadronic Atoms}\\[0.5cm]
Akaki Rusetsky\\[0.3cm]
Universit\"{a}t Bonn, Helmholtz-Institut f\"{u}r
Strahlen- und Kernphysik (Theorie),\\
Nu\ss allee 14-16, D-53115 Bonn, Germany\\[0.3cm]
\end{center}

The discussions in the working group were focused on the following questions:

{\em 1. Atoms containing strange hadrons.} Nemenov (DIRAC coll.)
has presented a new proposal to measure the energy-level 
shifts and the decay widths of both $\pi\pi$ and $\pi K$ systems in 
the external magnetic field \cite{Nemenov}. The preliminary
result for the $\pi\pi$ atom lifetime measurement
$\tau=(3.1^{+0.9}_{-0.7}~(stat)\pm 1.~(syst))\cdot 10^{-15}~s$ has been 
also reported. The preliminary results for the 
strong energy shift and the width of the kaonic hydrogen 
$\epsilon_{1s}=-175\pm 26\pm 15~eV$ and $\Gamma_{1s}=120\pm 100\pm 15~eV$
were reported by Cargnelli (DEAR coll. \cite{Cargnelli}). However, as stressed
by Gasser, the precise extraction of the $KN$ $\sigma$-term and strangeness 
content of the nucleon by using the DEAR data still poses a big challenge
for the theory \cite{Gasser}.

{\em 2. Isospin-breaking corrections to the energy levels and decay 
widths of the hydrogen-like hadronic atoms.} 
Zemp has discussed a systematic 
use of the non-relativistic effective Lagrangian approach
to study the decay of the $\pi^-p$ atom in the presence of the electromagnetic
$n\gamma$ decay channel. Full calculation of the $O(\alpha,m_d-m_u)$ 
corrections to the known leading-order result is forthcoming \cite{Zemp}. 
Selected isospin-breaking corrections in the pionic and kaonic hydrogen
have been addressed within the relativistic framework by Ivanov \cite{Ivanov}.

{\em 3. ChPT for heavy nuclei.}
Using ChPT to systematically study the properties
of the deeply bound states of pions has been discussed by Girlanda 
\cite{Girlanda}. Calculation of the 
pion-nucleus optical potential at $O(p^5)$, with the inclusion of all
electromagnetic and strong isospin-breaking effects, has been presented.

\newabstract 
\label{abs:wg4}

\begin{center}
{\large\bf Summary of Working Group IV:  
Chiral Lattice Dynamics}\\
{\large\em or$\ldots$}\\
{\large\em Why we need the lattice and why the lattice needs chiral EFT}
\\[0.5cm]
Elisabetta Pallante$^1$ and Martin Savage$^2$\\[0.3cm]
$^1$S.I.S.S.A. and INFN, Sezione di Trieste, Via Beirut 2-4, 34013 Trieste, 
Italy\\[0.3cm]
$^2$Department of Physics, University of Washington
Seattle, WA 98195, USA\\[0.3cm]
\end{center}

Lattice QCD is presently the only known rigorous method of computing strong 
and electroweak matrix elements between hadronic states.
This involves formulating QCD on a euclidean lattice with 
$1/L\ll m_\pi\ll \Lambda_\chi \ll 1/a$ 
($L$ is the lattice size, $a$ is the lattice spacing,
and $\Lambda_\chi$ the scale of chiral symmetry breaking).
Presently, typical parameters for lattice simulations are $L\sim 3~{\rm fm}$,
$a\sim 0.1~{\rm fm}$ and  light quark masses $m_q\sim m_{strange}/2$, still
quite far from the ideal regime. 
The goal of many near future simulations is to move toward 
the $L\to\infty$,  $m_q\to m_{u,d}^{phys}$ and $a\to 0$ limits.
Deviations from each of the three limits 
can be perturbatively described in terms of a  systematic expansion within an
effective field theory (EFT). 
Recent progress in the optimization of lattice fermion formulations was
described  by R. Mawhinney, with emphasis on domain wall (DW) fermions.
Nice tests of the chiral behavior of overlap fermions were presented 
by C. H\"obling, who reported on recent results for the pseudoscalar mass, 
the chiral quark condensate, $B_K^{\overline{MS}}(2\,\mbox{GeV})$, and  
$B_{7,8}^{3/2}$. 
Some problems due to operator mixing for particular fermions can 
be mitigated by a {\em twisted mass transformation} (Sint).
We were all excited to learn that simulations on $\geq$ Tflop machines with 
largely reduced lattice masses will be feasible during the next year or so.
Questions concerning tests of locality for  
overlap fermions were asked.

For different size lattices two alternative expansions can be 
implemented: the $p$-expansion, when $mL\gg 1$ and the $\epsilon$-expansion,
for $mL\ll 1$, with the product $m\Sigma V$ kept fixed. 
As discussed by   Chiarappa and D\"urr,
the two regimes offer alternative ways of constructing 
hadronic matrix elements and extracting the low-energy constants (LEC)
of $\chi$PT.

Shoresh stressed that, as partially-quenched QCD (PQQCD)
contains QCD when $N_{sea}=3$ and $m_{sea}=m_{valence}$, 
one can use $N_{sea}=3$ 
simulations with unphysical masses to determine 
the LEC of QCD.
Shoresh also presented results for the Wilson PQ$\chi$PT lagrangian 
in the meson sector up to ${\cal O}(a^2)$ in the lattice spacing.
Fits to PQQCD lattice data (with $N=2$) using Wilson PQ$\chi$PT are under 
study, as presented by Montvay.
Very nice results were reported by C. Aubin, using an EFT formulation 
for staggered fermions. The fit to lattice data with $N=3$
provides preliminary values of the strong LEC, 
$L_5 = 1.9(3)\left ({}^{+6}_{-2}\right )$, 
$L_4 = 0.3(3)\left ({}^{+7}_{-2}\right )$, 
$2L_8 - L_5 = -0.1(1)\left ({}^{+1}_{-3}\right )$, 
$2L_6 - L_4 = 0.5(2)\left ({}^{+1}_{-3}\right )$
and the ratio $f_K/f_\pi = 1.201(8)(15)$, in good agreement
with  nature.

Kaon weak matrix elements, in particular $K\to\pi\pi$ decays, are
a hot topic for lattice QCD, as $\varepsilon'/\varepsilon$
provides a crucial test of the Standard Model and its   
extensions. G. Villadoro explained how (partial) quenching affects 
in a crucial way $\Delta I=1/2$ amplitudes. 
The size of non-factorizable vs factorizable contributions to penguin matrix 
elements was discussed (S. Peris). 
They find a value of $\langle Q_8\rangle^{3/2}$ 
that is much larger than the lattice value, making the situation still 
controversial.

Another exciting topic studied in  lattice QCD is B meson physics.
Quenched lattice data of the $B\to\pi$ transition that is necessary for the 
extraction of $\vert V_{ub}\vert$ was analyzed using Q$\chi$PT in an effort
to estimate systematic errors in  the chiral extrapolation 
(Becirevic).
It appears that partially quenched simulations 
at smaller quark masses are necessary for a more quantitative understanding.

The next few years promise to be 
a very exciting time for strong interaction physics as we are starting 
to see lattice calculations of the nucleon properties, such as
the magnetic moments, matrix elements of the $n\le 3$ twist-2 operators,
masses, and also the $NN$ scattering lengths.
Nucleon form factors were discussed in our working group, 
with presentations of quenched calculations of the isovector anomalous
 magnetic moment and charge radii
(G\"ockeler) and the strange form factors (Lewis).
Significant work has been accomplished in examining the properties of nucleons
in quenched and partially-quenched chiral perturbation theory, Q$\chi$PT and
PQ$\chi$PT respectively (Beane and Arndt).
In addition, first calculations of the ${\cal O}(a)$ corrections to
nucleon properties arising from both the Sheikholeslami-Wohlert term in the
Symanzik action and also from ${\cal O}(a)$ corrections to the inserted
operators themselves, which are relevant for actions using  unimproved Wilson
fermions, were presented.  The importance of studying the NN scattering lengths
on the lattice was stressed by Beane.

Chiral extrapolations have received attention recently.  
A presentation by Procura on the extrapolation of the nucleon mass and
$g_A$ was interesting.  One would like to understand much better why some of
the forms used in chiral extrapolations can recover the experimental value,
despite extrapolations from pion masses that are clearly outside the range of
convergence.
It was reiterated that 
results must be regulator independent up to higher order terms in the 
expansion.

Maarten Golterman lead the working group in a very lively and interesting
discussion concerning what one can expect in the next five years and what we
should be thinking about.
Five areas emerged for scrutiny: finite volume effects, finite lattice spacing
effects, staggered fermions, (partial-) quenching and chiral perturbation
theory.

In conclusion, the chiral lattice dynamics working group was very stimulating
and a good addition to the chiral dynamics meeting.
The next five years will see significant progress in numerical lattice 
simulations, algorithm development and in the development of the 
appropriate  effective field theories.
We are entering a time period where rigorous statements about strong
interaction observables can be made directly from first principles.

\newpage
\section*{Working Group I: Goldstone Boson Dynamics}

\centerline{Convenors: Johan Bijnens and Ada Farilla}

\vskip5mm
{
\begin{tabular}{lp{9.9cm}r}
\stalk{S. Giudici}{ NA48 Results on $\chi PT$}{abs:giudici}
\stalk{G. Ecker}{ $K \to 2 \pi$, $\epsilon' / \epsilon$ 
and Isospin Violation}{abs:ecker}
\stalk{J. Bijnens}{ $K \to 3 \pi$ in Chiral Perturbation Theory}{abs:bijnens1}
\stalk{A.A. Poblaguev}{ E865 Results on $K_{e3}$}{abs:poblaguev}
\stalk{V. Obraztsov}{ High Statistics Study of 
the $K^- \to e \nu \pi^0$ and\newline
 $K^- \to \pi^- \pi^0 \pi^0$ 
Decays with ISTRA+ Setup}
{abs:obraztsov}
\stalk{V. Anisimovsky}{ Measurement of the $K_{e3}$ Form Factors Using Stopped
Positive Kaons}{abs:anisimovsky}
\stalk{J. Bijnens}{ $K_{l3}$ at Two Loops in ChPT}{abs:bijnens2}
\stalk{E. Frlez}{ Pion Beta Decay, $V_{ud}$ and $\pi \to l \nu\gamma$}
{abs:frlez}
\stalk{M.A. Moinester}{ A New Determination of the $\gamma\pi \to \pi\pi$ 
Anomalous\newline Amplitude Via 
$\pi^-  e \to \pi^- e \,\pi^0$ Data}{abs:moinester}
\stalk{A. Gasparian}{ The $\pi^0$ Lifetime Experiment 
and Future Plans at JLab}{abs:gasparian}
\stalk{M. B\"uchler}{ Renormalization Group Equations for Effective Field
Theories}{abs:buechler}
\stalk{P. Dhonte}{ Scalar Form Factors to $O(p^6)$  in SU(3) Chiral\newline 
Perturbation Theory}{abs:dhonte}
\stalk{I. Scimemi}{ Hadronic Processes and Electromagnetic Corrections}
{abs:scimemi}
\stalk{R. Unterdorfer}{ The Generating Functional of Chiral SU(3)}
{abs:unterdorfer}
\stalk{G. Colangelo}{ Hadronic Contributions to $a_\mu$ below 
1 GeV}{abs:colangelo1}
\stalk{F. Nguyen}{ The Measurement of the Hadronic Cross Section 
with KLOE
Via the Radiative Return}{abs:nguyen}
\stalk{M. Davier}{ 
Hadronic\,Vacuum\,Polarization:\,the\,\,$\mu$ Magnetic\,Moment}
{abs:davier}
\stalk{B. Shwartz}{ CMD-2 Results on $e^+ e^- \to$ hadrons}{abs:shwartz}
\stalk{P. B\"uttiker}{ Roy--Steiner Equations for $\pi K$ 
Scattering}{abs:buettiker}
\stalk{F.J. Yndur\'ain}{ Comments on some Chiral-Dispersive Calculations
 of $\pi\pi$ Scattering}{abs:yndurain}
\stalk{G. Colangelo}{ On the Precision of the Theoretical 
Calculations for $\pi \pi$ Scattering}{abs:colangelo2}
\stalk{J. Stern}{ From Two to Three Light Flavours}{abs:stern}
\stalk{S. Descotes-G.}{ Bayesian Approach to the Determination of\newline 
$N_f$=3 Chiral Order
Parameters}{abs:descotes}
\stalk{B. Borasoy}{ Hadronic Decays of $\eta$ and $\eta'$}{abs:borasoy}
\stalk{E. Oset} { $\Phi$ Radiative Decay to Two Pseudoscalars}{abs:oset2}
\stalk{B. Di Micco}{ $\eta$ Decays Studies with KLOE}{abs:dimicco}
\stalk{L. Roca}{ $\eta \to \pi^0 \gamma \gamma$  Decay within a Chiral
Unitary Approach}{abs:roca}
\end{tabular}
}
\newpage

\newabstract 
\label{abs:giudici}
 
\begin{center}
    {\large\bf NA48 Results on {\boldmath $\chi PT$}}\\[0.5cm]
    {\bf Sergio Giudici} for Na48 coll.\\[0.3cm]
    Scuola Normale Superiore \\
    7, P.za dei Cavalieri, Pisa, Italy
  \end{center}
\vspace{0.1cm}
  The experiment Na48, located at the CERN SPS accelerator, has originally 
  been designed to precisely measure direct CP violation in the decays of
  neutral kaon to pions pair. Apart from the important result concerning
  the CP violation parameter $Re(\epsilon^{\prime}/\epsilon$) \cite{eps}, the 
  experiment investigated some neutral kaon rare decays as well.
  
  \begin{itemize}  
  \item { $K_{L,S} \rightarrow e^+e^-\pi^+\pi^-$: the branching ratios are
      measured to be: $BR(K_{L} \rightarrow e^+e^-\pi^+\pi^-) = (3.08 \pm 0.2)
      \times 10^{-7}$ and $BR(K_{S} \rightarrow e^+e^-\pi^+\pi^-) = (4.69 \pm
      0.3) \times 10^{-5}$.  The expected adronic-leptonic angle asymmetry for
      $K_L$ has been measured to be $A_{\phi}(K_L) = (14.2 \pm 3.6) \% $
      \cite{lai1} }
    
  \item {$K_S \rightarrow \gamma \gamma$: the branching ratio is measured to
      be $BR(K_S \rightarrow \gamma \gamma = (2.78 \pm 0.07)$ revealing a
      $30\%$ discrepancy with respect to $\chi$PT at \cal{O}$(p^4)$
      calculation \cite{lai2}.}
    
  \item {$K_{S,L} \rightarrow \pi^0 \gamma \gamma$}: a first observation
    \cite{lai3} of the decays $K_S \rightarrow \pi^0 \gamma \gamma $ has been
    obtained yielding $BR(K_S \rightarrow \pi^0 \gamma \gamma)_{(z>0.2)} =
    (4.9 \pm 1.8) \times 10^{-8}$ in agreement with the $\chi$PT prediction
    \cite {theo1}.  The chiral structure of the vertex need more statistics to
    be proved.  The branching ratio for the $K_L$ is measured to be $BR(K_L
    \rightarrow \pi^0 \gamma \gamma) = (1.36 \pm 0.05) \times 10^{-6}$
    \cite{lai4}.  The corresponding VMD coupling found is $a_v = -0.46 \pm
    0.1$.
  \item {$K_S \rightarrow \pi^0 e^+ e^-$}:
    NA48 observed 7 events publishing \cite{sub} the result
    $BR(K_S \rightarrow \pi^0 e^+ e^-) = 5.8 ^{+2.8}_{-2.3}\times 10^{-9}$     
    in a remarkable agreement with the prediction \cite{theo2}
  \end{itemize}

\newabstract 
\label{abs:ecker}

\begin{center}
{\large\bf \mbox{\boldmath$K \to 2 \pi$}, 
  \mbox{\boldmath$\epsilon^\prime/\epsilon$}
  and Isospin Violation}\\[0.5cm]
Gerhard Ecker \\[0.3cm]
Institut f\"ur Theoretische Physik, Universit\"at Wien\\
Boltzmanngasse 5, A-1090 Vienna, Austria \\[0.3cm]
\end{center}

This talk is based on joint work with Vincenzo Cirigliano, Helmut
Neufeld and Toni Pich \cite{CENP}. 

We have performed a complete
analysis of isospin breaking in $K \to 2 \pi$
amplitudes in chiral perturbation theory, including both strong
isospin violation ($m_u \neq m_d$) and electromagnetic corrections to
next-to-leading order in the low-energy expansion.  The unknown chiral 
couplings are estimated at leading order in the $1/N_c$ expansion,
except for the basic weak couplings $G_8, G_{27}$ that are fitted to
the experimental branching ratios. For these couplings we find
small deviations from the isospin-limit case.\\[.2cm] 
The main results are:
\begin{itemize} 
\item Isospin breaking reduces the $\Delta I= 1/2$ ratio $Re\,A_0 /
  Re\,A_2$ by about 10 $\%$.
\item Isospin violation in the phases leads to a difference
  of about 13$^\circ$
  between the isospin-limit values extracted from $K \rightarrow \pi \pi$
  branching ratios and from $\pi\pi$ scattering data \cite{CGL}. This
  difference cannot easily be accommodated by the usual
  uncertainties of large-$N_c$ determinations of low-energy
  constants. 
\item Direct CP violation in $K^0 \rightarrow \pi \pi$ decays is 
  parametrized by the parameter $\epsilon^\prime$. We locate three
  different sources of isospin violation for $\epsilon^\prime$. For the
  overall measure of isospin violation  we find
$$
\Omega_{\rm eff} = \left\{\begin{array}{ll}
0.16 \pm 0.05 & \qquad {\rm strong~IB}\\
0.06 \pm 0.08 & \qquad {\rm strong+elm.~IB}~, \end{array}   \right.
$$
  depending on whether or not electromagnetic corrections are included 
  in the decay amplitudes.
\end{itemize}

\newabstract 
\label{abs:bijnens1}

\begin{center}
{\large\bf {\boldmath $K\to 3\pi$} in Chiral Perturbation Theory}\\[0.5cm]
{\bf Johan Bijnens}$^1$, Fredrik Borg$^{1}$ (formerly Persson)
and Pierre Dhonte$^1$\\[0.3cm]
$^1$Dep. Theor. Phys. 2, Lund University,\\
S\"olvegatan 14A, S22362 Lund, Sweden\\[0.3cm]
\end{center}

We discuss the treatment of $K\to3\pi$ in Chiral perturbation theory to
one loop \cite{k3pi1}. This has earlier been calculated
and various relations discussed in \cite{KMW}. The formulas from that
calculation have unfortunately been lost prompting a full new calculation
to take into account the new data. The $K\to2\pi$ amplitudes had already been
recalculated in \cite{BPP} and were confirmed in \cite{k3pi1} once
more. We found a simple parametrization in terms of one parameter functions
for the full amplitude and have compared our full expression to the data.

A full new fit of the Dalitz plot expansion parameters was also done. This
fit has no indication of isospin breaking and has an acceptable $\chi^2$.
The fit of the ChPT expression to the data has a fairly large $\chi^2$,
mainly due to the fact that the various curvatures in the data are poorly
fitted. It should be noted that there are large discrepancies between
various experiments and that the new data from ISTRA+ have again significantly
different curvatures \cite{ISTRA}. A full new fit is required
to see if the latter would solve some of the discrepancies.

We have embarked on a full calculation of all isospin breaking effects
on presented preliminary results on the local isospin breaking, i.e.
quark mass effects and local electromagnetic effects. The radiative corrections
from photon loops are under further investigation. The indications are
that the local isospin breaking effects do not solve the discrepancy with
the curvatures, but basically only slightly change the fitted ChPT parameters
\cite{k3pi2}.

The ChPT calculation of \cite{k3pi1} has been fully confirmed by \cite{GPS}.
There also a study of the CP asymmetries can be found.

\newabstract 
\label{abs:poblaguev}

\begin{center}
{\large\bf E865 Results on \boldmath$K_{e3}$}\\[0.5cm]
A.A.~Poblaguev for the E865 Collaboration\\[0.3cm]
Physics Department, Yale University, New Haven, CT 06511, USA and\\
Institute for Nuclear Research of Russian Academy of Sciences, Moscow, Russia
\\[0.3cm]
\end{center}

E865 \cite{E865} at the Brookhaven National Laboratory AGS collected about
70,000 $K^+\to\pi^0e^+\nu$ $(K^+_{e3})$ decays in flight with the purpose of
measuring the $K^+_{e3}$ branching ratio relative to the observed
$K^+\to\pi^+\pi^0$ ($K_{\pi2}$), $K^+\to\pi^0\mu^+\nu$ ($K_{\mu3}$), and
$K^+\to\pi^+\pi^0\pi^0$ ($K_{\pi3}$) decays \cite{Ke3}.  The $\pi^0$ in all
the decays were identified using the $e^+e^-$ pair from $\pi^0\to
e^+e^-\gamma$ decay and no photon detection was required. The branching ratio
was measured to be
$$\mathrm{BR}(K^+_{e3(\gamma)})/
[\mathrm{BR}(K_{\pi2})+\mathrm{BR}(K_{\mu3})+\mathrm{BR}(K_{\pi3}) =
0.2002\pm0.0008_\mathrm{stat}\pm0.0036_\mathrm{sys}],$$
where
$K^+_{e3(\gamma)}$ includes the effects of virtual and real photons.  Using
the Particle Data Group (PDG) braching ratios for the normalization decays we
obtain
$$\mathrm{BR}(K^+_{e3(\gamma)}) =
(5.13\pm0.02_\mathrm{stat}\pm0.09_\mathrm{sys}\pm0.04_\mathrm{norm})\%$$
This
result is about $2.3$ standard deviations higher than the current PDG value.

Radiative corrections for the decays inside the $K^+_{e3}$ Dalitz plot
boundary were estimated to be $-1.3\%$ using the procedure of Ref.
\cite{Cirigliano}, while the $K^+_{e3\gamma}$ decays outside the Dalitz plot
boundary led to a $+0.5\%$ correction. Using the total radiative correction of
$-0.8\%$, the PDG value for the form factor $f_+$ slope
$\lambda_+=0.0282\pm0.0027$, and the short distance enhancement factor
$S_{EW}(M_\rho,M_Z)=1.0232$ \cite{Cirigliano}, we obtain for the element
$V_{us}$ of CKM matrix
$|V_\mathrm{us}f_+(0)|=0.2242\pm0.0020_\mathrm{rate}\pm0.0007_{\lambda_+}$
which leads to
$$|V_\mathrm{us}|
=0.2272\pm0.0020_\mathrm{rate}\pm0.0007_{\lambda_+}\pm0.0018_{f_+(0)}$$
if $f_+(0)=0.9874\pm0.0084$ \cite{Cirigliano}. 

With this value of $V_\mathrm{us}$ and $V_\mathrm{ud}$ from superallowed
nuclear Fermi beta decays we obtain
$|V_\mathrm{ud}|^2+|V_\mathrm{us}|^2+|V_\mathrm{ub}|^2 = 0.9999\pm0.0016$.

While this result is in perfect agreement with CKM unitarity, our $K^+_{e3}$
decay rate disagrees with previous experimental measurements and increases the
discrepancy with $V_\mathrm{us}$ from $K^0_{e3}$ decay if extracted under
conventional theoretical assumptions about symmetry breaking.

\newabstract 
\label{abs:obraztsov}

\begin{center}
{\large\bf High Statistics Study of the 
{\boldmath $K^- \rightarrow e \nu \pi^0$} and\\[0.1cm]
{\boldmath $K^- \rightarrow \pi^- \pi^0 \pi^0$} 
Decays with ISTRA+ Setup}\\[0.5cm]
 {Vladimir Obraztsov}\\[0.3cm]
Institute for High energy Physics,\\
142284 Protvino,Moscow region, Russia\\[0.3cm]
\end{center}
ISTRA+ is a setup which has been operating in a negative unseparated secondary
beam of the U-70 PS \cite{setupo}.  The main statistics of the experiment was
accumulated during two runs in 2001. A simple trigger, which selected wide
class of K$^{-}$ decays was used. About 700M events were logged on tapes. A
sequence of selection cuts results in 550K events of the $K^- \rightarrow e
\nu \pi^0$ with remaining background of $\leq 1.5\%$ and 252K completely
reconstructed events of the $K^- \rightarrow \pi^- \pi^0 \pi^0$ decay with a
background of $\leq 0.05\%$. This is the highest statistics for these
decays in the world. \\
The Dalitz-plot analysis of the first decay was performed in terms of V-A
$f_{+}(t)$ formfactor, which is assumed to be , at most, quadratic function of
t: $f_{+}(t)=f_{+}(0)(1+\lambda_{+}t/m_{\pi}^{2}+
\lambda_{+}^{'}t^{2}/m_{\pi}^{4})$ and exotic scalar ($f_{S}$) and tensor
($f_{T}$) formfactors which assumed to be constant.  Different fits lead to
following results:
\begin{center}
$\lambda_{+}= 0.0286 \pm 0.0008 (stat) \pm 0.0006(syst)$;
$\lambda^{''}_{+} = -0.002^{+0.0031}_{-0.0066}$;\\
$f_{T}/f_{+}(0)=0.021^{+0.064}_{-0.075} (stat) \pm 0.026(syst) ; $ \\
$f_{S}/f_{+}(0)=0.002^{+0.020}_{-0.022}(stat) \pm 0.003(syst) $\\
\end{center}
Detailed description of the analysis and of the results can be found in
\cite{Paper1o}.\\
The second decay is analyzed in terms of the "Dalitz plot slopes" g,h,k
\begin{equation}
|A(K^{\pm} \rightarrow 3\pi)|^2 ~ \propto ~ 1+g\,Y+h\,Y^2+k\,X^2+...~,
\end{equation}
where $X = (s_1-s_2)/m_{\pi\,}^2$ and $Y = (s_3-s_0)/m_{\pi\,}^2$ ,
 $s_i = (p_K-p_i)^2$, $s_0 = {1\over3}(s_1+s_2+s_3)$, $p_K$ and $p_i$ are
the $K^{\pm}$ and $\pi_i$ four-momenta ($\pi_3$ is the odd pion). The values
obtained from the fit are: 
\begin{center}
$g=0.627\pm0.004(stat)\pm0.010(syst)$,\\
$h=0.046\pm0.004(stat)\pm0.012(syst)$,\\
$k=0.001\pm0.001(stat)\pm0.002(syst)$.
\end{center}
 The details can be found in \cite{Paper2o}.

\newabstract 
\label{abs:anisimovsky}

\begin{center}
{\large\bf Measurement of the {\boldmath $K_{e3}$} Form Factors Using 
Stopped\\[0.1cm] 
Positive Kaons}\\[0.5cm]
Valery Anisimovsky\\[0.3cm]
Institute for Nuclear Research RAS,
60th October Revolution prosp., 7a, Moscow, 117312 Russia\\[0.3cm]
(For KEK-PS E246 Collaboration)\\[0.3cm]
\end{center}

The E246 set-up was designed to measure T-violating muon polarization in the 
$K_{\mu 3}$ decay and consisted of the following main
elements: superconducting toroidal spectrometer, the CsI(Tl) photon
calorimeter, the muon polarimeter, active target, charged particle
detectors, and TOF system \cite{e246setup}.
The set-up allowed us to extract $K_{e3}$ events with background 
fraction $<0.55\%$ and measure $K_{e3}$ kinematics completely.
We were able to extract the natural 
Dalitz density by deconvolving the convolution equation (radiative 
corrections were subtracted using formulae of Ref. \cite{ginsberg}) and 
then we determined the form factors by maximizing the logarithmic likelihood 
between the MC and experimental Dalitz distributions.

We analyzed $102\times 10^3$ $K_{e3}$ events for toroidal field of 0.65T and 
obtained:
\begin{eqnarray*}
\lambda_+ &=& 0.0278 \pm 0.0017(stat) \pm 0.0016(syst) \\
|f_S/f_+(0)| &=& 0.0040 \pm 0.0160(stat) \pm 0.0078(syst) \\
|f_T/f_+(0)| &=& 0.019  \pm 0.080(stat)  \pm 0.040(syst)
\end{eqnarray*}
This result is published in \cite{levchenko}.
The values of exotic form factors are consistent with zero  predicted by
the  Standard Model and are in a good agreement with ISTRA+ results. 
The 90\% C.L. 
limits for the form factors are $|f_S/f_+(0)|<0.033$ and $|f_T/f_+(0)|<0.166$.
The limit for $|f_S/f_+(0)|$ allows us to put a constraint on the 
two Higgs doublet 
model \cite{godina}: $|\lambda_{us}\lambda_{e\nu}|<5.2\times 10^{-3}$.
The obtained $\lambda_+$ value is consistent with the ChPT prediction 
0.031 \cite{maiani}.

\newabstract 
\label{abs:bijnens2}

\begin{center}
{\large\bf\boldmath $K_{\ell3}$ at Two Loops in ChPT}\\[0.5cm]
{\bf Johan Bijnens}$^1$ and Pere Talavera$^2$\\[0.3cm]
$^1$Dep. Theor. Phys. 2, Lund University,\\
S\"olvegatan 14A, S22362 Lund, Sweden\\[0.3cm]
$^2$Departament de F{\'\i}sica i Enginyeria Nuclear,\\
Universitat Polit\`ecnica de Catalunya,\\
Jordi Girona 1-3, E-08034 Barcelona, Spain
\\[0.3cm]
\end{center}

We have calculated the decays $K\to\pi\ell\nu$ ($K_{\ell3}$) to order
$p^6$ in Chiral Perturbation Theory\cite{Kl3}. This calculation has been
performed in the isospin limit.

The full discussion can be found in \cite{Kl3} as well as in the two
published talks \cite{talks}. There are two main conclusions, one
experimental and the other theoretical.

The experimental conclusion is that in order to obtain $V_{us}$ with
a precision better than 1\% it becomes necessary to take the effect of
curvature in the form factor $f_+(t)$ into account. This curvature
can be predicted from ChPT by using the pion electromagnetic
form factor data \cite{emform}. This inclusion has also a rather strong effect
on the measured slope $\lambda_+$, typically moving it outside the error
bars quoted on $\lambda_+$ from the experiment when assuming a linear
dependence on $t$ only.

The main theoretical result relevant for $V_{us}$ is that the $p^6$
contribution to $f_+(0)$ only depends on two new parameters. These
can in turn be determined from the $K\pi$ scalar form factor slope and
curvature, thus allowing for a model independent determination of $f_+(0)$.
The curvature can be determined from other scalar form factors, as is discussed
in \cite{scalform}.

Work on the $p^6$ isospin breaking contribution is under way.

\newabstract 
\label{abs:frlez}

\begin{center}
{\large\bf Pion Beta Decay, \boldmath $V_{ud}$ 
and $\pi\to\ell\nu\gamma$}\\[0.5cm]
Emil Frle\v{z}, for the PIBETA Collaboration\\[0.3cm]
Department of Physics, University of Virginia,\\
Charlottesville, VA 22904-4714, USA\\[0.3cm]
\end{center}

We have used a non-magnetic pure CsI calorimeter (the PIBETA
detector~\cite{pb}) and stopped $\pi^+$ beam technique at PSI to collect, over
three years, a high statistics data set of (rare) $\pi^+$ and $\mu^+$ weak
decays.  >From the background-free data set of $\sim 60,000$ pion beta
($\pi\beta$) decay events we have extracted the preliminary $\pi^+\to \pi^0
e^+\nu_e$ branching ratio of
\begin{equation}
\label{eq1}
\Gamma_{\pi\beta}= (1.038\pm 0.004{\rm (stat)}
\pm  0.007{\rm (sys)})\times 10^{-8},
\end{equation}
by normalizing to the known rate of the $\pi^+\to e^+\nu_e$ 
decays~\cite{Mar93}.
This result implies the quark mixing matrix element $V_{ud}=0.9737(40)$,
in the agreement with the most recent Particle Data Group
average $V_{ud}=0.9734(8)$~\cite{PDG}. Our experiment tests for 
the first time the calculation of the $\pi\beta$ radiative 
corrections which stand at 
${\rm RC}_{\pi\beta}\sim (+3.3\pm 0.1)$\,\%~\cite{rc}.

In the framework of the Standard Model (SM) and the conserved vector current
hypothesis (CVC) the radiative pion decay process $\pi^+\to e^+\nu_e\gamma$ 
is described by the weak axial and vector form factors, $F_A$ and 
$F_V$~\cite{Bry82}. By fitting the absolute partial branching ratios deduced
from the $\sim 42,000$ radiative events' Dalitz plot we find the parameter
$\gamma\equiv F_A/F_V$ to be
\begin{equation}
\label{eq2}
\gamma_{\rm EXP}=0.475\pm 0.015{\rm (stat)} 
\pm 0.014{\rm (sys)},
\end{equation}
consistent with the present chiral symmetry 
phenomenology~\cite{Hol86}.

\label{abs:moinester}

\begin{center}
{\large\bf A New Determination of the 
{\boldmath $\gamma \pi \rightarrow \pi\pi$}
Anomalous\\[0.1cm] Amplitude Via 
{\boldmath $\pi^- e \rightarrow \pi^- e\, \pi^{0}$} Data}
\\[0.5cm] I.~Giller$^1$, T.~Ebertsh\"auser$^2$, {\bf M.~A.~Moinester}$^1$,
S.~Scherer$^2$ \\[0.3cm]
$^1$ School of Physics and Astronomy, R. and B. Sackler Faculty
of Exact Sciences,\\ Tel Aviv University, 69978 Tel Aviv, Israel\\[0.3cm]
$^2$ Institut f\"ur Kernphysik, Johannes Gutenberg-Universit\"at, D-55099
Mainz, Germany \\[0.3cm]
\end{center}

We discuss the reaction $\pi^- e \rightarrow \pi^- e \pi^{0}$ with 
the purpose of obtaining information on the  $\gamma \pi \rightarrow \pi\pi$ 
anomalous amplitude ${\cal F}_{3\pi}$.
We compare a full calculation at ${\cal O}(p^6)$ in chiral perturbation
theory and various phenomenological predictions 
\cite {Holstein:1995qj}
with the existing data of 
Amendolia {\em et al} \cite {Amendolia:1985bs}.
By integrating our theory results using Monte Carlo techniques and 
comparing them 
with the experimental cross section of $\sigma= (2.11\pm 0.47)$ nb we obtain 
an experimental value ${\cal F}_{3\pi}^{(0)\rm exp} \approx 10.0 \pm 1.0~ 
GeV^{-3}$, in good agreement with theoretical expectations.
We emphasize the need for new data to allow comparison of experimental and 
theoretical distributions, and to obtain 
${\cal F}_{3\pi}^{(0)\rm exp}$ with smaller uncertainty.

\newabstract 
\label{abs:gasparian}

\begin{center}
{\large\bf The {\boldmath $\pi^0$} 
Lifetime Experiment and Future Plans at JLab}\\[0.5cm]
A. Gasparian (for the PrimEx Collaboration)\\[0.3cm]
Department of Physics, NC A\&T State University, 
Greensboro, NC USA\\[0.3cm]
\end{center}

The three neutral light pseudoscalar mesons, $\pi^0$, $\eta$ and
$\eta^{\prime}$, represent one of the most interesting systems in low energy
QCD.  This system contains fundamental information about the effects of SU(3)
and isospin breaking by the $u$, $d$, and $s$ quark mass differences, leading
to important mixing effects among the mesons, as well as two types of chiral
anomalies. Therefore, precision measurements of the two-photon decays of the
$\pi^0$, $\eta$ and $\eta^{\prime}$ will allow one to determine the light
quark mass ratios and a direct extraction of the $\eta$, $\eta^{\prime}$
mixing angle.  Thus, providing a stringent test of the low energy limit of QCD
in a relatively clean setting.

Because of the small mass of $\pi^0$ meson, the prediction of the chiral
anomaly for the $\pi^0 \rightarrow \gamma \gamma$ decay width is more accurate
and exact in the massless quark limit:

\vspace{-0.40cm}
\[
\Gamma (\pi^0 \rightarrow \gamma\gamma)
=\frac{\alpha^2m_{\pi}^3}{64\pi^3F_\pi^2} = 7.725 ~{\rm eV}
\]

However, the real world current-quark  masses are of the order of 5-7 MeV. 
Calculations of the chiral corrections have been performed in the combined 
framework of chiral perturbation theory (ChPT) and the $1/N_c$ expansion up 
to ${\cal{O}}(p^6)$ and ${\cal{O}}(p^4\times 1/N_c)$ in the decay amplitude. 
It was shown that the next-to-leading order corrections increase the pion 
decay width by about 4\%, with an estimated uncertainty of less than 1\%.
The PrimEx collaboration at Jefferson Lab is planning to perform a precision 
measurement of the neutral pion lifetime using the small angle coherent 
photoproduction of $\pi^0$'s in the Coulomb field of a nucleus, {\em i.e.}, 
the Primakoff effect. This measurement will be a state-of-the-art experimental 
determination of the lifetime with a precision of less than 1.5\% \cite{prop1}.
Currently, we are in the final stages of the preparation of the experiment and 
expect to have the first data in the Spring of 2004.

With the planned 12 GeV CEBAF energy upgrade, we will be able to perform the
second part of this experimental program: radiative width and electromagnetic
transition form factor measurements of the more massive $\eta$ and
$\eta^{\prime}$ mesons \cite{prop2}. This comprehensive experimental program
will provide fundamental tests of both QCD and QCD inspired models.

\noindent
(This project is supported under NSF grant PHY-0079840.)

\newabstract 
\label{abs:buechler}

\begin{center}
{\large\bf Renormalization Group Equations for Effective Field\\[0.1cm] 
Theories}\\[0.5cm]
{\bf M. B\"uchler }$^1$ and G. Colangelo $^2$\\[0.3cm]
$^1$ Department of Physics,\\
University of Washington, Seattle, WA 98195-1550, USA\\[0.3cm]
$^2$ Institut f\"ur theoretische Physik,\\
Universit\"at Bern, Sidlerstr. 5, 3012 Bern, Switzerland \\[0.3cm]
\end{center}

We derive and discuss the renormalization group equations 
for an arbitrary non-renormalizable quantum field theory \cite{Buchler:2003vw}.
We show that with these equations one can get the structure of 
the leading divergences at any loop order in terms of sole 
one loop diagrams. In the framework of chiral perturbation 
theory, this allows to calculate the series of leading 
chiral logs by an iteration of one loop calculations. The 
renormalization group equations can also be used for the 
calculation of sub(sub,..)-leading divergences; these can be written 
in terms of one-loop,two-loop,(three-loop,..) diagrams.\\
Another important issue we discuss is the role of counterterms 
which vanish at the solution of the equation of motion.
Due to the freedom of choosing the couplings of these terms, one can 
find a basis of counterterms for which all divergences are generated 
only by one-particle-irreducible diagrams. This simplifies loop 
calculations considerably.   \\
Finally, we show that the renormalization group equations we obtained
apply equally well to  renormalizable theories.\\
With the help of these equations, we calculate the double chiral 
logs for the weak nonleptonic chiral lagrangian \cite{bctbp}.

\newabstract 
\label{abs:dhonte}

\begin{center}
{\large\bf Scalar Form Factors to {\boldmath ${O}(p^6)$}
in SU(3)\\[0.1cm] Chiral Perturbation Theory}\\[0.5cm]
Johan Bijnens$^1$ and {\bf Pierre Dhonte}$^1$\\[0.3cm]
$^1$Dep. Theor. Phys. 2, Lund University,\\
S\"olvegatan 14A, S22362 Lund, Sweden\\[0.3cm]
\end{center}

We present the results of a full $\mathcal{O}(p^6)$ calculation of all 
scalar form factors for the eight lightest pseudo-scalar mesons
in the framework of $SU(3)$ Chiral Perturbation Theory. This work is part of
a larger project aiming at developing and testing $SU(3)\,\,\chi$PT up to 
two-loop.

We worked in the
isospin limit and focused on the pion and kaon form factors. As a first
result we find several loop corrections to be large, as expected from the
Feynman-Hellman  theorem (see \cite{BD}). We also compared these
expressions with the behaviour extracted in the dispersive analysis of 
\cite{Moussallam1}. Using the procedure developed in \cite{ABT3} for the
analysis  of $K_{l4}$ and in the limit where the 
contribution from the $\mathcal{O}(p^6)$ low
energy constants 
(LEC) to the form factors at zero
vanishes, the matching of the scalar radii could be used to study 
the $\mathcal{O}(p^4)$ constants $L_4$ and $L_6$ while comparison of the 
curvatures gave information on the $\mathcal{O}(p^6)$ constants $C_{12}$ and
$C_{13}$. $C_{12}$ is of particular interest for the determination of $V_{us}$
via the $K_{l3}$ study performed in \cite{BT2}. The precision obtained 
on this constant was somewhat limited, but we could still
obtain new constraints
for $L_4$ and $L_6$. Thirdly, we extended to the scalar case the relation 
established by Sirlin in \cite{Sirlin} for the vector
form factors. This relation is valid at all non-special momentum transfers.

\newabstract 
\label{abs:scimemi}

\begin{center}
{\large\bf Hadronic Processes and Electromagnetic Corrections}\\[0.5cm]
Ignazio Scimemi,\\[0.3cm]
ECM-University of Barcelona,\\
 Diagonal 647, E-08028, 
Barcelona, Spain\\
[0.3cm]
\end{center}
The inclusion of electromagnetism in a low energy effective theory is worth
further study in view of the present high precision experiments (muon $g-2$,
$\pi_0\rightarrow \gamma \gamma$, $\tau$ decays, etc. , see also many talks of
this conference).  In particular in many applications of chiral perturbation
theory, one has to purify physical matrix elements from electromagnetic
effects.  The theoretical problems that I want to point out here are
following: the splitting of a pure QCD and a pure electromagnetic part in a
hadronic process is model dependent: is it possible to parametrise in a clear
way this splitting?  What kind of information (scale dependence, gauge
dependence,..) is actually included in the parameters of the low energy
effective theory?  In the recent work of ref.\cite{GRS:03} we attempt to
answer these questions introducing a possible convention to perform the
splitting between strong and electromagnetic parts in some examples.  We
conclude the following.  The splitting that is proposed is done order by order
in the loop expansion.  The strong part of a quantity depends only on
couplings defined in a theory with $e=0$ (up to the desired perturbative order
in $e$) and it has no running proportional to the electromagnetic coupling $e$
(still, up to the perturbative order in $e$ which is considered).  In order to
proceed correctly in the construction of an effective theory it is important
to characterise the relevant scales of the problem: $\mu$ (the renormalisation
scale of the underlying theory), $\mu_{\rm eff}$ (the renormalisation scale of
the effective theory) and $\mu_1$ (the scale at which the strong part of a
quantity is defined).  The splitting ambiguities are parametrised by the scale
$\mu_1$.  The uncertainty related to $\mu_1$ can be of numerical relevance as
in the pion form factor, $ F(\mu_1=1\,\mbox{\small{GeV}} ) =
F(\mu_1=500\,\mbox{\small{MeV}}) -0.1\, {\mbox{MeV}}\ .$ The error induced on
$F$ by $\mu_1$ is of the order of the PDG error~\cite{Hagiwara}.  Another
advantage of the proposed splitting, is that in the effective Lagrangian the
parameters in the strong sector are expressed through the ones of the
underlying theory in its strong sector. This makes the matching between the
underlying and the effective theory more transparent.  Finally the LECs of the
effective theory also contain all information about scale and gauge dependence
of the Green functions in the underlying theory with e.m.

\newabstract 
\label{abs:unterdorfer}
\begin{center}
{\large\bf The Generating Functional of Chiral SU(3)\footnote{Rene Unterdorfer
acknowledges the financial support provided through the European Community's
Human Potential Programm under contract HPRN-CT-2002-00311 (EURIDICE).}}
\\[0.5cm]
Rene Unterdorfer \\[0.3cm]
INFN, Laboratori Nazionali di Frascati,\\
P.O. Box 13, I 00044 Frascati, Italy.\\[0.3cm]
\end{center}
The generating functional of chiral SU(3) for strong
and nonleptonic weak interactions
\citetwo{gass2}{kambor} was extended to include up to three
propagators \cite{su3}. An application of the SU(2) version of the functional
was the analysis of four-pion production where also resonance exchange was
taken into account \cite{eu}.

Another interesting example is the decay $K_L \to \mu^+\mu^-$
that proceeds through two distinct mechanisms: the long-distance contribution
generated by the two-photon exchange and the short-distance part due to W and
Z exchange. The long-distance contribution contains
one combination of chiral couplings,
called $\chi$. Using the long-distance amplitude and the measured
branching ratio one can extract information on the Cabibbo-Kobayashi-Maskawa 
matrix element $V_{td}$ that occurs in the short-distance contribution.

Our aim was to calculate the coupling $\chi$.
We applied the leading order
large--$N_C$ expression of the $K_L \to \gamma^*\gamma^*$ form factor
\cite{KPPD} where we included two terms, the $\rho$ term and one additional
term with an effective mass $\Lambda_H$.
The constants were determined with the help of experimental data on $K_L \to
\gamma\, l^+l^-$ and by matching of the large--$N_C$ expression with
perturbative QCD \cite{iu}.
Our final result for the long-distance part of $\chi$ is:
$$ \chi_{\gamma\gamma}(M_\rho)=5.8\pm 1.1
$$

\newabstract 
\label{abs:colangelo1}

\begin{center}
{\large \bf Hadronic Contributions to {\boldmath $a_\mu$} below 1 GeV}\\[0.5cm]
Gilberto Colangelo\\[0.3cm]
Institut f\"ur Theoretische Physik der Universit\"at Bern\\[0.3cm]
Sidlerstr. 5, 3012 Bern
\end{center}

The hadronic vacuum polarization contribution to $a_\mu$ is dominated by
the $\pi \pi$ contribution (see, e.g. \cite{DEHZ}), which is given by the
pion vector form factor. The Omn\'es representation expresses the latter in
terms of its phase which, below 16 $M_\pi^2$ exactly, and up to 4 $M_K^2$
to a good approximation, coincides with the $\pi \pi$ P-wave phase shift.
The latter is strongly constrained by analyticity, unitarity and chiral
symmetry \cite{CGL}. We call the Omn\'es function constructed with the $\pi
\pi$ P-wave phase shift $G_1(s)$:
\begin{equation}
G_1(s)=\exp \left[ \Delta(s) \right] \qquad 
\Delta(s)=\frac{s}{\pi}\int_{4M_\pi^2}^\infty dx\,\frac{\delta^1_1(x)}
{x\,(x-s)} 
\end{equation}
and represent the physical vector pion
form factor as a product of three terms:
\begin{equation}
F_\pi(s)= G_1(s) G_2(s) G_\omega(s)
\label{eq1co1}
\end{equation}
where $G_2(s)$ accounts for inelastic effects (i.e. it is the Omn\'es
function constructed with the difference between the phase of the form
factor and the P-wave $\pi \pi$ phase shift), and $G_\omega(s)$ takes into
account the contribution of the $\omega$ through its interference with the
$\rho$. Both functions $G_2$ and $G_\omega$ can be described in terms of a
small number of parameters which are fixed by fitting the data. 

I have presented preliminary results for a calculation of the hadronic
vacuum polarization contribution to $a_\mu$ below 1 GeV obtained on the
basis of Eq.~(\ref{eq1co1}). The analysis is currently in progress and final
results will be presented in a forthcoming publication \cite{forthcoming}.

\newabstract 
\label{abs:nguyen}

\begin{center}
{\large\bf The Measurement of the Hadronic Cross Section with\\[0.1cm] KLOE
Via the Radiative Return}\\[0.5cm]
Federico Nguyen$^\star$ on the behalf of the KLOE
Collaboration~\footnote{The full authors' list can be found in~\cite{KLOE}.}
\\[0.3cm]
$^\star${\small Dipartimento di Fisica, 
Univ. Roma TRE and INFN, Sez. Roma III,\\
Via della Vasca Navale 84, I-00146 Roma, Italy}\\[0.3cm]
\end{center}
Estimates of the hadronic contributions to the magnetic anomaly
of the muon, $a_\mu^{had}$, need data from $\tau$ decays and the reaction
$e^+\,e^-\to\mathrm{\ hadrons}$.
At DA$\Phi$NE this latter measurement is done at $\sqrt{s}=M_\phi$ using
Initial State Radiation (ISR) events~\cite{BIN}. We measured
the differential cross section $d \sigma_{e^+e^-\to\pi^+\pi^-\gamma}
/d M_{\pi\pi}^2=\sigma_{e^+e^-\to\pi^+\pi^-}\: H(M_{\pi\pi}^2,
\theta_{\pi\pi}^{min})$, where $H$ is the radiation function and
$\theta_{\pi\pi}$ is the polar angle of the dipion system.
The KLOE detector~\cite{PRI}
consists in a large acceptance spectrometer and a fine grained calorimeter,
able to provide very good momentum
and vertex
resolution for pions and electrons.

The measurement of $d \sigma_{e^+e^-\to\pi^+\pi^-\gamma}
/d M_{\pi\pi}^2$ consists of the following steps:\\
1) the detection of two charged tracks with polar angle
$50^\circ\!<\!\theta\!<\!130^\circ$,\\
2) the requirement $\theta_{\pi\pi}\!<\!15^\circ$ (or
$\theta_{\pi\pi}\!>\!165^\circ$) for enhancing ISR with respect
to Final State Radiation (FSR) events,\\
3) the discrimination of pions from electrons by means of calorimetric
variables and the subtraction of $\mu^+\,\mu^-\,\gamma$
and $\pi^+\,\pi^-\,\pi^0$ events by suitable kinematic cuts,\\
4) the luminosity measurement using Bhabha events
($55^\circ\!<\!\theta_{e^\pm}\!<\!125^\circ$), with a systematic uncertainty
of $\delta L/L = 0.5\%_{th} \oplus 0.4\%_{exp}$.

After the whole selection we have \textit{ca.} $1.5\times 10^6$ events out of
140.7 pb$^{-1}$ of data collected in year 2001, the experimental systematic
error is $1.2\%$.  We derive the $\pi\pi$ contribution to the muon
anomaly~\footnote{These
  numbers supersede the results presented at this Workshop:\\
  $10^{10}\times a_\mu^{\pi\pi}=374.1\pm 1.1_{stat}\pm 5.2_{syst}\pm 3.0_{th}
  {}_{-0}^{+7.5} |_{FSR}$.}, $a_\mu^{\pi\pi}$, for
$M_{\pi\pi}^2\in[0.65,0.93]\mbox{ GeV}^2$:
$$10^{10}\times a_\mu^{\pi\pi}=378.4\pm 0.8_{stat}\pm
4.5_{syst}\pm 3.0_{th}\pm 3.8_{FSR}~.$$

Our data are
in agreement with those from CMD-2~\cite{SWA}, confirming
the discrepancy in $a_\mu^{had}$ using $e^+\,e^-$ or $\tau$ data.

\newabstract 
\label{abs:davier}

\begin{center}
{\large\bf Hadronic Vacuum Polarization: the {\boldmath $\mu$} 
Magnetic Moment}\\[0.5cm]
Michel Davier\\[0.3cm]
Laboratoire de l'Acc\'el\'erateur Lin\'eaire,\\
IN2P3-CNRS et Universit\'e de Paris-Sud, 91898 Orsay, France \\
\end{center}

A new evaluation of the hadronic vacuum polarization contribution 
to the muon magnetic moment is presented~\citetwo{dehz}{dehz03}. 
We take into account the reanalysis of the low-energy $e^+e^-$ 
annihilation cross section into hadrons by the CMD-2 Collaboration~\cite{cmd}.
The agreement between $e^+e^-$ and $\tau$ spectral functions in the
$\pi\pi$ channel is found to be much improved. Nevertheless, significant 
discrepancies remain in the center-of-mass energy range between 0.85 and 
$1.0\;{\rm GeV}$, so that we refrain from averaging the 
two data sets. The values found for the 
lowest-order hadronic vacuum polarization contributions are
\begin{eqnarray*}
a_\mu^{\rm had,LO} = \left\{
\begin{array}{ll}
         (696.3\pm6.2_{\rm exp}\pm3.6_{\rm rad})~10^{-10}
        & ~~~[e^+e^-{\rm -based}]~,\\[0.1cm]
         (711.0\pm5.0_{\rm exp}\pm0.8_{\rm rad}
                                \pm2.8_{\rm SU(2)})~10^{-10}
     & ~~~[\tau {\rm -based}]~, \\
\end{array}
\right.
\end{eqnarray*}
where the errors have been separated according to their sources: 
experimental, missing radiative corrections in $e^+e^-$ data, and 
isospin breaking. 
The corresponding Standard Model predictions for the muon 
magnetic anomaly read
\begin{eqnarray*}
a_\mu = \left\{
\begin{array}{ll}
        (11\,659\,180.9\pm7.2_{\rm had}
                \pm3.5_{\rm LBL}\pm0.4_{\rm QED+EW})~10^{-10}
     & ~~~[e^+e^-{\rm -based}]~, \\[0.1cm]
        (11\,659\,195.6\pm5.8_{\rm had}
                \pm3.5_{\rm LBL}\pm0.4_{\rm QED+EW})~10^{-10}
     & ~~~[\tau {\rm -based}]~, \\
\end{array}
\right.
\end{eqnarray*}
where the errors account for the hadronic, light-by-light (LBL) scattering
and electroweak contributions. The deviations from the  
measurement at BNL are found to be $(22.1 \pm 7.2 \pm 3.5 \pm 8.0)~10^{-10}$ 
(1.9~$\sigma$) and $(7.4 \pm 5.8 \pm 3.5 \pm 8.0)~10^{-10}$ (0.7~$\sigma$) 
for the $e^+e^-$- and $\tau$-based estimates, respectively, where the second 
error is from the LBL contribution and the third one from 
the BNL measurement~\cite{bnl}.

Preliminary results are given on the radiative return process 
$e^+ e^- \rightarrow \gamma \; 2\pi^+ 2\pi^-$ measured with BABAR.
They are shown to significantly improve the contribution of the $4\pi$ channel
to the muon magnetic anomaly.

\newabstract 
\label{abs:shwartz}

\begin{center}
{\large\bf CMD-2 Results on {\boldmath $e^+e^- \to$} hadrons}\\[0.5cm]
Boris Shwartz$^1$ for CMD-2 collaboration\\[0.3cm]
$^1$
Budker Institute of Nuclear Physics SB RAS \\
630090 Novosibirsk, Russia

\end{center}

This report presents the results obtained by the CMD-2 collaboration
in the years 2002-2003.
The CMD-2 \cite{cmd} detector had been collecting data from 1992 to
2000 at VEPP-2M collider in the energy range from 0.4
to 1.4 GeV. Although the data collection was finished more than
two years ago, the data processing is continued providing the
physics yield.

The new results were obtained for the several processes
\citethree{kskl}{ompi}{pipigam}.
Here, we mention briefly the results which are not 
published yet.

The process $e^+e^- \rightarrow \pi^0 e^+e^-$ was 
studied around the $\omega-$meson peak. The measured value of
$Br(\omega \rightarrow \pi^0e^+e^-) = (8.7 \pm 0.9 \pm 0.5)
\times 10^{-4}$ has the best accuracy and it's in agreement with 
the previous experimental result as well as with theoretical
predictions.

Study of the $\phi \rightarrow \pi^+\pi^-\pi^0$ decay dynamic.
The fit of the Dalitz-plot distribution with the probability
density function including the $\rho\pi$ final state amplitude
and three pion point like amplitude, 
$ A_n ae^{i\varphi} + A_{\rho\pi}$,
gives the values $a= 0.103\pm 0.028$ in a good agreement with
\cite{kloe} and $\varphi= -2.0 \pm 0.3$ that is rather different
from the mentioned paper.  

Very important subject studied at CMD-2 is the total hadron
cross section in the available energy range. Recently, some
uncertainty in the code for the calculation of the radiative
corrections for $e^+e^- \rightarrow e^+e^-$ was found.
Since this process is used for the luminosity measurement,
the total cross section changed by 2-3\% after reanalysis
\cite{rean}. 
The most intriguing value of the hadron vacuum polarization
contribution to muon (g-2) value became closer to theoretical
one, 
$a_{\mu}(exp)-a_{\mu}(th) = (22.1 \pm 11.3)\times 10^{-10}
(1.9\sigma)$.

\newabstract 
\label{abs:buettiker}

\begin{center}
{\large\bf Roy--Steiner Equations for \boldmath{$\pi K$} Scattering}\\[5mm]
{\bf Paul B\"uttiker}$^1$, Sebastien Descotes--Genon$^2$, Bachir
Moussallam$^3$\\[3mm]
$^1$ HISKP, Universit\"at Bonn, D--53115 Bonn,
$^2$ LPT, Universit{\'e} Paris--Sud, F-91406 Orsay,
$^3$ IPN, Universit{\'e} Paris--Sud, F-91406 Orsay
\end{center}
$\pi K$ scattering is the most simple $SU(3)$--process involving
strange quarks. It is therefore an ideal place to test chiral
predictions with non--vanishing strangeness. While this process is
interesting by itself, a detailed knowledge of $\pi K$ 
scattering also contributes to the understanding of the flavour
dependence of the order parameters of Chiral Perturbation Theory
(ChPT) \cite{stern}.
%
In our work we use Roy--Steiner equations  to analyze the available
experimental $\pi K$ and $\pi\pi\to K\bar{K}$ data and to explore the
low energy region where no such data are available \cite{ABM}.
\begin{floatingfigure}{7.0cm}
\includegraphics[scale=0.55]{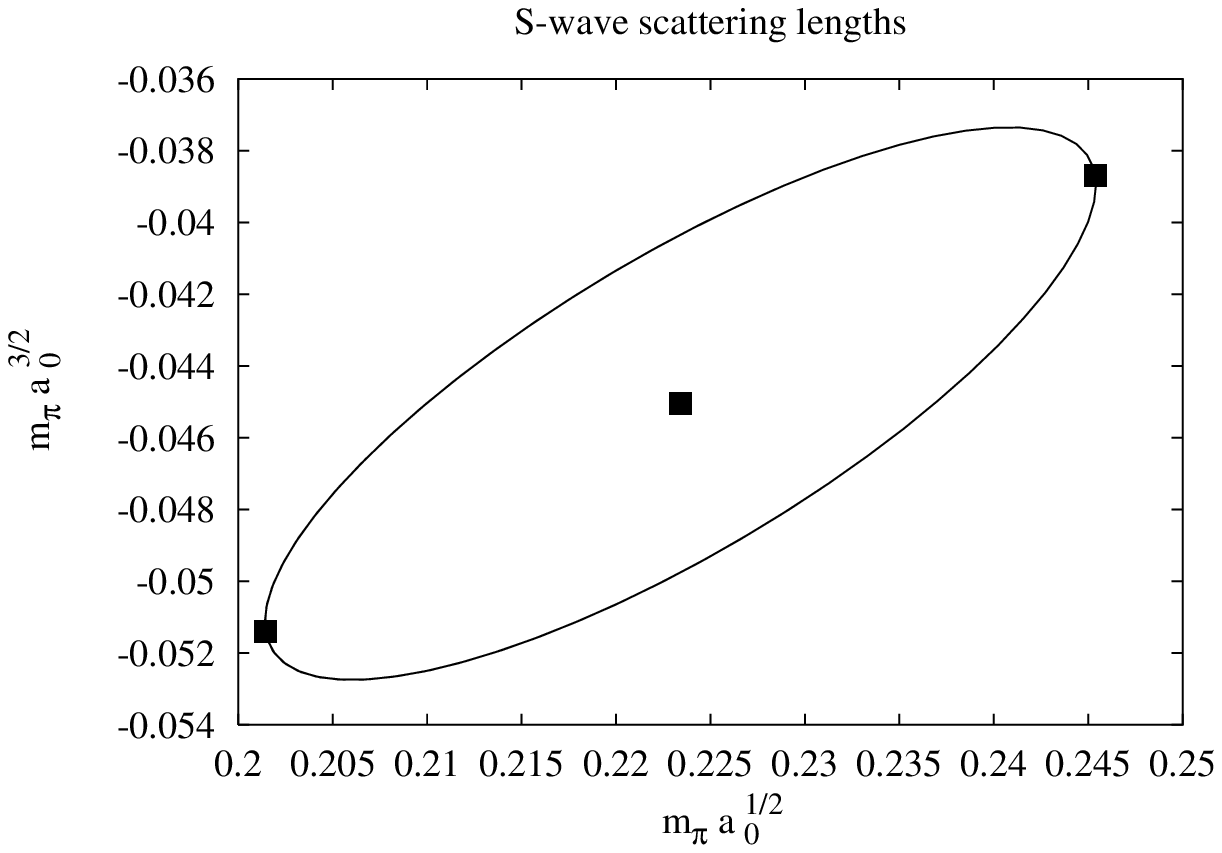}\\
Fig:~$1$--$\sigma$ error ellipse for the scattering\\\phantom{Fig:~}lengths
  $a^{1/2}_0$ and $a^{3/2}_0$.
\end{floatingfigure}
Taking the higher partial waves and all $S$-- and $P$--waves above
$\approx$ 1 GeV as input, we have found unique solutions for the 
$\pi K$ $S$--waves and the $I=1/2$ $P$--wave in
the energy region from threshold to $\approx$ 1 GeV. Our solutions for
the low--energy region turn out to be in general poor
agreement with the available experimental data.
Hence, the mass for the $K^*(892)$ is shifted by
$\approx$ 10 MeV from the published value to 905 MeV.
The solutions of the Roy--Steiner equations also yield
predictions for the scattering lengths $a^{1/2}_0$ and $a^{3/2}_0$. In
contradiction to $\pi\pi$ scattering the allowed values are strongly
constrained by the data, so that there is no universal band for the
two scattering lengths. We find
$    m_\pi a^{1/2}_0 = 0.224 \pm 0.022,\quad m_\pi a^{3/2}_0 
                          = -0.0448 \pm 0.0077\nonumber$
Finally, the matching of the subthreshold parameters in the chiral and
the dispersive framework yields the following estimates for the
low--energy constants (at the scale $\mu = m_\rho$ in units of $10^{-3}$):
$   L_1^r = 1.05\pm 0.12, L_2^r = 1.32\pm 0.03, L_3^r = -4.53\pm 0.14,
   L_4^r = 0.53\pm 0.39$ and $2 L_6^r + L_8^r = 3.66\pm 1.52$.

\newabstract 
\label{abs:yndurain}

\begin{center}
{\large\bf Comments on some Chiral-Dispersive Calculations
 of\\[0.1cm] {\boldmath $\pi\pi$} Scattering}\\[0.5cm]
F. J. Yndur\'ain\\[0.3cm]
Departamento de F\'{\i}sica Te\'orica, C-XI\\
 Universidad Aut\'onoma de Madrid,\\
 Canto Blanco,\\
E-28049, Madrid, Spain\\[0.3cm]
\end{center}

In two  recent papers, Ananthanarayan, Colangelo, Gasser and Leutwyler
and Colangelo, Gasser and Leutwyler (to be referred to as, respectively, 
ACGL and CGL) have used 
experimental information, analyticity and unitarity (in the form of  
the Roy equations) and, in CGL, chiral perturbation theory, 
to construct the  $\pi\pi$ 
scattering amplitude at low energy, 
$s^{1/2}\leq0.8$ GeV.

We will mainly discuss here the second paper, CGL, in which an outstanding
precision (at the percent level) is claimed but which, unfortunately, presents
a number of drawbacks.  First of all, the input scattering amplitude at high
energy ($s^{1/2}\geq1.42\,$ GeV) which ACGL, CGL use is not physically
acceptable; and it is also clear that the errors these authors take for some
of their experimental input data (in particular, for the phase shift
$\delta_0^{(0)}$ at 0.8 GeV) are excessively optimistic.  What is more, the
CGL threshold parameters, and the low energy S0, S2 and P waves below 0.8 GeV,
are displaced from what one gets by direct fits to experimental data, disagree
with what other authors (Descotes et al., and Kami\'nski, Le\'sniak and
Loiseau) find using also Roy equations, and fail to pass a number of
consistency tests.  Among these last, and by 2 to 4 $\sigma$, we find failure
of nonsubtracted forward dispersion relations for the amplitude with isospin 1
excahange (Olsson's sum rule), inconsistency of the D wave scattering length
for $\pi^0\pi^0\to\pi^+\pi^-$ scattering, and disagreement of the P wave
effective range parameter, $b_1$ with results from the pion form factor.
Moreover, we remark that some chiral parameters used by CGL are probably
displaced from their correct values.  As a consequence of this, we conclude
that the error estimates of CGL are too optimistic.

The details of all this, based on a recent   
 paper by the author and J.~R.~Pel\'aez, may be found in the report  
by the author, ``Comments on some chiral-dispersive calculations
 of $\pi\pi$ scattering", FTUAM 03-14 (hep-ph/0310206).

\newabstract 
\label{abs:colangelo2}

\begin{center}
{\large \bf  On the Precision of Theoretical Calculations for\\[0.1cm]
  {\boldmath $\pi \pi$} Scattering}\\[0.5cm]
Gilberto Colangelo\\[0.3cm]
Institut f\"ur Theoretische Physik der Universit\"at Bern\\[0.3cm]
Sidlerstr. 5, 3012 Bern
\end{center}

In a recent publication \cite{PY}, Pel\'aez and Yndur\'ain have raised
objections against our work on Roy equations \cite{ACGL} and the subsequent
matching between this representation and the chiral one, which led to
remarkably sharp predictions for the $\pi \pi$ scattering amplitude at low
energy \cite{CGL}. In particular, they claim that the asymptotic
representation we used is incorrect and that this affected all our results.
This latter conclusion was based on an indirect reasoning: they have
evaluated some of the low energy observables of $\pi\pi$ scattering and
obtained flat disagreement with our earlier results. In their opinion this
disagreement had to be attributed to the asymptotic representation for the
$\pi \pi$ amplitude we had used. In this talk I have shown that:
\begin{enumerate}
\item
if we take the asymptotic representation proposed by Pel\'aez and
Yndur\'ain the solution of the Roy equations barely changes;
\item
the indirect reasoning that led them to citicize our results is flawed --
there is a logical leap between the observed discrepancies among the two
calculations and the conclusion that the asymptotics we used is incorrect;
\item
the asymptotic representation proposed by Pel\'aez and Yndur\'ain is in
many respects worse than the one we used. In particular, it does not follow
from factorization applied to data on high energy total cross sections of
proton--proton and pion--proton.
\end{enumerate}

A full account of our answer to the criticism in \cite{PY} can be found in
\cite{CCGL}.

\newabstract 
\label{abs:stern}

\begin{center}
{\large\bf From Two to Three Light Flavours}\\[0.5cm]
 Jan Stern\\[0.3cm]
 Groupe de Physique Theorique, IPN, 91405 Orsay Cedex, France\\[0.3cm]
\end{center}

In $N_f=2$  ChPT one expands in powers of $m_u$ and $m_d$ keeping $m_s$
fixed at its physical value, whereas $N_f=3$ ChPT treats all three light
quark masses $m_u, m_d$ and $m_s$ as expansion parameters. The corresponding
$N_f=2$ and $N_f=3$ chiral limits are characterized by different order
parameters: $QCD$ suggests paramagnetic inequalities $\Sigma(3)<\Sigma(2)$
and $F^2(3)<F^2(2)$, where $\Sigma(N_f)$ denotes the $N_f=2$ or $N_f=3$
chiral limits of the condensate $-\langle\bar uu\rangle$ and  $F^2(N_f)$ stands
for the corresponding chiral limits of $F^2_{\pi}$. Hence, on general
grounds, one expects a suppression of the two fundamental three-flavour
order parameters relative to their two-flavour analogs. This effect is due
to massive strange sea quark pairs which induce  extra positive
contributions to $\Sigma(2)$ and $F^2(2)$ . The latter vanish for $m_s=0$
i.e. they do not show up in $\Sigma(3)$ and $F^2(3)$. In practice,the resulting
difference between $N_f=2$ and $N_f=3$ order parameters may be important
since $m_s \sim \Lambda_{QCD}$ and furthermore, due to fluctuations of the
lowest eigenvalues of Euclidean Dirac operator,  $QCD$ vacuum enjoys
OZI-rule violating and large $N_c$ suppressed correlations between scalar
massive $\bar ss$ and massless non strange $\bar uu + \bar dd$ pairs
\cite{DGS} .In ChPT , this phenomenon is encoded
in the values of the large $N_c$ suppressed LEC's $L^r_6(\mu)$ and
$L^r_4(\mu)$. Dominant effects of vacuum fluctuations of massive $\bar ss$
pairs are described by the scale independent combinations
$\Delta L_6 = L^r_6(M_{\rho}) + 0.26 \times 10^{-3}$ and
$\Delta L_4 = L^r_4(M_{\rho}) + 0.51 \times 10^{-3}$ which should remain
positive due to the paramagnetic inequalities. They are usually multiplied by
$m_s$ and often appear with large numerical coefficients.  Consequently,
rather small positive values of $\Delta L_6$ and/or $\Delta L_4$ ( close to
their upper bounds of order $0.001$ and $0.002$ respectively which reflect the
positivity of $\Sigma(3)$ and of $F^2(3)$)      could be
sufficient to produce an important suppression of the three-flavour condensate
and/or of the decay constant
and to destabilize the perturbative relation between order parameters
and Goldstone boson observables \cite{DFGS} . Such instability need not
affect the standard two-flavour ChPT characterized by a large condensate
$\Sigma(2)$ and it need not reflect a too large value of $m_s$ but rather
an important OZI-rule violating fluctuation of vacuum $\bar ss$ pairs.

\newabstract 
\label{abs:descotes}

\begin{center}
{\large\bf Bayesian Approach to the Determination of\\[0.1cm]
 {\boldmath $N_f$}=3 Chiral Order Parameters}\\[0.5cm]
S\'ebastien Descotes-Genon\\[0.3cm]
Laboratoire de Physique Th\'eorique, 91405 Orsay Cedex, France.\\[0.3cm]
\end{center}

Due to its light mass of order $\Lambda_{\rm QCD}$, the strange
quark can play a special role in Chiral Symmetry Breaking ($\chi$SB).
Differences in the pattern of $\chi$SB in the limits $N_f=2$ ($m_u,m_d\to 0$, 
$m_s$ physical) and $N_f=3$ ($m_u,m_d,m_s\to 0$) may arise due to
vacuum fluctuations of $s\bar{s}$ pairs in relation to the Zweig-rule 
violation in the scalar sector~\cite{fluct}. The latter is in particular
described by the $O(p^4)$ low-energy constants $L_4$ and 
$L_6$ of the strong chiral lagrangian. Recent
estimates of these constants~\citetwo{scalar}{kpi} suggest sizeable
fluctuations of $s\bar{s}$ pairs, which could lead to a significant 
suppression of the $N_f=3$ quark condensate and pseudoscalar constant 
with respect to their $N_f=2$ counterparts.

In the case of large fluctuations, 
the customary treatment of $SU(3)\times SU(3)$ chiral
expansions generates instabilities upsetting their convergence~\cite{TBP}.
We develop a systematic program to cure these instabilities 
by resumming nonperturbatively vacuum fluctuations of $s\bar{s}$ pairs,
in order to extract information about $\chi$SB from experimental
observations even in the presence of large fluctuations. The resulting
Bayesian framework is exploited to analyse the low-energy $\pi\pi$
scattering amplitude~\cite{DFGS} 
and to obtain a lower bound on the quark mass
ratio $2m_s/(m_u+m_d)\geq 14$ at 95\% confidence level. 
On the other hand, $\pi\pi$ scattering does not constrain
the quark condensate or the pseudoscalar decay constant in the $N_f=3$
chiral limit.

\newabstract 
\label{abs:borasoy}

\begin{center}
{\large\bf Hadronic Decays of $\mbox{\boldmath$\eta$}$ and
$\mbox{\boldmath$\eta'$}$ }\\[0.5cm]
B. Borasoy\\[0.3cm]
Physik Department, TU M\"unchen,\\
85747 Garching, Germany\\[0.3cm]
\end{center}

The hadronic decays $\eta \rightarrow \pi \pi \pi$, $\eta' \rightarrow \pi \pi
\pi$ and $\eta' \rightarrow \eta \pi \pi$ are investigated within a $U(3)$
chiral unitary approach \cite{BB}.  To this end, the most general chiral
effective Lagrangian which describes the interactions between members of the
lowest lying pseudoscalar nonet $(\pi, K, \eta, \eta')$ up to fourth chiral
order is utilized and the pertinent tree diagram amplitudes are derived.
Final state interactions are included by deriving the effective $s$-wave
potentials for meson meson scattering from the chiral effective Lagrangian and
iterating them in a Bethe-Salpeter equation.  With only a small set of
parameters we are able to explain both rates and spectral shapes of these
decays.

The decay $\eta \rightarrow \pi \pi \pi$ may be used to constrain the double
quark mass ratio $Q^2 = \frac{\textstyle{m_s - \hat{m}}}{\textstyle{m_d -
    m_u}} \, \frac{\textstyle{m_s + \hat{m}}}{\textstyle{m_d + m_u}}$ with $
\hat{m}= m_u + m_d \;$ from which we obtain $Q= 23.4 \pm 0.8$, a value which
is in agreement with the requirement by Dashen's theorem and indicates that
higher order corrections to this low-energy theorem may be small.

We can furthermore confirm the importance of the $a_0(980)$ resonance for
$\eta' \rightarrow \eta \pi \pi$---as has been 
claimed in the literature---without
including it explicitly in the theory.

The presented method can also be applied in the study of the decays $\eta /
\eta' \rightarrow \gamma \gamma$ \cite{BN}, $\eta / \eta' \rightarrow \pi^+
\pi^- \gamma$ and $\eta / \eta' \rightarrow \pi^0 \gamma \gamma$.

\newabstract 
\label{abs:oset2}

\begin{center}
{\large\bf {\boldmath $\Phi$} Radiative Decay to Two Pseudoscalars}\\[0.5cm]
J. Palomar, L. Roca, {\bf E. Oset} and M.J. Vicente Vacas\\[0.3cm]
Department of Theoretical Physics and IFIC, 
University of Valencia, Spain\\[0.3cm]

\end{center}

We study the radiative $\phi$ decay into $\pi^0 \pi^0 \gamma$ and $\pi^0 \eta
\gamma$ taking into account mechanisms in which there are two sequential
vector-meson-pseudoscalar or axial-vector-vector-pseudoscalar steps followed
by the coupling of a vector meson to the photon, considering the final state
interaction of the two mesons \cite{Palomar:2003rb}. There are other
mechanisms in which two kaons are produced through the same sequential
mechanisms or from $\phi$ decay into two kaons and then undergo final state
interaction leading to the final pair of pions or $\pi^0 \eta$.  The latter
mechanism is the leading one but the other mechanisms provide appreciable
contributions. The results of the parameter free theory together with the
theoretical uncertainties, are compared with the latest experimental KLOE
results at Frascati \cite{frascati} in the two figures below, which also show
the theoretical errors. It was shown during the Conference that the data for
$\pi^0 \pi^0 \gamma$ decay around 500 MeV have changed and are now in
agreement with the theory.
 
 \begin{figure}[hbp]
\centerline{\hbox{\psfig{file=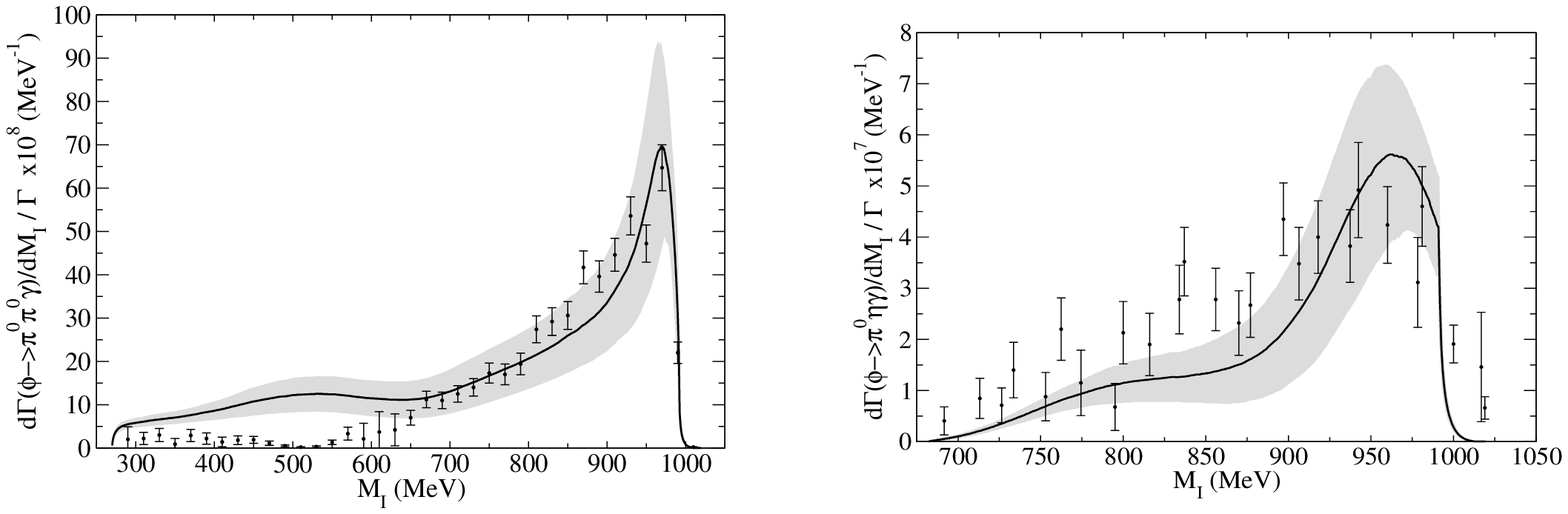,width=16cm,angle=0}}}
\end{figure}

\newabstract 
\label{abs:dimicco}

\begin{center}
{\large\bf {\boldmath $\eta$} Decays Studies with KLOE}\\[0.5cm]
Biagio Di Micco$^\dag$  
(on the behalf of the KLOE collaboration)\footnote{See 
\cite{miopreprint} for the complete authors list}\\[0.3cm]
{\small $^\dag$Dipartimento di Fisica, 
Universit\`a degi Studi di Roma Tre and INFN, Sez. Roma III,\\
Via della Vasca Navale 84, I-00146 Roma, Italy}\\[0.3cm]
\end{center}

We present the status of the analyses on $\eta$ meson decays with the KLOE
detector at DA$\Phi$NE $\Phi$ factory\cite{Primavera}. The KLOE detector has
collected about 20 milions of $\eta$ events, produced by the $\Phi \to \eta
\gamma$ decay process. Here we present the status of the analyses of the decay
process: $\eta \to 3 \gamma$, $\eta \to \pi^+ \pi^- \pi^0$, $\eta \to 3
\pi^0$, $\eta \to \pi^0 \gamma \gamma$ and $\eta \to \pi^+ \pi^- \gamma$.

The $\eta \to 3 \gamma$ decay is a C violating decay. The SM prediction of the
branching ratio is of the order of $10^{-12}$\cite{Eta Handbook}, far below
the present experimental limit ($5 \times 10^{-4}$ \cite{PDG}), so any
measurement of a larger branching ratio would be a clear signal of new
physics. KLOE has analyzed $17 \times 10 ^6$ $\eta$ decays \cite{miopreprint}.
No signal has been found, so we have evaluated an upper limit: $Br(\eta \to 3
\gamma) < 1.6 \times 10 ^{-5}$ (95 \% C.L.).

The $\eta \to \pi ^0 \gamma \gamma$ is interesting for testing ChPT to
O($p^6$) level \cite{Eta Handbook}. We expect to have from 5000 to 14000 such
decays in our data\cite{PDG}\cite{Eta Handbook}.  The main background is the
$\eta \to 3 \pi ^0$ channel when there is a merging of the clusters in the
calorimeter.

The Dalitz plot analysis of the $\eta \to \pi ^+ \pi ^- \pi ^0$ channel is
interesting for C violation and for ChPT testing \cite{Eta Handbook}. We have
analized about 400000 $\eta \to 3 \pi ^0$ events, producing the following very
preliminary results for the Dalitz plot parameters: (a,b,c) = $(-1.05 \pm
0.01,0.20 \pm 0.03,-0.009 \pm 0.009)$ (only statistical error reported).  The
c parameter is compatible with 0, so no C violation is observed.

For the $\eta \to 3 \pi^0$ the overall selection efficiency is 16\%.  So we
expect to have a sample of 1 million events with the full KLOE statistic. The
analysis reproduces well the Dalitz plot parameter put in the MC generator.

The $\eta \to \pi ^+ \pi ^- \gamma$ is an interesting decay to check the box
anomaly \cite{Eta Handbook} studying the energy distribution of the $\gamma$.
We have selected 18508 among 1 million of $\eta$ events with a preliminary
analysis.

\newabstract 
\label{abs:roca}

\begin{center}
{\large\bf {\boldmath $\eta \to \pi^0 \gamma \gamma$} Decay within a 
Chiral Unitary Approach}\\[0.5cm]
E. Oset$^1$, J. R. Pel\'aez$^2$ and {\bf L. Roca}$^1$\\[0.3cm]
$^1$Departamento de F\'{\i}sica Te\'orica and IFIC\\
 Institutos de Investigaci\'on de Paterna, 22085,
 46071, Valencia, Spain\\[0.3cm]
$^2$
   Departamento de
F\'{\i}sica Te\'orica II,  Universidad Complutense.\\
 28040 Madrid, Spain.\\[0.3cm]
\end{center}

The $\eta \to \pi^0 \gamma \gamma$ decay has attracted much theoretical 
attention since, in Chiral Perturbation Theory (ChPT), the
calculation at $O(p^2)$ vanishes and at $O(p^4)$ is very small.
The first sizeable contribution comes at $O(p^6)$. This makes
this reaction to be, in principle, a good test of ChPT
at $O(p^6)$ and higher. But the parameters involved in the ChPT
Lagrangians at these higher orders are not easy to determine.
On the other hand models using Vector Meson Dominance (VMD) have
been used, but expanding the vector meson propagator to obtain
the $O(p^6)$ chiral coefficients gives results $\sim 1/2$ of
those obtained by
keeping the full vector meson propagator which is supposed to
contain higher orders. This implies that $O(p^8)>O(p^6)$.
This makes us to conclude that a strict chiral counting has to be
abandoned and we have to think in terms of 'relevant mechanisms'
instead of 'diagrams contributing up to a certain order'.

At this point is where unitary extensions of ChPT (UChPT) can
help
to solve the problem. UChPT is based on the implementation of
unitary techniques to resum loops and has allowed to extend the
predictions of ChPT up to $\sim 1.2\,GeV$ even generating many
resonances dynamically without including them explicitly.
  
Considering this background, we improve the calculations of the 
$\eta \to \pi^0 \gamma \gamma$ decay within
the context of meson chiral lagrangians. We use a chiral unitary
approach for the meson-meson  interaction,
thus generating the $a_0(980)$ resonance and fixing
the longstanding sign ambiguity on its contribution.
This also allows us to calculate the loops
with one vector meson exchange, thus removing a former source of uncertainty. 
In addition we ensure the consistency of the approach
with other processes. First, by
using vector meson dominance
couplings normalized to agree with radiative vector meson decays.
And, second, by checking the consistency of the calculations with the
related $\gamma \gamma \to \pi^0 \eta$ reaction.
We find an $\eta \to \pi^0 \gamma \gamma$ decay width of $0.47\pm 0.10$ eV, 
in clear disagreement with published data but
in remarkable agreement with the most recent measurement.

\section*{Working Group II: Meson-Baryon Dynamics}

\centerline{Convenors: Reinhard Beck and Thomas R. Hemmert}

\vskip3mm
{
\begin{tabular}{lp{10.1cm}r}
\multicolumn{3}{c}{\bf Nucleon Spin Structure / GDH}  \\
\stalk{D. Drechsel}{GDH, Nucleon Spin Structure 
and Chiral Dynamics: Theory}{abs:drechsel}
\stalk{A. Braghieri}{ Helicity Dependence of Pion Photo-Production and the
GDH Sum Rule}{abs:braghieri}
[1mm]
\multicolumn{3}{c}{\bf Real Compton Scattering / VCS} \\
\stalk{H. Fonvieille}{ Virtual Compton Scattering at Low Energy and 
the\newline
  Generalized  Polarizabilities of the Nucleon}
{abs:fonvieille}
\stalk{R. Hildebrandt}{ Dynamical Polarizabilities from (Polarized) 
Nucleon  Compton
Scattering}{abs:hildebrandt}
\stalk{L. Fil'kov}{ Radiative Pion Photoproduction and Pion Polarizability}
{abs:filkov}
[1mm]
\multicolumn{3}{c}{\bf Strangeness Production / SU(3)}\\
\stalk{K.-H. Glander}{ Strangeness Production from ELSA}{abs:glander}
\stalk{B. Raue}{ Strangeness Production at Jefferson Lab}{abs:raue}
\stalk{E. Oset}{ Unitarized Chiral Dynamics: SU(3) and Resonances}{abs:oset1}
[1mm]
\multicolumn{3}{c}{\bf Pion (Electro-) Production and Chiral Dynamics}\\
\stalk{H. Merkel}{ Threshold Pion Electroproduction: Update on\newline
 Experiments}
{abs:merkel}
\stalk{M. Kotulla}{ Two $\pi^0$ Production at Threshold}{abs:kotulla}
\stalk{S.N. Yang}{ Threshold $\pi^0$ Photo- and Electro-Production}{abs:yang}
[1mm]
\multicolumn{3}{c}{\bf Hadronic Reactions / Sigma Term}\\
\stalk{R.A. Arndt}{ Partial Wave Analysis of Pion-Nucleon
Scattering below T$_{\rm LAB}$ = 2.1 GeV}{abs:r.arndt}
\stalk{P. Piirola}{ Towards the Pion-Nucleon PWA}{abs:piirola}
\stalk{M.F.M. Lutz}{ Meson-Baryon Scattering}{abs:lutz}
\stalk{M. Frink}{ Order Parameters of Chiral Symmetry Breaking 
}{abs:frink}
[1mm]
\multicolumn{3}{c}{\bf Chiral Dynamics and Nucleon Resonances}\\
\stalk{A.M. Bernstein}{ Hadron Deformation and Chiral Dynamics}{abs:bernstein}
\stalk{V. Pascalutsa}{ Chiral Dynamics in the $\Delta$(1232) Region}
{abs:pascalutsa}
[1mm]
\multicolumn{3}{c}{\bf Recent Formal Developments}\\
\stalk{R. Lewis}{Lattice Regularized Chiral Perturbation Theory}{abs:lewis2}
\stalk{S. Scherer}{ Renormalization and Power Counting in 
Manifestly Lorentz-invariant
Baryon Chiral Perturbation Theory}{abs:scherer}
\stalk{J. Gegelia}{ Baryon ChPT with Vector Mesons}{abs:gegelia}
\stalk{P.J. Ellis}{ Baryon ChPT with Infrared Regularization: Update}
{abs:ellis}
\stalk{U.-G. Mei{\ss}ner}{ Cutoff Schemes in Chiral Perturbation Theory and
the Quark Mass Expansion of the Nucleon Mass}{abs:meissner} 

\end{tabular}
}

\newabstract 
\label{abs:drechsel}

\begin{center} {\large\bf GDH, 
  Nucleon Spin Structure and Chiral Dynamics:\\[0.1cm] Theory} \\[0.5cm] 
  {Dieter Drechsel} \\[0.3cm] 
   Institut f\"ur Kernphysik, Universit\"at Mainz,\\ 
   55099 Mainz, Germany\\[0.3cm] \end{center} 
The Gerasimov-Drell-Hearn sum
rule ist based on low-energy theorems and forward dispersion
relations~\cite{Dre03}. It connects the anomalous magnetic moment to an
integral over the helicity difference $\sigma_{1/2} - \sigma_{3/2}$ of the
photoabsorption cross sections weighted with the inverse power of the photon
lab energy $\nu$. This cross section difference was recently measured by the
GDH Collaboration~\cite{Sch03} at MAMI for 200 MeV $<\nu<$ 800 MeV and at ELSA
for 700 MeV $<\nu<$ 2.8 GeV. The experiments confirmed the sum rule within a
10\% deviation, and most of the still missing strength is expected at the
higher energies on the basis of Regge models. In the mean time the
collaboration has also taken data for the deuteron, which holds the promise to
solve the ``neutron puzzle'': Theoretical analysis using photoproduction
multipoles results in a 20-30\% discrepancy between the sum rule and the
dispersion integral for the neutron. At the same time the GDH experiments have
provided a value for the forward spin polarizability~\citetwo{Dre03}{Sch03} by
weighting the cross section difference by $\nu^{-3}$. In the case of virtual
photons there appears an additional inclusive cross section $\sigma_{LT}$ as a
longitudinal-transverse interference, and all cross sections and related
integrals are also functions of the four-momentum transfer $Q^2>0$. These
integrals provide various generalizations of the GDH sum rule, the
Burkhardt-Cottingham sum rule, and values for generalized polarizabilities. A
wealth of new precision data has been obtained at the Jefferson
Lab~\cite{Kor}, which can be compared to the results of chiral
theories~\cite{Ber} and dispersion analysis~\cite{Dre03}. These ongoing
activities hold the promise to fully determine the spin structure in the
resonance region and thus to bridge the gap between low-energy coherent
processes and deeply inelastic incoherent scattering.

\newabstract 
\label{abs:braghieri}

\begin{center}
{\large\bf Helicity Dependence of Pion Photo-Production 
and the\\[0.1cm] GDH Sum Rule}\\[0.5cm]
Alessandro Braghieri\\[0.3cm]
INFN, Sezione di Pavia,\\
Via Bassi 6, 27100 Pavia, Italy\\[0.3cm]

\end{center}

The helicity dependence of the total photo-absorption cross section on the
proton was measured in the energy range 0.2$<$ E$_\gamma <$ 3.1 GeV. These
data provide the first experimental check of the Gerasimov-Drell-Hearn sum
sule \cite{orig} and measurement of the forward spin polarizability
$\gamma_0$.

The experiment was carried out in two steps: at MAMI (Mainz) up to 800 MeV and
at ELSA (Bonn) at higher energy.  First results from Mainz have already been
published \cite{gdh.1} and preliminary results from Bonn are available
\cite{gdh.2}. An estimate of the GDH sum rule was deduced combining these
measurements with model predictions in the unmeasured photon energy range and
it was found in agreement with the theoretical value. A preliminary estimate
of the forward spin polarizability gives $\gamma_0=(104\pm 8\pm 10) \cdot
10^{-6}$ fm$^4$.

In addition, all partial channels up to E$_\gamma$=800 MeV were measured and
the helicity dependence of both total and differential cross sections are now
available. First results have been published: the single $\pi^+$ and $\pi^0$
production in the $\Delta$ region \cite{gdh.3}; the double pion production
($\pi^+\pi^0$) \cite{gdh.4}; the $\eta$ production \cite{gdh.5}; the $\pi^0$
production in the second resonance region \cite{gdh.6}.

These measurements provide a strong constraint for multipole analysis
\cite{xaid} of pion photo-production and new information on excitation of
baryon resonances.

\newabstract 
\label{abs:fonvieille}

\begin{center}
{\large\bf Virtual Compton Scattering at Low Energy
and the\\[0.1cm] Generalized Polarizabilities of the Nucleon}\\[0.5cm]
H\'el\`ene Fonvieille\\[0.3cm]
Laboratoire de Physique Corpusculaire IN2P3-CNRS\\
Universit\'e Blaise Pascal Clermont-II, 63170 Aubi\`ere Cedex, France\\[0.3cm]
{\it For the Jefferson Lab Hall A Collaboration}\\
{\it and the VCS Collaboration}\\[0.3cm]
\end{center}

Virtual Compton Scattering has opened  a new field of investigation
of nucleon structure. At low center-of-mass energies,
the process \ $\gamma^* p \to \gamma p$ \
allows the determination of  the 
Generalized Polarizabilities (GPs) of the proton~\cite{gui9598}.
These observables generalize the concept of nucleon 
polarizabilities to any photon virtuality $Q^2$. 
The GPs are predicted by many models at low $Q^2$, including 
Heavy Baryon Chiral Perturbation 
Theory~\cite{chpt}. 
A first generation of experiments studying photon 
electroproduction \ $ e p \to e p \gamma$ \
have been performed at MAMI~\cite{mami1}, 
Jefferson Lab~\cite{e93050prop} and Bates~\cite{bates}.
They measure the unpolarized VCS structure functions
\ $P_{LL}-P_{TT}/\epsilon$ \ and \ $P_{LT}$ \ which are
linear combinations of the lowest order dipole GPs.
Analysis methods are based on the Low Energy Theorem~\cite{gui9598}
or the Dispersion Relation formalism~\cite{drmodel}.
Results of the MAMI~\cite{mami1} and JLab~\cite{e93050result}
experiments are presented, together with the future prospects in 
the field.

\newabstract 
\label{abs:hildebrandt}

\begin{center}
{\large\bf Dynamical Polarizabilities from (Polarized) 
           Nucleon\\[0.1cm] Compton Scattering}\\[0.5cm]
{\bf Robert P. Hildebrandt}$^1$, Harald W. Grie{\ss}hammer$^1$ and \\
Thomas R. Hemmert$^1$\\[0.3cm]
$^1$Theoretische Physik T39, Physik-Department, TU M\"unchen,\\
D-85747 Garching, Germany\\[0.3cm]
\end{center}

Nucleon polarizabilities 
describe the response
of the charged constituents of the nucleon to an external electromagnetic 
field. 
In \cite{hgriehemmert} the concept of \textit{dynamical}, i.e. 
energy-dependent polarizabilities, derived from Compton multipoles, has been 
introduced. 
These quantities 
gauge the response of the internal nucleonic degrees of freedom to an 
external, real photon 
field of definite multipolarity and non-zero energy. 
In this work we present a projector formalism to extract the nucleon 
polarizabilities 
from a multipole expansion of the structure part of the six Compton 
amplitudes. We calculate the 
dynamical polarizabilities up to leading-one-loop order in 
an Effective Field Theory 
with (Small Scale Expansion) and without (HB$\chi$PT) explicit $\Delta(1232)$ 
degrees of freedom and compare our result to 
dispersion relation calculations \cite{Compton1}, 
finding good agreement in most multipole channels and the 
possibility to determine the static 
spin-independent dipole polarizabilities $\bar{\alpha}_{E1}$ and 
$\bar{\beta}_{M1}$ 
from low energy Compton data via the SSE-amplitudes. To include the $\Delta$ 
resonance as explicit degree of freedom is crucial for the resonant 
polarizabilities (\textit{e.g.} $\beta_{M1}(\omega)$) and for a correct 
description of backward Compton scattering data \cite{Compton1}. 
We demonstrate that quadrupole polarizabilities 
give negligibly small contributions to Compton cross sections -- spin-averaged
as well as polarized ones -- i.e. the multipole expansion converges very 
fast. This leaves us with only the six dipole polarizabilities as 
free functions of energy 
which one therefore may be able to fit to experimental Compton scattering data.
This hope is confirmed, as we show that the dynamical spin 
polarizabilities give sizeable contributions even to spin-averaged Compton 
scattering and that they are absolutely dominant in certain 
polarization configurations \cite{Compton2}. 
Finally we present one possible, 
model-independent way to determine the dynamical spin dipole polarizabilities 
directly from experiment, proving the principle by fitting only two of them 
to spin-averaged Compton data.

\newabstract 
\label{abs:filkov}

\begin{center}
{\large\bf Radiative Pion Photoproduction and Pion Polarizability}\\
[0.5cm]
Lev Fil'kov$^1$ \\
(for A2 and TAPS collaborations of MAMI)\\[0.3cm]
$^1$Lebedev Physical Institute, Leninsky Prospect 53, 119991 Moscow, Russia\\
[0.3cm]
\end{center}

The process of the radiative $\pi^+$ meson photoproduction from the
proton $(\gamma p\to\gamma\pi^+ n)$ is studied with the aim to determine 
the $\pi^+$ meson polarizabilities \citetwo{drec}{ahr}. For this purpose
an experiment on the process under study has been carried out at MAMI-B 
in the kinematical region 537 MeV$<E_{\gamma}<$817 MeV,
$140^{\circ}\le\theta^{cm}_{\gamma\gamma}\le 180^{\circ}$ where
$\theta^{cm}_{\gamma\gamma}$ is the photon emission angle in $cms$ of the
$\gamma \pi$ scattering. The $\pi^+$
meson polarizabilities were determined from the comparison of the
experimental data with two different theoretical models (without \cite{unk} 
and with baryon resonances). In order to decrease the contribution of the 
baryon resonances, the differential cross section of the process under
study has been integrated over $\theta^{cm}_{\gamma\gamma}$ from
$140^{\circ}$ to $180^{\circ}$ and over $\varphi$ from $0^{\circ}$ to
$360^{\circ}$ (where $\varphi$ is the Treiman-Yang angle between the 
planes formed by the momenta of the initial photon and the proton
and the initial photon and the emitted pion in $cms$ of the
$\gamma\pi\to\gamma\pi$ reaction). 
To decrease a model dependence of the results obtained,
the kinematical region of this process has been chosen where the 
difference between the predictions of two considered models does not
exceed 3\% when $(\alpha-\beta)_{\pi^+}=0$.

The setup efficiency has been normalized by the comparison of the 
theoretical model predictions with the experimental data in the kinematical 
region where the pion polarizability contribution is negligible
$(s_1<5\mu^2)$.

In the region, where the pion polarizability contribution is essential
$(5<s_1/\mu^2<15, \; -12<t/\mu^2<-2)$, the difference of the electric
and magnetic $\pi^+$ meson polarizabilities has been determined from the 
comparison of the experimental data on the total cross section of the 
radiative pion photoproduction
with the predictions of two theoretical models under consideration. 
As a very preliminary result we have obtained:
$$
(\alpha-\beta)_{\pi^+}=(11.6\pm 1.5\pm 5.1)\times 10^{-4}\; {\rm fm}^3
$$
where the errors are statistical and systematic ones including the difference 
between the results found with help of the considered models, respectively.

\newabstract 
\label{abs:glander}

\begin{center}
{\large\bf Strangeness Production from ELSA}\\
[0.5cm]
{K.-H. Glander}
\\[0.2cm]
Physikalisches Insitut, Universi\"at Bonn\\
[0.1cm]
Nu{\ss}allee 12, D-53115 Bonn,
Germany\\[0.1cm]
\end{center}

\vspace{1cm}

The reactions $\gamma p \to K^+ \Lambda$ and 
$\gamma p \to K+ \Sigma^0$ were measured in the energy range from 
threshold up to a photon energy of 2.6 GeV \cite{saphir}. 
The data were taken with 
the SAPHIR detector at the electron stretcher facility, ELSA. 
Results on cross sections and hyperon polarizations are presented as 
a function of kaon production angle and photon energy. The total cross 
section for $\Lambda$ production rises steeply with energy close to threshold, 
whereas the $\Sigma^0$ cross section rises slowly to a maximum at about 
$E_\gamma = 1.45\,$GeV. Cross sections together with their 
angular decompositions into Legendre polynomials suggest 
contributions from resonance production for both reactions. 
In general, the induced polarization of the $\Lambda$ has negative 
values in the kaon forward direction and positive values in the backward 
direction. The magnitude varies with energy. The polarization of 
the $\Sigma^0$ follows a similar angular and energy dependence as 
that of the $\Lambda$, but with opposite sign. A detailed discussion 
of the data, their analysis and the results can be found in Ref.~\cite{saphir}.

\newabstract 
\label{abs:raue}

\begin{center}
{\large\bf Strangeness Production  at Jefferson
 Lab}\\[0.5cm]
Brian Raue \\[0.3cm]
Florida International University,\\
Miami, FL, USA.\\[0.3cm]
\end{center}

The Thomas Jefferson National Accelerator Facility has an extensive program
of studying the electromagnetic production of strange particles.  One of
the main components of this program has been the study of both photo- and
electroproduction of $K^++\Lambda^0$ and $K^++\Sigma^0$ final states.
Experiments are being, or have been conducted in all three of Jefferson
Lab's experimental halls measuring a wide range of observables at
kinematics from threshold up to $W\approx 3.0$ GeV and $Q^2$ from 0.4 up
to 5 (GeV/$c)^2$.

The largest effort in this endeavor is taking place in Hall B using the
CEBAF Large Acceptance Spectrometer (CLAS).  Data have been taken at about ten
different polarized electron beam energies and are currently being
analyzed. 

Preliminary results \cite{feuerbach} for one beam energy exist wherein the
unpolarized cross section has been separated into three components:
$\sigma_T+\epsilon_L\sigma_L$, $\sigma_{TT}$, and $\sigma_{LT}$. The data
indicate a $t$-channel dominance for the $\Lambda^0$ production and a
strong $s$-channel dominance for $\Sigma^0$ production.  The large
$\sigma_{LT}$ structure function for $\Lambda^0$ suggests a large coupling
to the longitudinal component of the virtual photon. The $W$ dependence of
the structure functions for both final states indicate the likely influence
of various intermediate resonances.  Precision measurements of the
photoproduction cross section have been made using CLAS and may also
indicate the presence of previously unobserved intermediate resonances
\cite{mcnabb}.

These results will soon be complemented with a full analysis of all CLAS
data wherein $\sigma_T$ and $\sigma_L$ will be measured as well as the
fifth structure function $\sigma_{LT^\prime}$ \cite{carman1}.  Measurements
of the transferred $\Lambda^0$ polarization \cite{carman2} along with
high-precision $L$-$T$ separations from Hall C \cite{mohring} and Hall
A \cite{markowitz} will also provide additional constraints for understanding
the underlying strangeness electroproduction process.

\newabstract 
\label{abs:oset1}

\begin{center}
{\large\bf Unitarized Chiral Dynamics: SU(3) and Resonances}\\[0.5cm]
{\bf E. Oset}, A. Ramos, C. Bennhold, D. Jido, J. Antonio Oller and 
U.-G. Mei{\ss}ner\\[0.3cm]
Valencia, Barcelona, Washington, Osaka, Murcia and Bonn Collaboration\\[0.3cm]

\end{center}

Using a chiral unitary approach for the meson--baryon interactions, we show
that two octets of $J^{\pi}=1/2^-$ baryon states, which are degenerate in the
limit of exact SU(3) symmetry, and a singlet are generated dynamically.  The
SU(3) breaking produces the splitting of the two octets, resulting in the case
of strangeness $S=-1$ in two poles of the scattering matrix close to the
nominal $\Lambda(1405)$ resonance. These poles are combinations of the singlet
state and the octets.  We show how actual experiments see just one effective
resonance shape, but with properties which change from one reaction to
another. Suggestions of experiments are made to populate either of the two
$\Lambda(1405)$ resonances \cite{Jido:2003cb}. Members of these two octets are
also the I=0 $\Lambda(1670)$ state and I=1 $\Sigma(1620)$ state with
strangeness S=-1 \cite{Oset:2001cn}, plus the $\Xi(1620)$ state with S=-2
\cite{Ramos:2002xh}.  Some partial evidence is seen for another I=1 state
around the $K N $ threshold and hints are given that the differences in the
$\pi ^+ \Sigma ^-$ and $\pi ^- \Sigma ^+$ distributions in $\Lambda(1405)$
photoproduction in a recent experiment at Spring8/Osaka could unveil the
existence of this I=1 state \cite{nakano}.

\newabstract 
\label{abs:merkel}

\begin{center}
{\large\bf Threshold Pion Electroproduction: Update on Experiments}\\[0.5cm]
Harald Merkel\\[0.3cm]for the {\sc A1 Collaboration}\\[0.3cm]
Institut f\"ur Kernphysik, Johannes Gutenberg-Universit\"at Mainz\\
Becherweg 45, 55099 Mainz\\[0.3cm]
\end{center}

While in threshold pion photoproduction experiment and theory is
in good agreement, the current published data\cite{nikhef} on
threshold pion electroproduction show considerable disagreement with
predictions in the framework of Chiral Perturbation Theory\cite{bernard}.

To further investigate this disagreement, an experiment with extended
momentum acceptance was performed at MAMI to increase the
sensitivity to the small p wave contributions. Simultanously, a measurement
with polarized electron beam and out of plane acceptance was
performed to investigate the effects of the opening $n \pi^+$ channel
on the $p \pi^0$ s wave amplitude via the TL' interference.

The new dataset\cite{weis03} covers the energy range from threshold up
to a center of mass energy of 40\,MeV above threshold at a photon virtuality of
$Q^2=0.05\,\mathrm{GeV}^2/c^2$. The extracted cross sections $\sigma_0
= \sigma_T + \epsilon \sigma_L$, $\sigma_{LT}$, and $\sigma_{TT}$ are
below the predictions of ChPT, but in good agreement with the 
phenomenological models MAID and DMT \cite{maid}. The unitary cusp in the 
the asymmetry $A_{TL'}$ is only described by the DMT model.

\newabstract 
\label{abs:kotulla}

\newcommand{\Piz}{\ensuremath{\pi^0}}
\newcommand{\Eta}{\ensuremath{\eta}}
\newcommand{\Egam}{\ensuremath{E_\gamma}}
\newcommand{\xpt}{\ensuremath{\chi}pt}
\newcommand{\Rpipi}{\ensuremath{\gamma p \rightarrow \Piz{} \Piz{} p}}
\newcommand{\Roper}{P\ensuremath{_{11}}(1440)}
\newcommand{\Donethree}{D\ensuremath{_{13}}(1520)}
\newcommand{\Rpip}{\ensuremath{\gamma
p \rightarrow \Piz{} p}}
\newcommand{\Rhp}{\ensuremath{\gamma
p \rightarrow \Eta{} p}}
\newcommand{\GDDg}{\ensuremath{\Delta \rightarrow \Delta \gamma^\prime}}
\newcommand{\muD}{\ensuremath{\mu_{\Delta^+}}}
\newcommand{\kapD}{\ensuremath{\kappa_{\Delta^+}}}

\begin{center}
{\large\bf Two {\boldmath $\pi^0$} Production at Threshold}\\[0.5cm]
{\bf Martin Kotulla}$^1$ for the TAPS/A2 Collaborations\\[0.3cm]
$^1$Department of Physics and Astronomy,\\
      University of Basel, CH-4056 Basel (Switzerland)\\[0.3cm]
\end{center}

The reaction $\gamma p \rightarrow \pi^0 \pi^0 p$ has been measured 
using the TAPS BaF$_2$ calorimeter at the tagged photon facility of the
Mainz Microtron accelerator.
Close to threshold,
chiral perturbation theory (ChPT)
predicts that this channel is
significantly enhanced compared to double pion final states with
charged pions \cite{bernard:pipithres}.
A ChPT calculation of the
2\Piz{} channel up to order $M_\pi^2$ attributes the total 
strength dominantly to
pion loops at order $q^3$ \cite{bernard:2pi0thres} - a finding that  
opens new prospects for the test of 
ChPT. 

\begin{figure}[h]
\begin{minipage}{0.0cm}
\epsfysize=8.0cm \epsffile{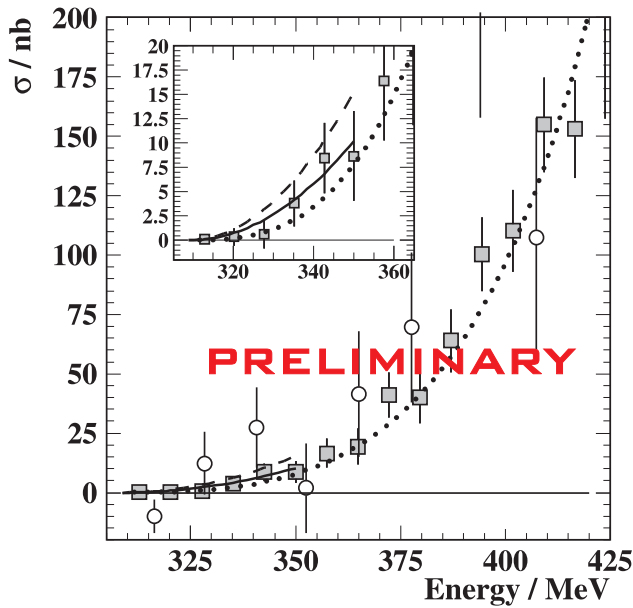}
\end{minipage}
\hspace*{8.5cm}
\begin{minipage}{5.5cm}
  \caption{
  Total cross section for the reaction \Rpipi{} (full squares) at
  threshold in comparison with \cite{wolf:pipi} (open circles).
  The prediction
  of the ChPT calculation \cite{bernard:2pi0thres} is shown (solid curve)
  and the prediction of Ref.
  \cite{roca:pipi} (dotted curve).
  The upper limit of the ChPT calculation stems from the uncertainty
  of the coupling $N^*N\pi\pi$ of the \Roper{} resonance (dashed curve).
}
\label{fig:kin}
\end{minipage}
\end{figure}

Our measurement of the cross section at threshold (see Fig~\ref{fig:kin})
is the first which is sensitive enough for a conclusive 
comparison with the ChPT calculation \cite{bernard:2pi0thres} 
and is in agreement with its prediction \cite{kottu:2pi0thres}.
The data are also in agreement with a calculation 
in the unitary chiral approach \cite{roca:pipi}.

\newabstract 
\label{abs:yang}

\begin{center}
{\large\bf Threshold $\pi^0$ Photo- and Electro-Production}\\[0.5cm]
{\bf S. N. Yang}$^1$, G.-Y. Chen $^1$, D. Drechsel$^2$,
S. S. Kamalov$^2$ and L. Tiator$^2$\\[0.3cm]
$^1$Dep. of Phys., National Taiwan University, Taipei 10617, Taiwan
\\[0.3cm]
$^2$Institut f\"ur Kernphysik, Universit\"at Mainz, 55099 Mainz, Germany
\\[0.3cm] \end{center}

In this talk we present the predictions of the Dubna-Mainz-Taipei (DMT)
dynamical model, based on meson-exchange picture, which we recently developed
in Ref. \cite{Kamalov99} for the threshold electromagnetic pion production and
compare them with the results of ChPT\cite{Bernard91}.  In the DMT model for
the pion photo- and electro-production the t-matrix is given as
\begin{eqnarray}
t_{\gamma\pi}(E)=v_{\gamma\pi}+v_{\gamma\pi}g_0(E)\,t_{\pi
N}(E)\,
\label{eq:tmatrix}
\end{eqnarray}
where $v_{\gamma\pi}$ is the $\gamma\pi$ transition potential,
$g_0$ and $t_{\pi N}$ are the $\pi N$ free propagator and
$t-$matrix, respectively, and $E$ is the total energy in the c.m.
frame. In the present study,  $t_{\pi N}$ is
obtained in a meson-exchange $\pi N$ model \cite{hung}.
Both $v_{\pi N}$ and $v_{\gamma\pi}$  are derived from an  effective
Lagrangian containing Born terms as well as $\rho$- and $\omega$-exchange
in the $t$-channel.

In such approach $t_{\pi N}$ contains the
effect of $\pi N$ rescattering to all orders. However, we have
found that only the first order rescattering contribution, i.e.
the 1-loop diagram, is important. In Table 1, the results obtained up
to tree, 1-loop, and 2-loop approximations for all four possible
pion photoproduction channels are listed and compared to the
experiments and ChPT results\cite{Bernard91}. We see that for $\pi^0$
production from both proton and neutron, it is necessary to include one-loop
contribution while tree approximation is sufficient for the
charged pion productions.

\begin{table}[htb]
Table 1. Threshold values of $E_{0+}$
(in units $10^{-3}/m_{\pi}$).
\newcommand{\m}{\hphantom{$-$}}
\begin{tabular}{ccccccc}
\hline
        & Tree & 1-loop & 2-loop & Full & ChPT & Exp\\
\hline
$\pi^0p$&$-2.26$ &$-1.06$ &$-1.01$ &$-1.00$&$-1.1$         &$-1.33\pm0.11$\\
$\pi^+n$&\m27.72 
&\m28.62 &\m28.82 &\m28.85&\m28.2\,$\pm$\,0.6 &\m28.3\,$\pm$\,0.3  \\
$\pi^0n$&\m0.46  &\m2.09  &\m2.15  &\m2.18 &\m2.13         &                \\
$\pi^-p$&$-31.65$&$-32.98$&$-33.27$&$-33.31$&$-32.7\pm0.6$&$-31.8\pm1.9$ \\
\hline
\end{tabular}
\vspace{-0.5cm}
\end{table}

\newabstract 
\label{abs:r.arndt}

\begin{center}
{\large\bf Partial Wave Analysis of Pion-Nucleon \\[0.1cm]
Scattering below {\boldmath ${\rm T}_{\rm LAB}$} = 2.1 GeV}\\[0.5cm]
{\bf R.A. Arndt}$^1$, W.J. Briscoe$^1$, M.M. Pavan$^2$, \\
I.I. Strakovsky$^1$ and R.L. Workman$^1$\\[0.3cm]
$^1$Dept. of Physics, The George Washington University, Washington, 
D.C., USA\\[0.3cm]
$^2$TRIUMF, Vancouver, B.C., Canada\\[0.3cm] 
\end{center}

We present results from a partial-wave analysis of pion-nucleon elastic
and charge-exchange data, covering the region from threshold to a lab
pion kinetic energy of 2.1 GeV, employing a coupled-channel formalism
to simultaneously fit $\pi^- p \to \eta n$ data to 0.8 GeV. The main result,
fit FA02, has been constrained by a complete set of dispersion-relation
constraints, from 20 MeV to 1 GeV, and from t = 0 to -0.4 (GeV/c)$^2$.
This represents an improvement, in the number and range of constraints,
as compared to our previous SM95 analysis\cite{SM95}.

We have applied a number tests to our solution, including fits with fewer or
no dispersion-relation constraints, fits excluding the charge-exchange
data base (to gauge the size of charge-symmetry violation), and fits 
using a different Coulomb correction schemes. A modified form of the
Nordita method has been used in the final FA02 result. We have also 
examined the sensitivity of the resonance extraction to a method commonly
used to determine Breit-Wigner parameters. We now fit data directly in
determining these BW parameters, as opposed to fitting single-energy
solutions, as was done for SM95. 

Our results for the pion-nucleon coupling constant and sigma term have not
changed significantly with the addition of further dispersion-relation
constraints or the inclusion of eta-production data. Somewhat surprisingly,
the S$_{11}$ partial-wave is also quite similar to that seen in SM95.
However, the revised fitting strategy has resulted in a width for the
S$_{11}(1535)$ 
which is now more in line with the PDG average.

\newabstract 
\label{abs:piirola}

\begin{center}
{\large\bf Towards the Pion-Nucleon PWA}\\[0.5cm]
{\bf Pekko Piirola}$^1$ and Mikko E. Sainio$^{1,2}$\\[0.3cm]
$^1$Dep. of Physical Sciences, University of Helsinki,\\
P.O.Box 64, 00014 University of Helsinki, Finland\\[0.3cm]
$^2$Helsinki Institute of Physics,\\
P.O.Box 64, 00014 University of Helsinki, Finland\\[0.3cm]
\end{center}

The last full Karlsruhe partial wave analysis of $\pi N$ scattering
was finished ca.\ 20 years ago \citetwo{kh80}{ka85}.  After that a huge
pile of new accurate $\pi N$ scattering measurements has emerged with
which the old analysis is not completely compatible.  We are making a
highly modernized version of the old analysis to all (recent
\emph{and} old) data, aiming to pay particular attention to the
discrepancies between different data sets and to the handling of the
experimental errors as well as to the electromagnetic corrections and
the effects of isospin breaking \cite{erice02}.

The most essential feature of our analysis is the use of fixed-$t$
dispersion relations as constraints.  Instead of working with
dispersion integrals directly, special series for the invariant $B$-
and $C$-amplitudes are used:
\[
   C(\nu,t) = C_N(\nu,t) + H(Z,t) \sum_{i=0}^n c_i Z^i(\nu,t) \ .
\]
Here the base functions $Z^i(\nu,t)$ are conformal mappings with the
proper analytic structure \cite{ctfm}, and $H(Z,t)$ is a factor
describing the expected asymptotic behaviour. When fitting the
coefficients $c_i$, convergence and smoothness of the fit are achieved
using the convergence test function method, \emph{i.e.}  minimizing
not only the pure $\chi^2$, but the sum $\chi^2 + \Phi$, where $\Phi$
is the penalty function of the CTF method \cite{ctfm}.

At the time of writing (\emph{i.e.} October 2003) we are testing the
machinery with the old KA85 solution \cite{ka85} excluding all of the
experimental input, in order to find out the possible internal
inconsistencies of our implementation.

\newabstract 
\label{abs:lutz}

\begin{center}
{\large\bf Meson-Baryon  Scattering}\\[0.5cm]
Matthias F.M. Lutz\\[0.3cm]
Gesellschaft f\"ur Schwerionenforschung (GSI)\\
Planck Str. 1, D-64291 Darmstadt, Germany \\[0.3cm]
\end{center}

We study $J^P=\frac{3}{2}^-$ baryon resonances as generated by chiral
coupled-channel dynamics in the $\chi-$BS(3) approach 
\citethree{LK00}{LK01}{LK02}.
Parameter-free results are obtained in terms of the Weinberg-Tomozawa term
predicting the leading s-wave interaction strength of Goldstone bosons with
baryon-decuplet states \cite{Copenhagen}. In the 'heavy' SU(3) limit with
$m_\pi = m_K \sim 500 $ MeV the resonances turn into bound states forming a
decuplet and octet representation of the SU(3) group. Using physical masses
the mass splitting are remarkably close to the empirical pattern.

Our effective field theory is based on the assumption that at subthreshold
energies the scattering amplitudes can be evaluated in standard chiral
perturbation theory. Once the available energy is sufficiently high to permit
elastic two-body scattering a further typical dimensionless parameter
$m_K^2/(8\,\pi f^2) \sim 1$ arises if strangeness is considered explicitly.
Since this ratio is uniquely linked to two-particle reducible diagrams it is
sufficient to sum those diagrams keeping the perturbative expansion of all
irreducible diagrams. We suggest to glue s- and u-channel unitarized
scattering amplitudes together at subthreshold energies
\citetwo{LK02}{Copenhagen}. 
This construction reflects our basic assumption that
diagrams showing a s- or u-channel unitarity cut need to be summed to all
orders at least at energies close to where the diagrams develop their
imaginary part. By construction, a glued scattering amplitude satisfies
crossing symmetry exactly at energies where the scattering process takes
place. At subthreshold energies crossing symmetry is implemented
approximatively only, however, to higher and higher accuracy when more chiral
correction terms are considered. Insisting that the scattering amplitude
reproduces the interaction kernel at subthreshold energies guarantees that
subthreshold amplitudes match smoothly and therefore the final glued
amplitudes comply with the crossing-symmetry constraint to high accuracy.

\newabstract 
\label{abs:frink}

\begin{center}
{\large\bf Order Parameters of Chiral Symmetry Breaking from\\[0.1cm] 
Meson -- Baryon Dynamics}\\[0.5cm]
Matthias Frink \\[0.3cm]
IKP, Forschungszentrum J\"ulich\\D-52425 J\"ulich, Germany\\[0.3cm]
\end{center}

In a theory with $N_f$ light quark flavours spontaneous breaking of chiral
symmetry $SU(N_f)_V\times SU(N_f)_A \rightarrow SU(N_f)_V$ entails $N_f^2-1$
(approximately) massless Goldstone bosons. Order parameters, i.\ e.\ vacuum
expectation values of operators transforming non--trivially under the broken
symmetry group, are a useful tool to examine symmetry breaking. Examples of
chiral order parameters are the quark condensate and Goldstone boson decay
constants.\\ Chiral order parameters $O$ are conjectured to behave
paramagnetically with an increasing number of light quark species, i.\ e.\ 
$O(N_{f+1})< O(N_f)$ \cite{DGJ}. Fields of relevance and possibel applications
are e.\ g.\ the determination of hadron masses, the applicability of
Generalized ChPT and an approach to OZI rule violation in the scalar sector
\cite{D}.\\ Descotes and Stern have analyzed the role of chiral order
parameters in the mesonic sector \cite{DGJ}, \cite{D}. A step towards a
further investigation is the inclusion of meson--baryon dynamics into the
analysis.\\ Basic input for such calculations is the complete and minimal
three--flavour meson--baryon Lagrangian. Its third--order portion has been
formulated by Krause \cite{K}. Building blocks for the Lagrangian to fourth
order are besides baryon and meson fields external source terms and
Clifford--algebra elements. Various reduction mechanisms can be used in the
process of minimization, e.\ g.\ relations between the fields, trace relations
for $SU(3)$, identities among Clifford--algebra elements, and equation of
motion eliminations. Work is in progress here.\\ As a first application we
plan to calculate octet baryon masses to fourth order in the chiral expansion,
using the infrared regularization scheme. The corresponding third--order
contributions are given in \cite{ET}. Further applications comprise e.\ g.\ 
scalar baryon form factors and their low--energy relation with scattering
amplitudes as well as sigma--term and strangeness content determinations.

\newabstract 
\label{abs:bernstein}

\begin{center}
{\large\bf Hadron Deformation and Chiral Dynamics}\\[0.5cm]
A.M.Bernstein\\
Physics Dept. and Laboratory for Nuclear Science,M.I.T.\\
Cambridge, Mass., U.S.A. 
\end{center}
The much conjectured deviation of the proton shape from spherical symmetry
cannot be directly observed since it has spin 1/2 and therefore no quadrupole
moment. This has led to an active field of measurements of the electromagnetic
transition of the proton to its first excited state, the $\Delta$ with spin
3/2. We have taken accurate data at $Q^{2} = 0.126 GeV^{2}$ at the MIT/Bates
accelerator with the out of plane spectrometer system (OOPS)\cite{OOPS}. These
and other photo- and elctro-pion production experiments\citetwo{OOPS}{AB} have
observed significant non-zero electric and Coulomb quadrupole matrix elements
which signify deviations from spherical symmetry in the proton and/or the
$\Delta$.  These observed quadrupole amplitudes are generally an order of
magnitude greater then predicted by the quark model\cite{AB}. As has been
pointed out for years in terms of the cloudy bag model, this should not be
surprising since quark models do not properly handle chiral symmetry or the
pion. More recently two dynamical models of the $\gamma^{*} p \rightarrow \pi
N$ reaction\cite{cloud} have shown that the contribution of the pion cloud is
the dominant one and that this maximizes in the vicinity of our Bates
data\cite{OOPS} .

In my view the dominance of pionic effects follows naturally from the
spontaneous breaking of chiral symmetry. The coupling of a pion (Goldstone
Boson) to a nucleon is given $L_{\pi N} = g\vec{\sigma}\cdot \vec{k}$ where g
is the $\pi-N$ coupling constant( predicted by chiral symmetry), $\sigma$ is
the nucleon spin, and k is the pion momentum. This interaction vanishes in the
s wave and is strong in the p wave. This is the basis of the deviation from
spherical symmetry in the nucleon since the pion cloud is emitted and absorbed
in the p wave. Furthermore this is the same physical effect as is observed in
the well known, long range, tensor force between nucleons.$V_{NN}=g^{2}\ 
\vec{\sigma_{1}}\cdot \vec{k} \ \frac{1}{k^{2}+m_{\pi}^{2}} \ 
\vec{\sigma_{2}}\cdot \vec{k} $.

It is an open problem to calculate the deformation in ways more strongly
linked to QCD such as ChPT or the lattice, and to perform more measurements to
determine the precise magnitude of the proton and $\Delta$ deformations.

\newabstract 
\label{abs:pascalutsa}

\begin{center}
{\large\bf Chiral Dynamics in the {\boldmath $\Delta(1232)$} Region}\\[0.5cm]
{\bf Vladimir Pascalutsa}$^{1,2}$ and Daniel R.~Phillips$^2$\\[0.3cm]
$^1$Department of Physics, College of William \& Mary,
Williamsburg, VA 23185\\[0.1cm]
{\it and} Theory Group, JLab, 12000 Jefferson Ave, Newport News, 
VA 23606\\[0.2cm]
$^2$Department of Physics and Astronomy, Ohio University, 
Athens, OH 45701\\[0.2cm]
\end{center}

The limit of applicability of ChPT in the baryon sector is set by the
excitation energy of the first resonance: $\Delta\equiv M_\Delta-M_N\approx
293$ MeV.  We have recently proposed a systematic scheme that extends ChPT
into the resonance region in a natural way \cite{PP2002} (see also
\cite{PP2003} for a more concise explanation).

Our power counting catches the most crucial feature of the resonance physics:
the need for the resummation of the self-energy insertions which result in
dynamical generation of the width.  This is achieved by distinguishing the
resonance-excitation energy from the other low-energy hadronic scales, such as
$m_\pi$. In terms of small expansion parameter $\delta$, we assume that
$\Delta=O(\delta)$ while the lower-energy scales are of higher order, e.g.,
$m_\pi=O(\delta^2)$. The power-counting then depends on kinematics, namely on
whether we are in the low-energy regime $p\sim m_\pi$ or in the resonance
region $p\sim\Delta$, where $p$ is a typical relative momentum.  It shows that
at low energies the $\Delta$-isobar contributions are suppressed with respect
to the nucleon ones by a power of the small parameter.  One the other hand, in
the resonance region one class of graphs --- the one-$\Delta$-reducible graphs
--- needs to be resummed, since each graph in this class goes as some power of
$1/(p-\Delta)$ and hence does not have a well-defined power counting for
$p\sim \Delta$.

After the resummation, the one-$\Delta$-reducible graphs obtain a definite
power-counting index, which is determined by the index of the self-energy.
The expansion of the $\Delta$ self-energy begins with the one $\pi N$-loop,
which goes of course as $p^3$. Since the $\Delta$-propagator in a {\it
  resummed} one-$\Delta$-reducible graph is $1/(p-\Delta-\Sigma)$, it counts
as $1/\delta^3$ in the resonance regime.  Thus, the one-$\Delta$-reducible
graphs are enhanced in the resonance region by two powers of the small
parameter.

The mechanisms of $\Delta$ excitation and its dynamical width generation are
of course present in many phenomenological models. Our goal was to incorporate
them into a systematic power-counting scheme which could extend ChPT.

In this talk we have also discussed several applications. Most notably, the
extraction of the magnetic moment of the $\Delta$(1232), where both $p\sim
m_\pi$ and $p\sim \Delta$ regimes are at play simultaneously!

\newabstract 
\label{abs:lewis2}

\begin{center}
{\large\bf Lattice Regularized Chiral Perturbation Theory}\\[0.5cm]
Bu\=gra Borasoy$^1$, {\bf Randy Lewis}$^2$ and Pierre-Philippe A. Ouimet$^2$
\\[0.3cm]
$^1$Physik Department, Technische Universit\"at M\"unchen, D-85747 Garching,
Germany\\[0.3cm]
$^2$Department of Physics, University of Regina, Regina, SK, S4S 0A2, Canada
\\[0.3cm]
\end{center}

The spacetime lattice is a regularization technique for quantum field theories.
Its application to chiral perturbation theory (ChPT) can be of interest to
lattice QCD studies since the effective theory, lattice ChPT, is then defined
in the same discrete spacetime as the underlying theory, lattice
QCD.\cite{matchQCD}
Lattice ChPT can also be viewed as yet another way to implement a cut-off
regularization scheme, somewhat comparable to Ref.~\cite{cutoffs},
and with the advantage that chiral symmetry and
gauge invariance are automatically perserved throughout any
lattice ChPT calculation.
Since lattice regularization does not rely on
perturbation theory, it might be valuable for
multi-nucleon effective field theories as well.\cite{multinuc}

The simplest lattice ChPT Lagrangian is obtained by writing the standard
ChPT Lagrangian in Euclidean spacetime and converting derivatives to finite
differences in such a way as to preserve the desired symmetries without
introducing doublers or ghost states.
In the continuum limit, ChPT observables
are independent of regularization scheme and continuum observables agree
exactly with dimensional regularization.
For more details, see Ref.~\cite{Cairns}.

\newabstract 
\label{abs:scherer}

\begin{center}
{\large\bf Renormalization and Power Counting in Manifestly\\[0.1cm]
Lorentz-invariant Baryon Chiral Perturbation Theory}\\[0.5cm]
Thomas Fuchs$^1$, Jambul Gegelia$^{1,2}$, George Japaridze$^3$
and {\bf Stefan Scherer}$^1$\\[0.3cm]
$^1$Institut f\"ur Kernphysik, Johannes 
Gutenberg-Universit\"at, Mainz, Germany\\[0.3cm]
$^2$ High Energy Physics Institute, 
Tbilisi State University, 
380086 Tbilisi, Georgia\\[0.3cm]
$^3$
CTSPS,
Clark Atlanta University, Atlanta, Georgia 30314, USA\\[0.3cm]
\end{center}
   The implementation of a successful effective field theory program 
requires two main ingredients, namely: (a) a knowledge of the most general 
effective Lagrangian and (b) an expansion scheme for observables in terms 
of a consistent power counting method.
   It was shown by Gasser, Sainio, and \v{S}varc \cite{GSS:1988} that,
in the baryonic sector, the application of dimensional 
regularization in combination with the modified minimal subtraction scheme 
of chiral perturbation theory (ChPT) does not suffice to produce a simple 
correspondence between the loop expansion and the chiral expansion.
   A solution to this problem was first obtained in the framework of the 
heavy-baryon formulation of ChPT \cite{HB}.
   More recently, the problem was also attacked in a manifestly 
Lorentz-invariant form \citetwo{MLIBCHPT}{FGJS:2003}.
   In Ref.\ \cite{FGJS:2003} we have proposed a new renormalization scheme 
which makes use of finite subtractions of dimensionally regularized 
diagrams beyond the standard $\overline{\rm MS}$ scheme of ChPT
to remove contributions violating the power counting.
  Such subtractions can be realized in terms of local counterterms in the 
most general effective Lagrangian.
   Our approach may be used in an iterative procedure to renormalize 
higher-order loop diagrams and also allows for implementing a 
consistent power counting 
when vector (and axial-vector) mesons are explicitly 
included \cite{Fuchs2:2003}.

\newabstract 
\label{abs:gegelia}

\begin{center}
{\large\bf Baryon ChPT with Vector Mesons}\\[0.5cm] T.~Fuchs$^1$,
M.~R.~Schindler$^1$, {\bf J. Gegelia}$^{1,2}$ and S.
Scherer$^1$\\[0.3cm] $^1$Institut f\"ur Kernphysik, Johannes
Gutenberg-Universit\"at, Mainz, Germany\\[0.3cm] $^2$High Energy
Physics Institute, Tbilisi State University, 380086 Tbilisi,
Georgia\\[0.3cm]
\end{center}

By considering characteristic self-energy and vertex Feynman
diagrams \cite{Fuchs:2003sh} we have demonstrated that the
inclusion of explicit virtual (axial) vector particles in
manifestly Lorentz-invariant baryon chiral perturbation theory
does not violate the power counting if one uses the extended
on-mass-shell (EOMS) renormalization scheme of
Ref.~\cite{Fuchs:2003qc}. This renormalization procedure can also
be (iteratively) applied to multiloop diagrams leading to a
consistent power counting.

We have reformulated the IR regularization of Becher und Leutwyler
\cite{Becher:1999he} in a form analogous to the EOMS
renormalization scheme \cite{Schindler:2003xv}. Within this new
formulation the subtraction terms are found by expanding the
integrands of loop integrals in powers of small parameters (small
masses and Lorentz-invariant combinations of external momenta and
large masses) with subsequent exchange of integration and
summation. Within this new formulation of IR regularization one
does not necessarily need to use dimensional regularization. The
renormalized results are regularization scheme independent.

The IR regularization in its new formulation can be easily applied
to diagrams with an arbitrary number of propagators with various
masses (e.g. resonances) and/or diagrams with several fermion
lines as well as to multiloop diagrams.

Applying IR regularization in our formulation to EFT with explicit
vector mesons in the antisymmetric tensor field representation and
analyzing the diagrams contributing to the electromagnetic form
factors of the nucleon to ${\cal O}(q^4)$, we observe that in
Ref.~\cite{Kubis:2000zd} all relevant loop integrals have actually
been taken into account. This is due to the fact that those loop
integrals involving vector mesons which formally contribute at a
given order vanish in IR regularization.

\newabstract 
\label{abs:ellis}

\begin{center}
{\large\bf Baryon ChPT with Infrared Regularization: Update}\\[0.5cm]
Paul J. Ellis\\[0.3cm]
School of Physics \& Astronomy, University of Minnesota\\
Minneapolis, MN 55455, USA
\end{center}

In order to preserve power counting, loop integrals which involve both 
baryon and meson propagators are often evaluated using the heavy baryon (HB) 
approximation. Infrared regularization (IR) is an alternative method proposed 
by Becher and Leutwyler \cite{bl} which retains only the infrared singular 
part of the integral; the remainder can be expanded in a Taylor series and 
absorbed in the low energy constants. The IR part includes the corresponding 
HB result, thus preserving power counting, and also sums an infinite series of 
(baryon mass)$^{-n}$ corrections which appear in higher orders in the HB 
scheme. This gives the correct analytic behavior in threshold regions
which is not always true in the HB case.

HB results for the baryon masses in SU(3) give third order contributions of
similar magnitude to second order, whereas using IR they are smaller by 
roughly a factor of a half so that the convergence is much improved
\cite{et}. The convergence of the chiral expansion for electromagnetic form 
factors through fourth order \cite{km} is also improved when IR is employed. 
Including vector 
mesons explicitly, IR yields a good account of the data. In $\pi-N$ 
scattering the IR representation does not allow a sufficiently 
accurate extrapolation \cite{bl2} from the threshold region to the 
unphysical Cheng-Dashen point. Further in the physical region \cite{te}
IR allows a fit to the phase shifts only over a much more limited region 
than the HB scheme and even then the parameters are unreasonable. 
An IR study of the baryon octet axial vector currents 
shows poor convergence even when the decuplet is included \cite{z}. 
(For discussion of the spin structure of
the nucleon see V. Bernard, Plenary Session 2). Overall the picture is mixed
at the moment.

\newabstract 
\label{abs:meissner}

\begin{center}
{\large\bf Cutoff Schemes in Chiral Perturbation  
Theory and the\\[0.1cm] Quark Mass Expansion of the Nucleon Mass}\\[0.5cm]
V{\'e}ronique Bernard$^1$, Thomas R. Hemmert$^2$ and
{\bf Ulf-G. Mei\ss ner}$^3$\\[0.3cm]
$^1$Laboratoire de Physique Th\'eorique, ULP, F-67084 Strasbourg, 
France\\
$^2$Physik Department T39, TU M\"unchen, D-85747 M\"unchen, Germany\\
$^3$HISKP (TH), Universit\"at Bonn, D-53115 Bonn, Germany\\[0.3cm]
\end{center}

We discuss the use of cutoff methods in chiral perturbation theory \cite{BHM}.
We develop a cutoff scheme based on the operator structure of the effective
field theory that allows to suppress high momentum contributions
in Goldstone boson loop integrals and by construction is free of the 
problems traditional
cutoff schemes have with gauge invariance or chiral symmetries. 
As an example, we discuss the chiral
expansion of the nucleon mass. Contrary to other claims in the literature 
we show that the mass of a nucleon in heavy baryon chiral perturbation theory 
has a well behaved chiral expansion up to effective Goldstone boson masses of 
400 MeV when one utilizes standard dimensional regularization techniques. 
With the help of the here developed cutoff scheme we can demonstrate 
a well-behaved chiral expansion for the nucleon mass up to 600 MeV 
of effective Goldstone Boson masses. We also discuss in detail the prize, 
in numbers 
of additional short distance operators involved, that has to be paid for this 
extended 
range of applicability of chiral perturbation theory with cutoff 
regularization, 
which is usually not paid attention to. We also compare the fourth order result
for the chiral expansion of the nucleon mass with lattice results and draw some
conclusions about chiral extrapolations based on such type of representation.
In particular, the recent CP-PACS data \cite{CPPACS} can 
be well represented by the fourth order
dimensionally regularized nucleon mass expression with LECs consistent
with values obtained from the analysis of pion-nucleon scattering,
see e.g. \citetwo{Paul}{FM}.

\newpage
\section*{Working Group III: Few-Body Dynamics\\ 
\centerline{and Hadronic Atoms}}

\centerline{
Convenors: Paulo Bedaque, Nasser Kalantar-Nayestanaki and  Akaki Rusetsky}

\vskip3mm
{
\begin{tabular}{lp{9.8cm}r}
\multicolumn{3}{c}{\bf Few-Body Dynamics I}  \\
\stalk{U. van Kolck}{ Charge 
Symmetry Breaking in Pion Production}{abs:vanKolck}
\stalk{E. Stephenson}{ Observation of the Charge Symmetry Breaking\newline 
$d$+$d$ $\to$ $^4$He+$\pi^0$
 Reaction near Threshold}{abs:stephenson}
\stalk{A. Nogga}{ Probing Chiral Interactions in Light Nuclei}{abs:nogga}
\stalk{F. Myhrer}{ Solar Neutrino Reactions and Effective Field Theory}
{abs:myhrer}
\stalk{S. Ando}{ Solar-Neutrino Reactions on Deuteron in 
EFT* and\newline Radiative Corrections of Neutron Beta Decay}
{abs:ando}
\stalk{M.C. Birse}{ EFT's for the Inverse-Square Potential\newline 
and Three-Body 
Problem}{abs:birse}
[1mm]
\multicolumn{3}{c}{\bf Hadronic Atoms}\\
\stalk{M. Cargnelli}{ DEAR - Kaonic Hydrogen: First Results}{abs:cargnelli}
\stalk{J. Gasser}{ Comments on $\bar K N$ Scattering}{abs:gasser}
\stalk{A.N. Ivanov}{ On Pionic and Kaonic Hydrogen}{abs:ivanov}
\stalk{P. Zemp}{ Deser-type Formula for 
Pionic Hydrogen}{abs:zemp}
\stalk{L. Nemenov}{ Atoms Consisting of $\pi^+\pi^-$ and $\pi K$ 
Mesons}{abs:nemenov}
\stalk{L. Girlanda}{ Deeply Bound Pionic Atoms: Optical Potential at $O(p^5)$ 
in ChPT}{abs:girlanda}
[1mm]
\multicolumn{3}{c}{\bf Few-Body Dynamics II}\\
\stalk{H. Grie{\ss}hammer}{ All-Order Low-Energy Expansion in the 
3-Body System}{abs:griesshammer}
\stalk{T. Black}{ New Results in the Three and Four Nucleon Systems} 
{abs:black}
\stalk{H. Schieck}{ $N d$ Scattering at Low Energies}{abs:schieck}
\stalk{W. Tornow}{Nucleon-Deuteron Scattering at Low Energies}{abs:tornow}
\stalk{S. Kistryn}{ Three-Nucleon System Dynamics Studied in the\newline
$d p$-Breakup
 at 130 MeV}{abs:kistryn}
\stalk{J. Messchendorp}{ Few-Body Studies at KVI}{abs:messchendorp}
\stalk{H. Sakai}{ $d p$ Elastic Scattering Experiments}{abs:sakai}
[1mm]
\multicolumn{3}{c}{\bf Few-Body Dynamics III}\\
\stalk{J.F. Donoghue}{ Nuclear Binding 
Energies and Quark Masses}{abs:donoghue2}
\stalk{F. Rathmann}{Proton-Proton Scattering Experiments}{abs:rathmann}
\stalk{C. Hanhart}{ Systematic Approaches to Meson Production in $N N$ and 
$d d$ Collisions}{abs:hanhart}
\stalk{D. Cabrera}{ Vector Meson Properties in Nuclear Matter}{abs:cabrera}
\stalk{A. Schwenk}{ A Renormalization Group Method for Ground State 
Properties
of Finite Nuclei}{abs:schwenk}
\stalk{L. Platter}{ The Four-Body System in EFT}{abs:platter}
\stalk{H. Krebs}{ Electroproduction of Neutral Pions 
Off the Deuteron Near Threshold} {abs:krebs}
\end{tabular}}
\newpage

\newabstract 
\label{abs:vanKolck}
\begin{center}
{\large\bf Charge Symmetry Breaking in Pion Production}
\footnote{Research supported in part by the US DOE and
the Alfred P. Sloan Foundation.}\\[0.5cm]
{U. van Kolck}$^{1,2}$\\[0.3cm]
$^1$Department of Physics, University of Arizona,\\
Tucson, AZ 85721, USA\\[0.3cm]
$^2$RIKEN BNL Research Center, Brookhaven National Laboratory,\\
Upton, NY 11973, USA\\[0.3cm]
\end{center}

Charge symmetry is a rotation of $\pi$ around the $y$ axis in isospin space.
One of the most important low-energy consequences of charge-symmetry breaking
is the proton-neutron mass splitting $\Delta m_N=m_p -m_n= -1.3$ MeV, 
which receives 
contributions from both the quark-mass difference 
---denoted $\delta m_N$--- and 
electromagnetism ---denoted $\bar\delta m_N$.
At least one other suitable CSB observable
is needed to separate these two low-energy constants.

The approximate chiral symmetry of QCD
implies \cite{vK} that there are two sets of pion-nucleon interactions
with strengths fixed by $\delta m_N$ and $\bar\delta m_N$,
respectively.
Assuming $\bar\delta m_N>0$ as in most models,
it was predicted in Ref. \cite{vKNM}
that the corresponding contributions
to the front-back asymmetry
in the reaction $pn\to d\pi^0$ near threshold
would be 1--2 times
larger, but in the opposite direction, than the sum of other,
more conventional mechanisms.
This prediction has recently
been confirmed at TRIUMF \cite{Oetal}.

In addition, an unambiguous signal 
was seen at IUCF for the CSB reaction $dd\to \alpha \pi^0$ near threshold, 
resulting in a cross-section
of about 10 pb \cite{Setal}.
A group of theorists is now working on 
the evaluation of the effects from these (and other) interactions \cite{Getal}.

The corresponding two-pion exchange two-nucleon potential has
also been calculated \cite{FvKPC}, but
its effects on CSB observables are generically small.

\newabstract 
\label{abs:stephenson}

\begin{center}
{\large\bf Observation of the Charge Symmetry Breaking \\[0.1cm]
{\boldmath ${\rm d}+{\rm d}\to{^4{\rm He}}+\pi^0$} 
Reaction near Threshold}\\[0.5cm]
{\bf E.J. Stephenson}, for the CSB Collaboration\\[0.3cm]
Indiana University Cyclotron Facility,\\
Bloomington, IN 47408 USA.\\[0.3cm]
\end{center}

We report the first observation of the isospin violating and charge
symmetry breaking (CSB) ${\rm d}+{\rm d}\to{^4{\rm He}}+\pi^0$
reaction near
threshold.  The
measurements were made with the Indiana University electron-cooled
storage ring.  A magnetic channel was constructed just downstream of a
$6^\circ$ bending magnet to separate the forward-going $^4$He recoil
nuclei from the circulating  beam.  A
deuterium gas jet target was placed upstream
of the bend and surrounded by two arrays of Pb-glass detectors to signal
the presence of photons from the decay of the $\pi^0$.   The formation of
a CSB $\pi^0$ was separated from the
continuum of isospin-allowed double radiative
capture
$^4{\rm He}+\gamma +\gamma$ events by measuring the $^4$He recoil
momentum and reconstructing the pion missing mass.  The 4-momentum
measurement depended upon the $^4$He
time of flight in the channel
and scattering angle.
A clean sample of candidate events was produced
by requiring the correct energy loss in the channel scintillators and the
presence of two
coincident 70-MeV photons in the Pb-glass arrays.
The CSB $\pi^0$s showed a clear peak above the double radiative capture
continuum.

Measurements were made at two energies to verify the algorithms for pion
reconstruction.  The total CSB cross section was found to be $12.7\pm
2.2$~pb
at 228.5~MeV and $15.1\pm 3.1$~pb at 231.8~MeV \cite{EJS}.
The systematic errors of
6.6\%\ were considerably less than the statistical errors from background
subtraction.  These measurements are consistent with S-wave pion production
and are proportional to $\eta =p_\pi /m_\pi$ with a slope of
$\sigma_{\rm TOT}/\eta =80\pm 11$~pb.  The integral of the double radiative
capture process for the 2~MeV range just below the kinematic limit in
missing mass was $6.9\pm 0.9$~pb at 228.5~MeV and $9.5\pm 1.4$~pb at
231.8~MeV.

Charge symmetry breaking effects arise fundamentally from the differences in
the masses of the down and up quarks and differences in their electromagnetic
interactions.  These two effects give rise to the neutron-proton mass
difference.  A detailed analysis of the ${\rm d}+{\rm d}\to{^4{\rm He}}
+\pi^0$ reaction is underway \cite{AG}.  It will include
entrance channel distortions, modern wavefunctions, and the contributions to
CSB from meson mixing ($\pi^0-\eta$ and $\rho^0-\omega$).

\newabstract 
\label{abs:nogga}

\begin{center}
{\large\bf Probing Chiral Interactions in Light Nuclei}\\[0.5cm]
Andreas Nogga\\[0.3cm]
Institute for Nuclear Theory, University of Washington, 
Box 351550, Seattle, WA 98195-1550, USA\\[0.3cm]
\end{center}

Recently, chiral perturbation theory was used to obtain a consistent 
model of NN and 3N interactions (see \cite{epelbaum}). The model is able to
describe nd scattering up to at least 65~MeV reasonably well. The chiral 3N
force has two {\it a priori} unknown constants, which were determined using
the 3N binding energy and the doublet nd scattering length. The resulting 3N
force could also describe the $^4$He binding energy. 

In Ref.~\cite{pieper} it was found that the binding energy of the $p$-shell
nuclei are sensitive to the 3NF structure. A simple 3NF model, which
is able to describe the 3N and 4N systems, at least at low energies, does not
provide the correct binding and excitation energies for $p$-shell nuclei.
This can be resolved by additional 3NF terms.
The aim of this study is to understand the effects of chiral NN and 3N
interactions on these spectra. The Schr\"odinger 
equation for $p$-shell nuclei is solved using the no-core shell model 
approach \cite{navratil}. Here we concentrate on $^6$Li.

We need to establish the convergence with respect to the chiral 
expansion. The calculation shows that the $p$-shell binding 
energies are affected by NNLO contributions, which is the highest order
in the chiral expansion, we take into account. 
We also find variations of the order of 1.5~MeV varying the
cut-off of our interactions. This indicates that our binding energies can not
be predicted to a higher accuracy at this order. The excitation energy is also
affected by NNLO contributions. The cut-off dependence indicates that 
the error of the NNLO prediction is 200~keV. The experimental value
is within the predicted range. We have 
also established that the 3NF terms can change the excitation energy and 
binding energy of $^6$Li, even if the $^3$H and $^4$He binding energies 
remain constant.  
The error estimates hint to visible N3LO contributions, 
which will be subject of a forthcoming study. 
More details of this work can be found in \cite{nogga}. 

The work was supported
in parts by grants from the NSF and DOE. The numerical calculations have been
made at NERSC.

\newabstract 
\label{abs:myhrer}

\begin{center}
{\large\bf Solar Neutrino Reactions and Effective Field Theory 
}\\[0.5cm]
Fred Myhrer \\[0.3cm]
Univ. South Carolina,
Columbia, SC 29208, USA .\\[0.3cm]
\end{center}

Many astrophysical phenomena
are governed by low-energy nuclear weak-interaction processes,
and effective field theory 
(EFT) is believed to be a useful framework 
for describing them.
I present here a comparison of EFT with 
the standard nuclear physics approach (SNPA),
focusing on the $\nu d$ reactions
relevant for 
the SNO experiments.
The low-energy $\nu d$ reactions
are dominated by the spatial component
of the axial current, 
$\vec{A} =\vec{A}_{one} + \vec{A}_{two}$, 
a sum of 
1-body and 2-body terms. 
$\vec{A}_{one}$ being well known,
one's task is to pin down 
$\vec{A}_{two}$.
In SNPA, $\vec{A}_{two}$ is derived from pion, 
rho and $A_1$ meson exchange diagrams \cite{Nak02}, 
devised to fulfill the low-energy theorems, 
current algebra, etc. SNPA 
contains some unsatisfactory aspects:
(1) There is no obvious link 
between SNPA and QCD;
(2) It does not offer a systematic way 
to estimate theoretical uncertainties; 
(3) Low-energy physics 
($E \!\ll \!m_\pi$) and high-energy physics 
($E\!\ge \!m_\pi$) are intermingled.
EFT can resolve these 
problems. 

The terms in the EFT lagrangian, ${\cal L}_\chi$,
form a perturbation series in powers of
$(Q/\Lambda_\chi)$, where $Q$ is 
a typical energy of the process in question.
In \cite{Nak02} we used EFT$^*$ 
wherein the operators 
are derived from ${\cal L}_\chi$
while the nuclear wave functions are 
generated from a realistic $NN$ potential.
The $\vec{A}_{two}$ in \cite{Nak02}
involves a pion-exchange and a four-nucleon contact term
with one LEC.
Since this LEC is determined from the 
triton $\beta$-decay rate,
EFT$^*$ allows a model-independent 
calculation of the $\nu d$ cross sections\cite{Nak02}.
The results agree with those 
obtained in SNPA to 1 \% .
Butler et al.'s EFT calculation\cite{butl01}
has one unknown LEC, $L_{1A}$, 
which was adjusted to reproduce the 
$\nu$-$d$ cross sections calculated in \cite{Nak02}.  
The attempts to constrain
$L_{1A}$ from experimental values  
of observables 
are not yet very stringent.

\newabstract 
\label{abs:ando}

\begin{center}
{\large\bf 
Solar-Neutrino Reactions on Deuteron in EFT*
and\\[0.1cm] Radiative Corrections of Neutron Beta Decay
}\\[0.5cm]
Shung-ichi Ando
\\[0.3cm]
TRIUMF, 4004 Wesbrook Mall, Vancouver, B.C. V6T 2A3, Canada 
\end{center}

The cross sections 
for low-energy neutrino-deuteron reactions 
are calculated 
within heavy-baryon chiral perturbation theory 
employing a cut-off regularization 
scheme\cite{Aetal}.
The transition operators 
are derived 
up to 
next-to-next-to-next-to-leading order 
in accord with the Weinberg counting rule, 
while the nuclear matrix elements 
are evaluated using wave functions 
generated by a high-quality 
phenomenological $NN$ potential.
%
Only one unknown low-energy constant 
appears in our calculation,
the axial-current-four-nucleon coupling
constant,
which is fixed by using data 
from tritium beta decay\cite{TSPetal}. 
Our results exhibit 
a high degree of stability 
against different choices of the cutoff parameter, 
a feature which indicates that, 
apart from radiative corrections,
the uncertainties in the calculated cross sections 
are less than 1\%.
We also discuss 
the feasibility of fixing
the low energy constant 
through the accurate measurement of 
the ordinary muon capture rate on 
the deuteron being planned at PSI,
which would avoid the complication of 
the three-body current in tritium beta decay\cite{dOMC}.

The lifetime and angular correlation 
coefficients of neutron beta decay 
are also evaluated up to next-to-leading order
in effective field theory\cite{Aetal2},
where pions are integrated out
because of the small typical energy.
Up to this order 
the nucleon recoil corrections, 
including weak-magnetism,
and the radiative corrections are calculated.
Our results agree with those of the
long-range and model-independent part 
of previous calculations.
Moreover, in the effective theory 
the model dependent ``inner'' radiative corrections 
are replaced by a well-defined low energy constant.
Based on the counting rule of the effective 
theory, we estimate the accuracy of our
results to be of the order of 10$^{-3}$.

\newabstract 
\label{abs:birse}

\begin{center}
{\large\bf EFT's for the Inverse-Square Potential\\[0.1cm] 
and Three-Body Problem}\\[0.5cm]
{\bf Michael C. Birse} and Thomas Barford\\[0.3cm]
Department of Physics and Astronomy, University of Manchester,\\
Manchester, M13 9PL, UK\\[0.3cm]
\end{center}

Compared with EFT's for the corresponding two-body systems, it has proved  
much more complicated to determine the power counting for three-body 
systems with short-range forces and large two-body scattering lengths.
This is particularly true of attractive systems such as three bosons or 
the triton. 

As shown by Efimov \cite{ef}, the wave functions at short distances satisfy a
2D Schr\"odinger equation with an attractive $1/r^2$ potential. These wave
functions are ill-defined without a boundary condition to ensure that no flux
is lost at the centre and to fix the phase of the wave functions for small $r$
(a self-adjoint extension). The resulting wave functions have the form
$\sin(s_0\ln r/R_0)$ as $r\rightarrow 0$.

Using the distorted-wave RG \cite{BB1}, we apply a cut-off at
$E=\pm\Lambda^2/M$ on the distorted waves and bound states of this potential,
and follow the RG flow of the three-body forces as $\Lambda\rightarrow 0$. The
oscillatory behaviour of the wave functions means that the potential tends to
a limit cycle.  Since the amplitude of these oscillations is constant as
$r\rightarrow 0$, the power counting for perturbations is the same as that for
a 2D system with zero angular momentum. This implies that the leading,
energy-independent interaction is marginal \cite{BB2}.

We find the following power counting for these systems. At LO ($\Lambda^0$)
there is the marginal three-body force. This corresponds to the RG limit cycle
and it determines the phase of the wave functions as $r\rightarrow 0$.  At NLO
($\Lambda^1$) the two-body effective range appears as a perturbation. Then at
NNLO ($\Lambda^2$) we have an energy-dependent three-body force, along with
the two-body effective range at second order.  This agrees with the counting
obtained by Bedaque et al.\ \cite{BGHR} using the Skorniakov--Ter-Martirosian
equation with a momentum cut-off.

\newabstract 
\label{abs:cargnelli}

\begin{center}
{\large\bf DEAR - Kaonic Hydrogen: First Results}\\[0.5cm]

{\bf M.~Cargnelli$^d$,} G.~Beer$^i$, A.M.~Bragadireanu$^{a,e}$,
C.~Curceanu (Petrascu)$^{a,e}$, J.-P.~Egger$^{b,c}$,
H.~Fuhrmann$^d$, C.~Guaraldo$^a$, M.~Iliescu$^a$,
T.~Ishiwatari$^d$, K.~Itahashi$^g$, M.~Iwasaki$^f$, P.~Kienle$^d$,
B.~Lauss$^h$, V.~Lucherini$^a$, L.~Ludhova$^b$, J.~Marton$^d$,
F.~Mulhauser$^b$, T.~Ponta$^{a,e}$, L.A.~Schaller$^b$,
R.~Seki$^{j,k}$, D.~Sirghi$^a$, F.~Sirghi$^a$,
P.~Strasser$^f$ , J.~Zmeskal$^d$ \\[0.3cm]

$^a$INFN - Laboratori Nazionali di Frascati; $^b$Universite de
Fribourg; $^c$Universite de Neuch$\hat{a}$tel; $^d$Institute for
Medium Energy Physics, Vienna; $^e$Institute of Physics and
Nuclear Engineering, Bucharest; $^f$RIKEN, Saitama; $^g$Tokyo
Institute of Technology; $^h$University of California and
Berkeley; $^i$ University of Victoria; $^j$California Institute of
Technology;
$^k$California State University \\[0.3cm]
\end{center}

The DEAR\footnote{DA$\Phi$NE Exotic Atom Research, conducted at
the Frascati electron positron collider} experiment \cite{rf:1}
measures the energy of X-rays emitted in the transitions to the
ground states of kaonic hydrogen. The shift $\epsilon$ and the
width $\Gamma$ of the 1s state are related to the real and
imaginary parts of the complex S-wave scattering length by the
Deser Trueman formula. \

\begin{figure} [h]
 \centering
 \begin{minipage}[c]{.45\textwidth}
   \centering
   \caption{Background subtracted energy spectrum of kaonic hydrogen.
   For the first time K$_\beta$, K$_\gamma$, K$_{high}$ are clearly resolved.}
   \label{fig:1}
 \end{minipage}
 \hfill
 \begin{minipage}[c]{.45\textwidth}
   \centering
   \includegraphics [scale=.30]{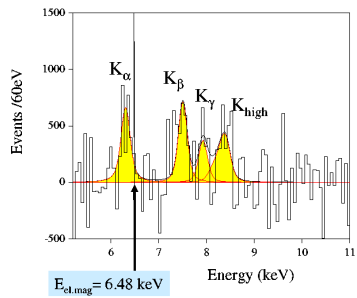}
 \end{minipage}
\end{figure}

The preliminary results are: $\epsilon$= - 183 $\pm$ 62 eV  and
$\Gamma$= 213 $\pm$ 138 eV. Both values are smaller then those
from the previous experiment \cite{rf:2} and consistent with
recent theoretical studies \cite{rf:3}. \\

Part of the work was supported by "Transnational access to
Research Infrastructure" (TARI) Contract No. HPRI-CT-1999-00088.

\newabstract 
\label{abs:gasser}

\begin{center}
{\large\bf Comments on {\boldmath $\bar K N$} Scattering}\\[0.5cm]
J\"urg Gasser\\[0.3cm]
Institute for  Theoretical Physics, 
University of Bern\\
Sidlerstrasse 5, CH-3012 Bern, Switzerland\\[.3cm]
\end{center}
The DEAR collaboration has investigated the 
ground state of $\bar Kp$ atoms, in particular, 
the energy level shift and width $\epsilon_{1s}$ 
due to strong interactions \cite{dear}. 
The quantity $\epsilon_{1s}$ 
is related to the elastic threshold amplitude 
$T_{\bar K p}^{\rm th}$ \cite{deser},
\begin{eqnarray}
\epsilon_{1s}=-2\alpha^3\mu_r^2 T_{\bar K p}^{\rm th}(1+X)\, .
\end{eqnarray}
Here, $T_{\bar K p}^{\rm th}$ denotes the amplitude  in pure
QCD  in the isospin symmetry limit $m_u=m_d$, 
and the quantity $X$
 stands for isospin symmetry breaking corrections. These are 
presently evaluated
  \cite{deserbonn} in the framework of 
effective non relativistic quantum field theories \cite{eff}.
To work out the implication of a precise measurement 
of $\epsilon_{1s}$ on the kaonic 
sigma terms \cite{bianco} requires a  complex theoretical 
analysis. 
 I have suggested in my talk a more modest aim 
as an intermediate step: Calculate the
amplitude $T_{\bar K p}^{\rm th}$ as precisely as possible, 
based on a framework that relies on QCD 
(incorporating chiral symmetry at least), and compare with (1).
 Available 
predictions \citetwo{kaiser}{oller} differ considerably - 
it would be instructive to investigate how 
these calculations  can be made more precise.

\newabstract 
\label{abs:ivanov}

\begin{center}
{\large\bf On Pionic and Kaonic Hydrogen}\\[0.5cm] {\bf A. N. Ivanov},
M. Cargnelli, M. Faber, A. Hirtl, J. Marton, N. I. Troitskaya, and
J. Zmeskal\\[0.3cm] Atominstitute, Vienna University of Technology and
Institute of Medium Energy Physics, Austrian Academy of Sciences,,
Vienna, Austria \\[0.3cm]
\end{center}

In \cite{IV1} the energy level displacement of the ground state of
hadronic hydrogen $A_{hp}$, where $h = \pi$ or $K$, has been expressed
in terms of the amplitude $M(h^-p \to h^-p)$ of $h^-p$ scattering,
weighted with the wave function $\Phi_{1s}$ of hadronic hydrogen in
the ground state
\begin{eqnarray}\label{label1}
- \epsilon_{1s} + i\,\frac{\Gamma_{1s}}{2} &=&
  \frac{1}{4m_{h^-}m_p}\int\frac{d^3k}{(2\pi)^3}
  \int\frac{d^3q}{(2\pi)^3}\,
  \sqrt{\frac{m_{h^-}m_p}{E_{h^-}(\vec{k}\,)E_p(\vec{k}\,)}}\,
  \sqrt{\frac{m_{h^-}m_p}{E_{h^-}(\vec{q}\,)
  E_p(\vec{q}\,)}}\nonumber\\ &\times&\Phi^{\dagger}_{1s}(\vec{k}\,)\,
  M(h^-(\vec{k}\,)p(-\vec{k},\sigma_p) \to
  h^-(\vec{q}\,)p(-\vec{q},\sigma_p))\, \Phi_{1s}(\vec{q}\,).
\end{eqnarray}
The knowledge of the amplitude of $h^-p$ scattering for arbitrary
relative momenta should give the possibility to calculate explicitly
the energy level displacement. The low--energy limit $k, q \to 0$,
justified by the wave functions $\Phi^{\dagger}_{1s}(\vec{k}\,)$ and
$\Phi_{1s}(\vec{q}\,)$, reduces (\ref{label1}) to the DGBT formula
\cite{SD54}. In \cite{IV1} (nucl--th/0310027) formula (\ref{label1})
has been generalized to arbitrary excited $n\ell$ states of the
hadronic atom $A_{hp}$. In \cite{IV1} (nucl--th/0310081) we have
suggested a model of low--energy $K^-p$ scattering near threshold of
the $K^-p$ pair. We have assumed that the amplitude of $K^-p$
scattering near threshold is fully defined by the contribution of
three resonances $\Lambda(1405)$, $\Lambda(1800)$ and $\Sigma(1750)$
and a smooth elastic background. We predict the energy level
displacement of the ground state of kaonic hydrogen \cite{IV1} $-
\epsilon_{1s} + i\Gamma_{1s}/2 = (-203 \pm 15) + i\,(113 \pm 8)\,{\rm
eV}$ fitting well preliminary experimental data by the DEAR
Collaboration $- \epsilon^{\exp}_{1s} + i\Gamma^{\exp}_{1s}/2 = (-183
\pm 62) + i\,(106 \pm 69)\,{\rm eV}$ \cite{DEAR4}.

\newabstract 
\label{abs:zemp}

\begin{center}
  {\large\bf Deser-type Formula for Pionic Hydrogen }\\[0.5cm]
  {Peter Zemp}\\[0.3cm]
  {
Institute for theoretical Physics, University of Bern, \\
    Sidlerstrasse 5, 3012 Bern, Switzerland}\\[0.3cm]
\end{center}
The decay width of the ground state of pionic hydrogen is dominated by
the two decay channels $(\pi^-p)_{1s}\to \pi^0n$ and $(\pi^-p)_{1s}\to
n\gamma$.  Using $\delta$ as a common counting for the light quark mass
difference $m_u-m_d$ and for the fine-structure constant $\alpha$, the leading
term of the former (latter) decay channel is of order $\delta^{7/2}$
$(\delta^4)$. The decay width through these two channels can be expressed by a
Deser-type formula
\begin{equation}
  \label{eq:deser-type}
  \Gamma_{1s}= 4\, \alpha^3 M_r^2  q_{\scriptscriptstyle 0} \left ( 1 +
    \frac{1}{P} \right ) [a_{\pi^-p \to \pi^0
    n}(1+\delta_{\scriptscriptstyle\Gamma})]^2 \;.
\end{equation}
Since the Panofsky ratio $P = \sigma(\pi^-p \to \pi^0 n) / \sigma(\pi^- p \to
n \gamma)$ is about $1.5$, the $n \gamma$ channel amounts to a 60\%
correction. To the best of my knowledge, this formula was first established in
the framework of potential models \cite{PotentialModels}. I want to verify its
validity in QCD~+~QED, using the technique of non-relativistic effective field
theories \cite{NREFT}.  Constructing the effective Lagrangian and solving the
master equation, one encounters the obstacle that the photon in the
intermediate state carries a hard momentum of order $M_{\pi}$.  
To circumvent this
problem, I do not incorporate $n\gamma$ intermediate states with effective
fields. Instead I introduce a non-hermitian contact term \cite{NREFT}
(cf.~also~\cite{NonhermitianCouplings}) in the Lagrangian with the coupling
$d_1^R + i d_1^I$,
\begin{equation}
  \label{eq:intLagrangian}
  \mathcal{L}^{n \gamma}_{\rm I} = (d_1^R + i d_1^I) \; \psi^\dagger(x)
  \psi(x) \; \pi^\dagger_-(x) \pi_-(x)\;.
\end{equation}
The imaginary part $d_1^I$ replaces the imaginary part generated by the
intermediate $n\gamma$ state.  With this contact term, I can prove that
formula (\ref{eq:deser-type}) is valid at order $\delta^4$. It remains to
evaluate $\Gamma_{1s}$ at order $\delta^{9/2}$ and to determine
$\delta_{\scriptscriptstyle \Gamma}$ in equation 
(\ref{eq:deser-type})\,\cite{WorkInProgress}.

\newabstract 
\label{abs:nemenov}

\begin{center}
{\large\bf Atoms Consisting of {\boldmath $\pi^+\pi^-$} 
and {\boldmath $\pi K$} Mesons}\\[0.5cm]
L.Nemenov$^1$\\[0.3cm]
$^1$JINR, Dubna\\[0.3cm]
\end{center}

Using experience obtained in the DIRAC experiment at CERN \cite{[1]}  
new experiments are proposed for CERN PS and J-PARC in Japan  \cite{[2]} 
to check precisely predictions of low energy QCD.

These experiments aim to measure the $\pi^+\pi^-$ atom 
($A_{2\pi}$) lifetime with precision better than 6\% and to determine the 
S-wave $\pi \pi$ scattering length difference $|a_0-a_2|$ at 3\%. 
Simultaneously with the $A_{2\pi}$ investigation 
DIRAC plans to observe $\pi K$ atoms ($A_{\pi K}$) and to measure their
lifetime at the 20\% level in order to evaluate the S-wave $\pi K$ 
scattering length difference $|a_{1/2}-a_{3/2}|$ at 10\%. 

Furthermore the observation of long-lived (metastable) $A_{2\pi}$ states 
is also envisaged to be done in the same setup. This allows 
to measure the energy difference between nS and nP states and 
to determine in a model-independent way $2a_0 +a_2$.

Low energy QCD \cite{[3]} predicts $\pi \pi$ scattering lengths at 
2\%~\citetwo{[4]}{[5]} and $\pi K$~scattering lengths at 
10\%~\citetwo{[6]}{[7]}. These scattering lengths have 
never been verified experimentally with the same accuracy as 
theoretically predicted.

The theoretical results have been obtained assuming strong
condensation of quark-antiquark pairs in the vacuum. Therefore 
the proposed experiments will test crucially low energy QCD
predictions and accordingly our understanding of the nature of the 
QCD vacuum \cite{[8]}.

\newabstract 
\label{abs:girlanda}

\begin{center}
{\large\bf Deeply Bound Pionic Atoms: Optical Potential 
at {\boldmath $O(p^5)$}}\\[0.1cm]{\large\bf  in ChPT}\\[0.5cm]
{\bf Luca Girlanda}$^1$, Akaki Rusetsky$^{2,3}$ 
and Wolfram Weise$^{1,4}$\\[0.3cm]
$^1$ECT*, Strada delle Tabarelle 286, I-38050 
Villazzano (Trento), Italy\\
$^2$HISKP (Theorie), University of Bonn, 
Nu\ss{}allee 14-16, 53115 Bonn, Germany\\
$^3$HEPI, Tbilisi State University, 
University St.~9, 380086 Tbilisi, Georgia\\
$^4$Physik-Department, TU M\"{u}nchen,
D-85747 Garching, Germany
\end{center}

Deeply bound pionic atoms can be considered as a
laboratory to test the effects of baryon density on chiral symmetry
breakdown~\cite{Weise:sg}.
The present effective field theory approach to this problem is
provided by the in-medium ChPT~\cite{inmedium}, which
treats the 
nucleus as an infinite and uniform Fermi-sea of protons and
neutrons. 
The results are then expressed through a local Fermi momentum
$\vec{k}(\vec{r})$~\cite{kolo} with some (arbitrary) prescription. Such
arbitrariness is worrisome  especially for the case at hand, in 
which the bound state wave function is picked at the nuclear surface,
due to a balance between  Coulomb attraction and  strong
repulsion. 

In Ref.~\cite{GRWpreparation} we give a consistent formulation of ChPT in a
non-uniform fermionic background, corresponding to the (finite) nucleus.  The
nuclear structure information is encoded in a set of nuclear matrix elements
of free-nucleon field operators. Chiral counting applied to the above matrix
elements allows to reduce considerably the nuclear input needed.  As an
application we calculate the charged pion self-energy in the background of a
heavy nucleus at $O(p^5)$ of the chiral expansion, including consistently all
isospin-breaking effects arising at this order.  Finite nuclear size effects
in the pion-nucleus optical potential are identified unambiguously, without
any ad-hoc prescription such as local density approximation.

However, in order to be relevant phenomenologically, this analysis has
to be extended to $O(p^6)$, beyond the linear density
approximation for the optical potential. We plan to address this issue
in future publications.

\newabstract 
\label{abs:griesshammer}

\begin{center}
  {\large\textbf{All-Order Low-Energy Expansion in the 3-Body System}}\\[0.5cm]
  Harald W.~Grie{\ss}hammer$^1$\\[0.3cm]
  $^1$ Institut f\"ur Theoretische Physik (T39), TU M\"unchen, \textit{and}
  ECT*, Trento
  \\[0.3cm]
\end{center}

In ``pion-less'' Effective Field Theory, the power-counting in the 3-body
system with a shallow two-body bound state was extended and systematised to
all orders in the low-energy expansion in Ref.~\cite{paper}. The analytical
considerations rest on the tenet that a 3-body force (3BF) is included if and
only if necessary to cancel cut-off dependences in observables at a given
order.  The one momentum-independent 3BF needed for cut-off independence up to
NLO in the ${}^2\mathrm{S}_\frac{1}{2}$ (triton) channel is determined by the
three-body scattering length.  One and only one new parameter enters at NNLO,
namely the Wigner-$SU(4)$-symmetric 3BF with two derivatives, fixed to
reproduce the triton binding energy.  Only the $SU(4)$-symmetric 3-body forces
are systematically enhanced over their contributions found from a na\"{\i}ve
dimensional estimate, with a $2n$-derivative 3BF entering at the $2n$th order.
A new, computationally convenient and simple scheme to perform higher-order
calculations in the three-body system iterates a kernel to all orders, which
has been expanded to the desired order of accuracy.  The $nd$ scattering phase
shifts agree well with phase shift analysis and modern potential model
calculations.

In contradistinction, Wigner-symmetric 3BFs are suppressed compared to
na\"{\i}ve dimensional analysis in all other partial waves~\cite{hgrie}: In
the $l$th partial wave of the spin doublet ($\lambda=1$) and quartet
($\lambda=-\frac{1}{2}$) channels, the amplitude converges for large off-shell
momenta $p$ as $p^{-s_0-1}$, with $s_0$ the solution to
\[
1\;=\;(-)^{{l}}\;\frac{2^{1-{l}}\;{\lambda}}{\sqrt{3\pi}} \;\frac{
  \Gamma[\frac{{l}+1+s_0}{2}]
  \;\Gamma[\frac{{l}+1-s_0}{2}]}{\Gamma[{l}+\frac{3}{2}]}
\;\;{}_2F_1\left[\frac{{l}+1+s_0}{2},\frac{{l}+1-s_0}{2};
  {l}+\frac{3}{2};\frac{1}{4}\right]\;\;.
\]
Except in the ${}^2\mathrm{S}_\frac{1}{2}$ channel, the asymptotics is always
weaker than in the na\"{\i}ve analysis ($s_0=1$). Thus, effective range
corrections in the two-body system containing $n$ external momenta need 3BFs
with $m$ derivatives to absorb UV divergences only when $n\ge
\textrm{Re}[m+2s_0]$. For example, the first Wigner-symmetric 3BF in the
${}^4\mathrm{S}_\frac{3}{2}$ channel is demoted from N$^4$LO to N$^7$LO,
suggesting that two-body physics rules this channel to very high accuracy.

Sponsored in part by DFG Sachbeihilfe GR 1887/2-1 and BMBF.

\newabstract 
\label{abs:black}

\begin{center}
{\large\bf New Results in the Three and Four Nucleon Systems}\\[0.5cm]
Timothy Black\\[0.3cm]
Department of Physics, University of North Carolina at Wilmington\\
Wilmington, NC, USA\\[0.3cm]
\end{center}

Recent measurements of the n-d and n-$^3$He bound coherent scattering lengths,
conducted at the Neutron Interferometer and Optics Facility at NIST, are in
significant disagreement with exact calculations of these observables using
modern NN+3NF potentials\citetwo{Wit03}{Hof03}.  The bound coherent n-d
scattering length was measured to be $b_{nd} = (6.665 \pm 0.004)$~fm, which
yields a new world average for this parameter of $b_{nd} = (6.669 \pm
0.003)$~fm\citetwo{Bla03}{Sch03}.  The bound coherent scattering length is
related to the free doublet and quartet scattering lengths by $b_{nd} =
\frac{m_D + m_n}{m_D}\left(\frac{1}{3} a_2 + \frac{2}{3} a_4\right)$.
Combining the new world average for $b_{nd}$ with calculated values of
$^4a_{nd}$ yields a value for the free doublet scattering length of $^2a_{nd}
= (0.645 \pm 0.003 [{\rm expt}] \pm 0.007 [{\rm thry}])$~fm.

In the n-$^3$He system, the bound coherent scattering length was measured as
$b_{n^3{\rm He}} = (5.857 \pm 0.007)$~fm, which yields a new world average for
this parameter of $b_{n^3{\rm He}} = (5.854 \pm 0.007)$~fm.  Combining these
measurements with Zimmer's result for the bound incoherent scattering length
of $b_{i} = (-2.365 \pm 0.020)$~fm\cite{Zim02}, one obtains the free singlet
and triplet scattering lengths for the n-$^3$He system; $a_0 = (7.456 \pm
0.026)$~fm, and $a_1 = (3.364 \pm 0.010)$~fm, respectively.  The extreme
precision of these new measurements in the n-d and n-$^3$He systems renders
them suitable for use in high-order, contemporary EFT calculations.

\newabstract 
\label{abs:schieck}

\begin{center}
{\large\bf Nd Scattering at Low Energies}\\[0.5cm]
H. Paetz gen. Schieck\\[0.3cm]
Institut f\"ur Kernphysik, Universit\"at zu K\"oln,\\
Z\"ulpicher Stra{\ss}e 77, D-50937 K\"oln, Germany\\[0.3cm]
\end{center}
Despite the tremendous progress made by few-body
theory --- by numerically exact calculations using precision meson-exchange
nucleon-nucleon potentials including different three-body forces and,
recently, in applying realistic effective-field theory --- a number of
long-standing low-energy puzzles are still unsolved. These are: the d+N
elastic cross section {\bf (Sagara) anomaly}, the Nd elastic {\bf A$_y$ (and
  iT$_{11}$) puzzle}, the Nd breakup anomaly in the {\bf space-star (SST)}
configuration, and the new Nd breakup anomaly in the {\bf quasi-free
  scattering (QFS)} configuration.

For the pd system the data base is much larger, with mostly better-quality
data, than for the nd system.  For a general survey see \cite{glo}.  In order
to compare low-energy pd data to realistic calculations inclusion of the
Coulomb force is mandatory. This has been achieved for Faddeev calculations of
elastic scattering observables (see e.g. Fig. 1) whereas for the pd breakup
this is still eagerly awaited.
Some breakup data (e.g. in the SST and QFS, but not in the final-state\\[0.9mm]
\noindent \begin{minipage}{40mm} interaction 
configurations) exhibit large Coulomb effects \cite{sch}. Fig. 2 shows the
low-energy systematics of nd vs. pd SST cross sections. The unsolved puzzles
call for new theoretical efforts with different three-body forces or
modifications of the two-nucleon input. Also measurements of new observables
and in larger phase-space areas may be useful.
\end{minipage}
\begin{figure}[htb] \hfill \begin{minipage}{95mm} 

\vspace{-95mm}

\epsfig{file=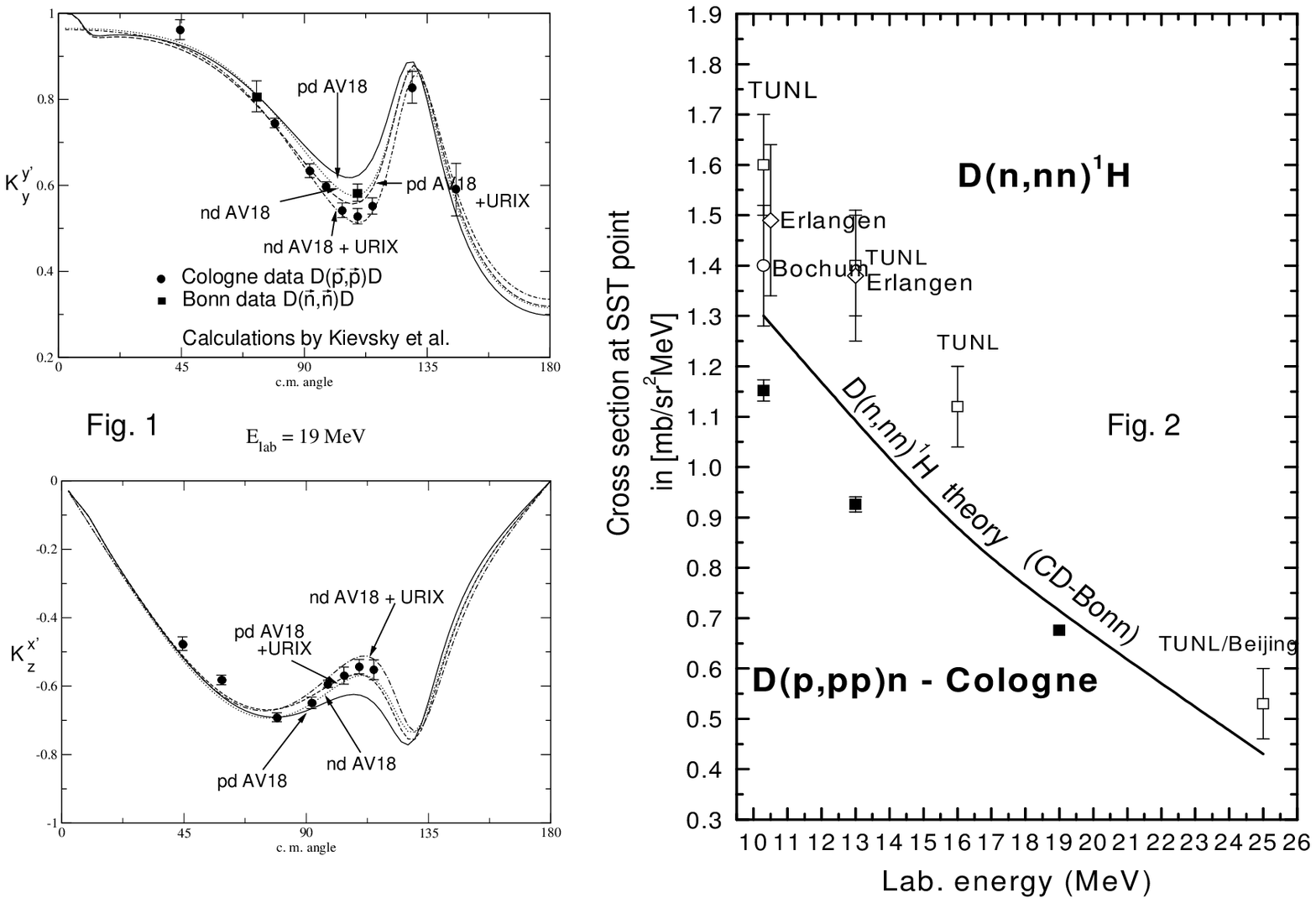,width=90mm}

\vspace{-8mm}

{\small Fig. 1: Polarization transfer coefficients compared to realistic
  calculations, also including the UR IX three-body and the Coulomb force
  \cite{kie}. Fig. 2: Comparison of nd vs. pd breakup cross sections in the
  symmetric space-star configuration at various energies.}
\end{minipage}
\end{figure}

\newabstract 
\label{abs:tornow}

\begin{center}
{\large\bf Nucleon-Deuteron Scattering at Low Energies}\\[0.5cm]
W. Tornow\\[0.3cm]
Department of Physics, Duke University and \\
Triangle Universities Nuclear Laboratory, Durham, NC 27708, USA
\end{center}

We discuss a new twist to the long-standing nucleon-deuteron analyzing 
power puzzle.  The new low-energy neutron-deuteron $A_y(\theta)$ data
at $E_n$ = 1.2 and 1.9~MeV \cite{Nei03} revealed a sizeable difference when
compared to proton-deuteron data.  This difference cannot be accounted for by
the Coulomb interaction (in the case of $p$-$d$ scattering) and it is 
orders of magnitude larger than expected from charge-symmetry breaking effects
in the $^3$P$_j$ NN interactions.  We speculate that the experimentally 
observed effect is due to the different magnetic moment interactions between
neutrons and protons and the deuteron target nuclei.  Recent calculations
support this conjecture \citetwo{Wit03}{Kie}.  The proton-deuteron angular 
distribution for $A_y(\theta)$ is considerably affected by the magnetic
moment interaction, while for neutron-deuteron scattering the effect is 
restricted to very forward angles.

We present proton-deuteron phase-shift analysis results \cite{Che} that clearly
show that the discrepancy between the experimental $^4$P$_J$ proton-deuteron
3N phase shifts and the theoretically predicted $^4P_J$ phase shifts 
(i.e., the $A_y$ puzzle) increases with increasing energy.  This is in contrast
to the observation that the discrepancy between $A_y(\theta)$ data and 
calculations decreases with increasing energy.  This finding suggests that
$A_y(\theta)$ is not dominated anymore at higher energies by the $^4$P$_J$
3N phase shifts.  Sensitivity calculations for $A_y(\theta)$ with varying
$^4$P$_J$ phase shifts support this conjecture.

We argue that the 3N $A_y(\theta)$ puzzle cannot be solved by concentrating
on the 3N system alone.  In the 4N system the $p$-$^3$He elastic scattering
$A_y(\theta)$ shows an even larger discrepancy (40\%) between data and 
calculations than observed in the 3N system (25\%).  $A_y(\theta)$ data
for $n$-$^3$He elastic scattering are needed to make further progress.

\newabstract 
\label{abs:kistryn}

\begin{center}
{\large\bf Three-Nucleon System Dynamics\\[0.1cm]
           Studied in the dp-Breakup at 130 MeV}\\[0.5cm]

R.~Bieber$^3$, 
A.~Biegun$^2$,
K.~Bodek$^1$, 
K.~Ermisch$^3$, 
W.~Gl\"ockle$^4$, 
J.~Golak$^1$, 
N.~Kalantar-Nayestanaki$^3$, 
{\bf St.~Kistryn}$^1$,
M.~Ki\v{s}$^3$,
J.~Kuro\'s-\.Zo{\l}nierczuk$^1$, 
A.~Micherdzi\'nska$^2$,
M.~Mahjour-Shafiei$^3$,
R.~Skibi\'nski$^1$, 
E.~Stephan$^2$, 
H.~Wita{\l}a$^1$, 
J.~Zejma$^1$ and 
W.~Zipper$^2$\\[0.3cm]
$^1$Institute of Physics, Jagellonian University,
    30059 Krak\'ow, Poland\\[0.3cm]
$^2$Institute of Physics, University of Silesia,
    40007 Katowice, Poland\\[0.3cm]
$^3$Kernfysisch Versneller Instituut,
    9747 AA Groningen, The Netherlands\\[0.3cm]
$^4$Institut f\"ur Theoretische Physik II,
    Ruhr Universit\"at Bochum,\\
    44780 Bochum, Germany\\[0.5cm]
\end{center}

The measurement of exclusive deuteron-proton breakup, employing a detection  
system covering large fraction of the available phase space and using
130~MeV polarized deuteron beam has been performed at KVI Groningen. 
Theoretical calculations~\cite{Kur02} predict large but generally model 
dependent three-nucleon force (3NF) effects in certain regions of the 
breakup phase space for different experimentally attainable observables.

In the first step of data analysis~\cite{Mic03} high precision fivefold 
cross-section values in 38 kinematical configurations have been extracted, 
using for normalization the simultaneously measured elastic dp-scattering 
data.  
The results have been compared to predictions of modern nuclear forces.  
To this aim the three-nucleon (3N) Faddeev equations have been solved 
rigorously using realistic NN potentials AV18, CD~Bonn, Nijm I and Nijm II 
alone, and combining them with the $2\pi$-exchange Tucson-Melbourne (TM) 
3NF and with its modified version TM99, more consistent with 
chiral symmetry. 
Global comparison of the measured cross sections to pairwise NN force 
predictions only and with 3NF's included clearly reveals the presence 
of 3NF effects.  

All details of our study, showing the usefulness of the kinematically 
complete breakup reaction measured in the full phase space to search 
for 3NF effects, are presented in Ref.~\cite{Kis03}.  Further insight
into details of the 3N interaction dynamics is expected from the 
undergoing analysis of the vector and tensor analyzing powers.

\newabstract 
\label{abs:messchendorp}

\begin{center}
{\large\bf Few-Body Studies at KVI}\\[0.5cm]
J.G. Messchendorp$^1$\\[0.3cm]
$^1$Kernfysisch Versneller Instituut, Zernikelaan 25,\\
9747 AA Groningen, The Netherlands\\[0.3cm]
\end{center}

The KVI facility allows a detailed study of few-nucleon interactions below the
pion-production threshold exploiting polarized proton and deuteron beams up to
energies of 190~MeV.  The experimental and theoretical studies of
bremsstrahlung in the two-nucleon scattering processes have a long-standing
tradition at KVI. The first experimental efforts in this field date from 1995
where the $pp\rightarrow pp+\gamma^{(*)}$ channel was measured with an
incident-proton beam of 190~MeV using a coincidence setup consisting of the
hadron detector SALAD~\cite{kal00} and the photon spectrometer
TAPS~\cite{gab94}. The setup allowed to determine the real
($\gamma$)~\cite{hui99} and virtual ($\gamma^*\rightarrow
e^+e^-$)~\cite{mes99} bremsstrahlung yields. Modern state-of-the-art
microscopic calculations based on a fully-relativistic framework are not able
to describe the measured bremsstrahlung data in the two-nucleon system.

The focus of the few-body program at KVI has recently shifted towards the
understanding of the three-nucleon system by exploiting $p+d$ and $d+p$
reactions with polarized proton and deuteron beams.  Different final states
have been observed which includes the elastic, break-up, and radiative capture
reactions.  A systematic study of the elastic $p+d$ scattering reaction has
been carried out in which high-precision cross sections and analyzing powers
have been determined for several bombarding energies up to 190~MeV using the
Big-Bite Spectrometer (BBS)~\cite{ber95}. The data show that effective models
based on a well-understood two-nucleon interaction are not sufficient to
describe the three-nucleon $pd$ system~\cite{erm01}.  Including the most
modern type of three-nucleon potentials in the calculations resolves only a
part of the discrepancies.

\newabstract 
\label{abs:sakai}

\begin{center}
{\large\bf \boldmath{$dp$} Elastic Scattering Experiments}\\[0.5cm]
Hide Sakai \\[0.3cm]
Dep. Phys./CNS, University of Tokyo,\\
Hongo 7-3-1, Bunkyo, Tokyo, Japan \\[0.3cm]
\end{center}

Recent experimental results to study the three-nucleon-force (3{\it NF})
effects in the $dp$ elastic scattering at intermediate energies are reported.

One of the new measurement at RIKEN is the deuteron-to-proton polarization
transfer (PT) coefficients for the $dp$ elastic scattering at 135 MeV/u by
Sekiguchi {\it et al.} \cite{kimiko_PT}. Very precise data for $K_y^{y'}$,
$K_{xx}^{y'}$, $K_{yy}^{y'}$ and $K_{xz}^{y'}$ are obtained for
$\theta_{cm}=90^{\circ}-180^{\circ}$.

Including those new data, comparisons between the existing data
\citetwo{sakai}{kimiko} at 135 MeV/u and the Faddeev calculations are made.
\begin{itemize}
\item $d\sigma/d\Omega$ and $iT_{11}$ 
are well reproduced by including 3{\it NF}, 
which indicates the clear signature of 3{\it NF}s.
\item Spin observables ($T_{20}$, $T_{21}$,$T_{22}$, $P^{y'}$ and PT
  coefficients) are not necessarily reproduced by including 3{\it NF}, which
  shows deficiency of present day 2$\pi$ exchange type 3{\it NF} models.
\item Particularly the TM 3{\it NF} does a poor job while updated versions
  TM'(99) 3{\it NF}s respecting chiral symmetry do much better jobs similar to
  UR-IX 3{\it NF}.
\item Comparisons are also made between data at 70 MeV/u \cite{kimiko} and
  recent calculations based on the chiral effective field theory
  \cite{chiral}, which does a reasonably good job for $d\sigma/d\Omega$ but
  not much for tensor analyzing powers.
\end{itemize}
Data for the $\vec{n}d$ \cite{yukie} and $\vec{p}d$ \cite{hatanaka} at 250
MeV/u at RCNP are also presented and compared with calculations.

\newabstract 
\label{abs:donoghue2}

\begin{center}
{\large\bf Nuclear Binding Energies and Quark Masses}\\[0.5cm]
John F. Donoghue\\[0.3cm]
Dept. of Physics, University of Massachusetts,\\
Amherst, MA 01003, USA\\[0.3cm]
\end{center}

In this talk, I described some of my recent work on the nuclear force and its
dependence on quark mass~\cite{nuclear}. My initial motivation was in
connection with a project on the equivalence principle and dilaton
couplings~\cite{dilaton}.  For a dilaton that couples in a way similar to a
quark mass term, the least understood effects of its violation of the
equivalence principle is the effects that comes from nuclear binding energies.
These can be understood if one understands the quark mass effects on binding.
However, the results obtained are interesting in their own right in the
context of the nuclear force.

The main innovation was to use the Omnes function technique~\cite{omnes} in
the dispersive integral that determines the scalar-isoscalar nuclear
potential~\cite{potential}. The potential is calculated in terms of its
imaginary part, and the low energy contributions to the imaginary part are
known in chiral perturbation theory. However, as one considers these at higher
energies, the chiral results quickly blow up. Here the Omnes technique becomes
useful. By matching the Omnes function to the chiral results at low energy one
can provide an extension to higher energies that is well behaved. The matching
is given by
\begin{equation}
\rho_S(\mu) = \rho_S^{NNLO} Re \Omega(\mu) + \rho_S^{NNNLO}|\Omega(\mu)|^2
\end{equation}
where $\Omega$ is the Omnes function. Interestingly this representation leads
to a nuclear potential which is very similar to a conventional $\sigma$
potential, and also emerges with about the right strength. This provides a
novel understanding of this ill-understood component of the nuclear force. In
addition, knowing these ingredients one can explore the quark mass dependence
in a way that was not possible before.

\newabstract 
\label{abs:rathmann}
\begin{center}
  {\large\bf Proton-Proton Scattering Experiments}\\[0.5cm]
  Frank Rathmann\\[0.3cm]
  Institut f\"ur Kernphysik, Forschungszentrum J\"ulich, 
  52425 J\"ulich, Germany\\[0.3cm]
\end{center}

The lack of high-quality elastic proton-proton scattering data at the
beginning of the 1990's hampered both the further improvement of potential
models of the $NN$ interaction and the refining of partial wave
analyses. A precise $pp$ data base was also needed to provide
polarimetry for the analysis of inelastic reaction channels.  Through
a series of measurements at the Indiana Cooler with stored polarized
protons impinging on a polarized hydrogen storage cell target, a new
method to measure spin correlation parameters could be established.
The measurements yielded full angular distributions of analyzing
powers $A_y$ and spin correlation parameters $A_{xx}$, $A_{yy}$ and
$A_{xz}$ at eight different energies between 197 and 449 MeV
\cite{haeb}, the fourth spin correlation parameter $A_{zz}$ was
measured at 197 MeV only \cite{lore}. At higher energies, without a
storage cell, the EDDA collaboration achieved high-quality data of
$pp$ excitation functions of cross sections, analyzing powers, and
spin correlation parameters between 200 MeV and 2.5 GeV at COSY
\cite{edda}.  Our theoretical understanding of the $NN$ system above 1
GeV is still unsatisfactory \cite{machleidt}. 

But understanding the $NN$ interaction involves {\it both} isospin
channels. While for the $pp$ system the data situation today can be
called fair, the $pn$ database up to 3 GeV is almost empty.  Although
not ideally suited for polarization experiments, some $pn$ observables
can be measured with the ANKE dipole spectrometer at COSY.  Through
the detection of spectator protons from a polarized deuterium gas
target that is bombarded by polarized protons, $pn$ elastic scattering
at cm angles below $\theta=30^\circ$ is accessible. There, the $pn$
data base is almost empty at all energies \cite{rentmeester}.

\newabstract 
\label{abs:hanhart}

\begin{center}
{\large\bf Systematic Approaches to Meson Production 
in {\boldmath $NN$} and\\[0.1cm] {\boldmath $dd$} Collisions}\\[0.5cm]
C. Hanhart\\[0.3cm]
IKP, Forschungszentrum J\"ulich,\\
52428 J\"ulich, Germany\\[0.3cm]
\end{center}

The goal of this presentation was to demonstrate that it is possible to
extract interesting physics from $NN$ and $dd$ induced meson production
reactions, regardless the large momenta involved. For illustration the
techniques/ideas to extract various charge symmetry breaking 
(CSB) matrix elements
was discussed. 

One can not expect the same methods to be applicable in reactions at different
energies. It was demonstrated recently that pion production can be analyzed
within the framework of chiral perturbation theory, once the scheme is
adjusted to the large momentum transfer \cite{chpt}. The expansion parameter
turned out to be $\chi=\sqrt{m_\pi/M_N}$. Within this framework an
extraction of CSB matrix elements from data on $pn\to d\pi^0$
as well as is straight forward. The corresponding calculations are
in progress \cite{biras}.

As we go to higher energies the straight forward application of chiral
perturbation theory can not work any more. Already at the $\eta$ production
threshold the corresponding expansion parameter is 0.7. Therefore at present
the
extraction of 
CSB matrix elements in a controlled way from the production of heavier mesons
is only possible in a few special situation. In the talk ideas were presented
on how is could be possible to extract the CSB amplitude $\eta \alpha \to
\pi^0\alpha$  from $dd\to \alpha \pi^0$ near the $\eta$ production
threshold. 

The situation is a lot more transparent when it comes to the extraction of the
CSB $f_0-a_0$ mixing matrix element from $pn\to d\pi^0\eta$ and $dd\to
\alpha\pi^0\eta$, for in this case the mixing effect is automatically enhanced
compared to other CSB effects as they might occur in the production operator,
since $a_0$ and $f_0$ are rather narrow overlapping resonances \cite{utica}.

\newabstract 
\label{abs:cabrera}

\begin{center}
{\large\bf Vector Meson Properties in Nuclear Matter}\\[0.5cm]
{\bf Daniel Cabrera}$^1$, E. Oset$^1$ and M.J. Vicente Vacas$^1$\\[0.3cm]
$^1$Departamento de F\'{\i}sica Te\'orica and IFIC, Univ. de Valencia-CSIC,\\
Inst. de Inv. de Paterna, Apdo. Correos 22085, E-46071 Valencia, Spain\\[0.3cm]
\end{center}

In this talk, we report on the properties of vector meson resonances in the
nuclear medium.  First, the $\phi$ meson spectrum due to the $\bar{K} K$
channel is obtained in terms of the $\phi$ selfenergy, which in vacuum is
provided by a chiral $SU(3)$ dynamics model. To consider the $K$ and $\bar{K}$
in-medium properties we use the results of previous calculations which account
for $S-$ and $P-$wave kaon selfenergies based on the lowest order meson-baryon
chiral effective Lagrangian \cite{Oset:2000eg}. The $S-$wave kaon selfenergy
is built from a selfconsistent coupled channel chiral unitary approach to the
$\bar{K}N$ interaction. The $P-$wave kaon selfenergy is driven by the
excitation of $\Lambda-h$, $\Sigma-h$ and $\Sigma^*(1385)-h$ components. In
addition, a set of vertex corrections is evaluated as demanded by gauge
invariance. Within this scheme the mass shift and decay width of the $\phi$
meson in nuclear matter are studied. We find an increase of the $\phi$ decay
width to around 30 MeV at normal nuclear density and a small mass shift to
lower energies which stands below 1 \% of the size of the $\phi$ mass in free
space \cite{Cabrera:2002hc}.

Second, the $\rho$ meson is studied by looking at the pion-pion scattering
amplitude in the vector-isovector channel. We start from a chiral unitary
approach to meson-meson scattering, based on the tree level contributions from
the lowest order $\chi PT$ Lagrangian including explicit resonance fields. Low
energy chiral constraints are considered by matching our expressions to those
of one loop $\chi PT$, and unitarity is fulfilled exactly in the frame of the
$N/D$ method. To account for the medium corrections, the pion propagators are
modified with a $P-$wave selfenergy, driven by the excitation of $p-h$ and
$\Delta -h$ components. The terms where the $\rho$ couples to the hadrons in
the $p-h$ or $\Delta-h$ excitations are also considered as requested by gauge
invariance. In addition, the $\rho$ is allowed to couple to baryonic
resonances, particularly to the $S-$wave $N^{*}(1520)$. The main visible
effect of the nuclear medium is an enhancement of the $\rho$ width as well as
a small shift of the peak to higher energies as the nuclear density increases.
We observe that an important source of strength appears in the $\rho$ spectral
function at low energies from the direct coupling to the $N^{*}(1520)-h$
components \cite{Cabrera:2000dx}.

\newabstract 
\label{abs:schwenk}

\begin{center}
{\large\bf A Renormalization Group Method for Ground State\\[0.1cm] 
Properties of
Finite Nuclei}\\[0.5cm]
Janos Polonyi$^{1, 2}$ and {\bf Achim Schwenk}$^3$\\[0.3cm]
$^1$Institute for Theoretical Physics, Louis Pasteur University,
Strasbourg, France\\
$^2$Department of Atomic Physics, Lorand E\"otv\"os University,
Budapest, Hungary\\[0.3cm]
$^3$Department of Physics, The Ohio State University,
Columbus, OH 43210, USA\\[0.3cm]
\end{center}

The renormalization group (RG) is a powerful method to study 
complex physical systems with a large number of scales. In nuclear 
physics, the RG is an ideal tool for constructing effective interactions
both in the two-nucleon system (leading to a model-independent NN interaction 
at low momenta, called $V_{\mathrm{low}\,k}$~\cite{Vlowk}) and in many-body 
systems, such as for Fermi liquids~\citetwo{RGnm}{tensor}.

In this talk, we propose a novel method for microscopic calculations of 
nuclear ground state (gs) properties within the framework of density functional
theory. Starting from non-interacting nucleons in a harmonic oscillator 
background potential $(1-\lambda) V_\Omega$ with a control parameter
$\lambda$, we use the RG to gradually turn on the two-body interaction 
$\lambda V_{\mathrm{low}\,k}$, while the background potential is removed. 
The RG evolution for $0 \leqslant \lambda \leqslant 1$ is constructed to 
follow the minimum of the density functional $\Gamma_\lambda[\rho]$. 
After separating off the background and Hartree 
contributions to the density functional, 
$\Gamma_\lambda[\rho] = (1-\lambda) V_\Omega \cdot \rho + \frac{\lambda}{2} 
\, \rho \cdot V_{\mathrm{low}\,k} \cdot \rho + \tilde\Gamma_\lambda[\rho]$, 
the RG equation for the exchange-correlation functional 
$\tilde\Gamma_\lambda[\rho]$ is given by~\cite{DFT}
\begin{equation}
\partial_\lambda\tilde\Gamma_\lambda[\rho] 
= \frac{1}{2} \, \mathrm{Tr} \left[ 
V_{\mathrm{low}\,k} \cdot \left(
\frac{\delta^2 \tilde\Gamma_\lambda[\rho]}{\delta\rho 
\, \delta\rho} + \lambda \, V_{\mathrm{low}\,k} \right)^{-1} \right] ,
\end{equation}
where the solution to the RG equation yields the gs density 
$\rho_{\mathrm{gs};\lambda}$ and gs energy 
$E_\mathrm{gs;\lambda} \sim \Gamma_\lambda[\rho_\mathrm{gs;\lambda}]$.
The method can be improved systematically, e.g., to include three-body forces, 
pairing correlations, external currents and a projection on center-of-mass 
momentum. We expect it to be applicable over a wide region of the nuclear 
chart.

\newabstract 
\label{abs:platter}

\begin{center}
{\large\bf The Four-Body System in EFT}\\[0.5cm]
Lucas Platter\\[0.3cm]
IKP, Forschungszentrum J\"ulich\\
D-52425 J\"ulich, Germany\\
Email: l.platter@fz-juelich.de~.\\[0.2cm]
\end{center}
The contact effective field theory provides an ideal framework to
compute observables for low-energy nuclear and atomic systems to very high
precision. The theory has been applied successfully to the three-body
system with large two-body scattering length \cite{hbvk}, 
where it was shown that
 a three-body force at leading order is necessary to renormalize the problem. 
The obvious next step is to apply this EFT to the four-body system. Answering
the question of the relevance of a four-body force in this system would
shed some light on the open problem of a general power-counting for
many-body forces in the contact theory. Furthermore, there is a vast amount
of four-body reactions where a high-precision theory would be very useful .
\\
We have set up the integral equations for the four-body system but
instead of solving these rather complicated integral equations
directly, we restricted ourselves so far to solve a simplified version 
of these 
equations to get some insight into the problem. We found that in this case no
four-body force is needed and speculate that this finding might hold
for the full problem.\\
In parallel, we are investigating a second approach. We make
use of the fact that we can write an effective potential which is equivalent
to the contact theory and use the Yakubovski-equations to solve for the
four-body binding energies. By varying the relevant parameters we hope to find
a quick answer to the to the question of the importance of a four-body
force at leading order\cite{lp}.\\[0.3cm]
This work is done in collaboration with H.-W. Hammer and
U.-G.~Mei\ss ner.

\newabstract 
\label{abs:krebs}

\begin{center}
{\large\bf Electroproduction of Neutral Pions 
Off the Deuteron Near\\[0.1cm] Threshold}\\[0.5cm]
{Hermann Krebs}\\[0.3cm]
{Universit\"at Bonn\\
Helmholtz--Institut f\"ur Strahlen- und Kernphysik~(Theorie)\\
Nussallee 14--16, D--53115 Bonn, Germany}
\end{center}

In the last twenty years chiral perturbation theory (CHPT) became a 
standard tool for the description of low energy physics. 
Based only on the symmetries of QCD, especially on chiral symmetry, this theory
gives a systematic and model--independent description of low energy processes.
Nowadays CHPT is not only formulated for the light meson sector but
also for interactions of Goldstone bosons with one or more nucleons and
even nuclei~\cite{Wein3}.
  
In this talk I presented our analysis of electroproduction of neutral pions 
on the deuteron 
near the threshold using the framework of heavy baryon CHPT~\cite{KBMElektro}. 
I firstly discussed the results of the third order calculation, where two
unknown low energy constants appear. They are related to 
$\pi^0$--electroproduction on the neutron, which contributes to the single
nucleon part~(these are 
diagrams with the scattering only on one nucleon. The second nucleon remains 
untouched during the scattering process.). They are fixed from the fit to the 
total cross section data obtained at MAMI~\cite{Ewald}. Finally 
I discussed our
new fourth order calculation of the three body part~(these are diagrams,
where both nucleons are involved in the scattering process). No new unknown
low energy constants appear at this order. I demonstrated the improvement
of the third order results for the prediction of the differential 
cross sections
and the S--wave amplitudes and concluded that our partial fourth order
calculation is in fair agreement with the experiment~\cite{Ewald}.

\newpage
\section*{Working Group IV: Chiral Lattice Dynamics}

\centerline{Convenors: Elisabetta Pallante and Martin Savage}

\vskip3mm
{
\begin{tabular}{lp{10.2cm}r}
\multicolumn{3}{c}{\bf Formal Lattice Issues}\\
\stalk{I. Montvay}{ Partially Quenched Chiral Perturbation Theory 
and\newline Numerical
Simulations}{abs:montvay}
\stalk{C. Aubin}{ Pion and Kaon Properties in Staggered Chiral\newline
Perturbation Theory}{abs:aubin}
\snotalk{R. Mawhinney}{ Recent Developments in Domain Wall Fermions}{ }
\stalk{S. Sint}{ Twisted Mass QCD and Lattice Approaches to the\newline
$\Delta$I = 1/2 Rule}{abs:sint}
\stalk{C. Hoelbling}{ Overlap Phenomenology}{abs:hoelbing}
[2mm]
\multicolumn{3}{c}{\bf Mesons Properties from the Lattice}\\
\stalk{N. Shoresh}{ Lattice QCD and Chiral Effective Lagrangians}{abs:shoresh}
\stalk{D. Be\a'cirevi\a'c}{ Chiral Perturbation Theory, Heavy-light Mesons 
and\newline 
Lattice QCD}{abs:becirevic}
\stalk{S. D\"urr}{ The Pion Mass in a Finite Volume}{abs:duerr}
\stalk{G. Villadoro}{ The Lattice Determination of $K\to \pi\pi$ 
in the\newline $I$=0 Channel}{abs:villadoro}
\stalk{T. Chiarappa}{ Investigation of the $ \epsilon$ Regime of ChPT: 
Meson\newline Correlation Functions}{abs:chiarappa}
\stalk{S. Peris}{ $\epsilon'/ \epsilon$ and the $\Delta $I=1/2 Rule in the 
$1/N_c$ Expansion}
{abs:peris}
[2mm]
\multicolumn{3}{c}{\bf Nucleon Properties from the Lattice}\\
\stalk{M. G\"ockeler}{ Nucleon Form Factors from the QCDSF Collaboration}
{abs:goeckeler}
\stalk{S.R. Beane}{ N and NN Properties in PQQCD}{abs:beane2}
\stalk{D. Arndt}{ Partially Quenched Chiral Perturbation Theory 
for the Baryon and
Meson Charge Radii}{abs:d.arndt}
\stalk{R. Lewis}{ Strange Matrix Elements in the Nucleon}{abs:lewis1}
\stalk{M. Procura}{ Nucleon Mass and Nucleon Axial Charge 
from Lattice QCD}{abs:procura}
[2mm]
\multicolumn{3}{c}{\bf Future Directions and Focus}\\
\stalk{M. Golterman}{ What should be addressed in the next five 
years}{abs:golterman}
\end{tabular}}

\newabstract 
\label{abs:montvay}

\begin{center}
{\large\bf Partially Quenched Chiral Perturbation Theory and \\[0.1cm]
           Numerical Simulations}                            \\[0.5cm]
Istv\'an Montvay\\[0.3cm]
Deutsches Elektronen-Synchrotron DESY   \\
Notkestr.\,85, D-22603 Hamburg, Germany \\[0.7cm]
\end{center}

 The Gasser-Leutwyler constants \cite{CHPT} of the low energy Chiral
 Lagrangian in QCD can be computed in lattice QCD by numerical
 simulations.
 This is facilitated by the fact that, besides the possibility of
 changing momenta, on the lattice one can also change the masses of the
 quarks.
 Chiral Perturbation Theory (ChPT) can also be extended by changing the
 {\em valence quark masses} in quark propagators independently from the
 {\em sea quark masses} in virtual quark loops.
 In this way one arrives at Partially Quenched Chiral Perturbation
 Theory (PQChPT) \cite{PQCHPT}.
 PQChPT can be directly formulated on the lattice at non-zero lattice
 spacing \cite{SHARPE-SINGLETON} and this can be used for correcting
 leading lattice artifacts in the results \cite{RUPAK-SHORESH}.

 In recent papers of the qq+q Collaboration
 \citefour{NF2TEST}{VALENCE}{SEA}{TSUKUBA} we reported on numerical simulations
 with two degenerate light quark flavours and investigated the
 dependence of the pseudoscalar meson masses and decay constants on sea
 and valence quark masses.
 The present simulations at small quark masses but on relatively coarse
 lattices give first estimates for the values of the Gasser-Leutwyler
 constants $L_4$, $L_5$, $L_6$ and $L_8$ which are within the expected
 range. 


\newabstract 
\label{abs:aubin}

\begin{center}
{\large\bf Pion and Kaon Properties in Staggered Chiral\\[0.1cm] 
  Perturbation Theory}\\[0.3cm]
\textbf{C. Aubin} and C. Bernard\\
Washington University,
St. Louis, MO 63130 USA\\[0.2cm]
\end{center}

Extracting physical quantities from lattice simulations requires an
extrapolation to physical quark masses as well as the continuum
limit. The first limit can be performed using Chiral Perturbation
Theory ($\raise0.4ex\hbox{$\chi$}$PT), while taking the continuum
limit has been uncontrolled until recently. Using dynamical staggered
fermions, scaling violations arising from taste symmetry breaking at
$O(a^2)$ are not negligible.  We must incorporate these taste
violations into a staggered $\raise0.4ex\hbox{$\chi$}$PT
(S$\raise0.4ex\hbox{$\chi$}$PT) to take the continuum limit
systematically.

This has been done for a single flavor (one staggered field with four
tastes) by Lee and Sharpe \cite{LEE-SHARPE}. In order to apply this to
lattice data, we need to generalize the Lee-Sharpe Lagrangian from
Ref.~\cite{LEE-SHARPE} to multiple flavors. Additionally, after
performing the one-loop calculations, we must remove the effects of
the extra staggered tastes.  The Lee-Sharpe Lagrangian for multiple
flavors adds to the continuum chiral Lagrangian a term
$a^2\mathcal{V}$, where $\mathcal{V}$ is the taste-symmetry breaking
potential, written in full for multiple flavors in
Ref.~\cite{AB-SCHPT}.

Using S$\raise0.4ex\hbox{$\chi$}$PT, we can calculate the masses and
decay constants of pions and kaons, shown in
Refs.~\citetwo{AB-SCHPT}{CB-POSTER}. Here we show the specific case of the
kaon mass in the $m_u=m_d\equiv m_l\ne m_s$ limit, including all
analytic terms. Writing $(m^{1-{\rm loop}}_{K^+_5})^2 =
\mu(m_l+m_s)\left( 1 + \delta_{m^2} \right)$, and denoting the chiral
logarithm as $\ell(m^2) = m^2 \ln m^2 + {\rm \it finite\ volume\
corrections}$:
\begin{eqnarray}
        \delta_{m^2}
        &\!\!\!=\!\!\!& \frac{1}{16\pi^2 f^2}\Biggl(
        \frac{2a^2\delta'_V\left( \ell( m^2_{\eta'_V})
        - \ell(m^2_{\eta_V})\right)}{m_{\eta'_V}^2 - m^2_{\eta_V}}
        +\bigl[ V\to A \bigr]+ \frac{2}{3}\ell(m_{\eta_I}^2)\Biggr)
        \nonumber\\*
        &\!\!\!\!\!\!&+\frac{16\mu}{f^2}\left(2L_8-L_5\right)
        \left(m_l+m_s\right)
        +\frac{32\mu}{f^2}
        \left(2L_6-L_4\right) \left(2m_l+m_s \right) +
        a^2 C.
\end{eqnarray}

Expressions for the pion and kaon properties can be used to
extract physical numbers to be compared with experiment. For example,
we find preliminary values for the pion decay constant: $f_\pi^{\rm
th} = 130.7(1.0)_{\rm stat}(3.5)_{\rm sys} {\rm Me\!V}$ ($f_\pi^{\rm
exp} =130.7(0.4) {\rm Me\!V}$) and for the kaon decay constant:
$f_K^{\rm th} =157.4(1.0)_{\rm stat}(3.7)_{\rm sys}{\rm Me\!V}$
($f_K^{\exp}=159.8(1.5) {\rm Me\!V}$). Also of interest is the
combination of low-energy constants $2L_8 - L_5 =
-0.3(1)(^{+1}_{-2})\!\times\! 10^{-3}$, since this number is well outside
the allowed range for $m_u = 0$ to be a solution of the strong CP
problem.


\newabstract 
\label{abs:sint}

\begin{center}
{\large\bf Twisted Mass QCD and Lattice Approaches to the\\[0.1cm]
 {\boldmath $\Delta I=1/2$} Rule}\\[0.5cm]
Carlos Pena$^1$, {\bf Stefan Sint}$^2$ and Anastassios Vladikas$^3$\\[0.3cm]
$^1$DESY Theory Group, Notkestrasse 85  D-20607
Hamburg, Germany,\\[0.3cm]
$^2$Departamento de F\'{\i}sica Te\'orica C-XI,
Universidad Aut\'onoma de Madrid,\\
E-28049 Cantoblanco, Madrid, Spain.\\[0.3cm]
$^3$INFN, Sezione di Roma II, Dipartimento di Fisica,
Universit\`a di Roma ``Tor Vergata'', Via della Ricerca Scientifica 1,
I-00133 Rome, Italy\\[0.3cm]
\end{center}
Twisted mass lattice QCD (tmQCD) was originally
designed to eliminate the problem of unphysical zero-modes
in lattice QCD with Wilson-type quarks~\citetwo{tmQCD_proc1}{tmQCD_1}.
This is achieved by adding a chirally twisted quark mass term which
sets a lower bound on the spectrum of the Wilson-Dirac
operator. Renormalised tmQCD can be transformed back
to standard QCD by a (non-singlet) chiral rotation of the
fields, and the renormalized tmQCD correlation functions 
can thus be interpreted as linear combinations
of the correlation functions in standard QCD. 
As the chiral field rotation is not a lattice symmetry,
the renormalization of bare lattice
operators can be quite different in the two cases.
This has already been used for the computation of 
$B_K$ where the usual renormalization problems with
Wilson fermions are avoided~\citetwo{tmQCD_1}{Dimopoulos:2003kc}.
Here we generalise tmQCD to four Wilson quark flavours, 
and study the computation of $K\rightarrow\pi$
matrix elements of the CP conserving 
weak Hamiltonian~\cite{current}.
We show that, with an active charm quark, the renormalization of the
$K\to\pi$ matrix elements requires at most the subtraction of
a linearly divergent counterterm. In 
the quenched approximation, the parameters can be chosen
such that only a finite counterterm needs to be subtracted.
As a result one may be able to match to lowest
order chiral perturbation theory and 
obtain the effective couplings $g_8$ and $g_{27}$,
which determine the $\Delta I=1/2$ rule in non-leptonic
kaon decays.

\newabstract 
\label{abs:hoelbing}

\begin{center}
{\large\bf Overlap Phenomenology}\\[0.5cm]
Christian Hoelbling\\[0.3cm]
Centre de Physique Th\'{e}orique de Marseille (CNRS)\\
163 Avenue de Luminy, Case 907
13288 Marseille cedex 9, France
\end{center}

Quenched overlap fermions \cite{Neuberger:1997fp} are used to study QCD
phenomenology. With the simulation technique described 
in\,\cite{Giusti:2001yw}, we found the mass of the strange 
quark\,\cite{Giusti:2001pk}
\begin{equation}
m_s^{\overline{MS}}(2 \mbox{GeV}) = 102(6)(18)  \mbox{MeV}
\end{equation}
and the chiral condensate
\begin{equation}
\frac{1}{N_f}\langle\bar{\psi}\psi\rangle
^{\overline MS}(2  \mbox{GeV}) = (267(5)(15)  \mbox{MeV})^3
\end{equation}
We determined $B_K$ \cite{Garron:2003cb} 
\begin{equation}
B_K^{RGI}=0.87(8)^{+2+14}_{-1-14}
\end{equation}
where the first error is statistical, the second is systematic and the third
is an estimate of the quenching error.

The calculation of other weak matrix elements including $B_7$ and $B_8$, of
baryon spectra and the charmed meson decay constant is underway and
preliminary results are reported in \cite{Berruto:2003rt}.

\newabstract 
\label{abs:shoresh}

\begin{center}
{\large\bf Lattice QCD and Chiral Effective Lagrangians}\\[0.5cm]
Oliver B\"ar$^1$, Gautam Rupak$^2$, and {\bf Noam Shoresh}$^3$\\[0.3cm]
$^1$Institute of Physics, University of Tsukuba,\\
Tsukuba, Ibaraki 305-8571, Japan\\[0.3cm]
$^2$Lawrence Berkeley National Laboratory,\\
Berkeley, CA 94720, U.S.A. \\[0.3cm]
$^3$Department of Physics, Boston University,\\
Boston, MA 02215,U.S.A.[0.3cm]
\end{center}

Chiral perturbation theory ($\chi$PT) is a necessary tool in
extracting information about QCD from lattice simulations, which are
done with relatively heavy quark masses. We consider several low
energy effective theories -- extensions of $\chi$PT -- which capture
the leading discretization artifacts for certain lattice actions.
The strategy is a two-step process. First one constructs the local
Symanzik action for a given lattice action. Then, in close analogy to
the derivation of $\chi$PT from QCD, one constructs a low energy
chiral effective action. The resulting chiral Lagrangian reduces to
the  familiar chiral Lagrangian in the limit $a\to0$. When $a$ is non-zero, 
new terms appear in the Lagrangian with corresponding
unknown coefficients (analogous to the Gasser-Leutwyler
coefficients). 

This method has been applied to the Wilson lattice action in
Ref.~\cite{Rupak:2002sm} to order $a$, and to order $a^2$ in
Refs.~\citetwo{Bar:2003mh}{Aoki:2003yv}. The chiral Lagrangian for
staggered fermions
has been derived in Refs.~\citetwo{Lee:1999zx}{Aubin:2003mg}. Another type
of lattice action which has been analyzed   
is one with Ginsparg-Wilson valence quarks and Wilson sea quarks
\citetwo{Bar:2002nr}{Bar:2003mh}.

\newabstract 
\label{abs:becirevic}

\begin{center}
{\large\bf Chiral Perturbation Theory, Heavy-light Mesons 
\\[0.1cm] and Lattice QCD}\\[0.5cm]
{\bf Damir~Be\'cirevi\'c}$^1$,  
Sa\v{s}a Prelov\v{s}ek$^{2,3}$, 
Jure~Zupan$^{2,4}$\\[0.3cm]
$^1$
Laboratoire de Physique Th\'eorique (B\^at 210)~\footnote{Unit\'e mixte de
Recherche du CNRS - UMR 8627.}, Universit\'e de Paris Sud, \\ 
Centre d'Orsay, 91405 Orsay-Cedex, France.\\[0.3cm]
$^2$J.~Stefan Institute, Jamova 39, P.O. Box 3000,\\ 
1001 Ljubljana, Slovenia.\\[0.3cm]
$^3$Department of Physics, University of Ljubljana, Jadranska 19,\\ 
1000 Ljubljana, Slovenia.\\[0.3cm]
$^4$Department of Physics,  
Technion--Israel Institute of Technology,\\ 
Technion City, 32000 Haifa, Israel.\\[0.3cm]

\end{center}
We study the chiral properties of the $B\to \pi$ and $B\to K$ transition form
factors in the static heavy quark limit by employing the quenched, partially
quenched and full (unquenched) chiral perturbation theory (ChPT).  From the
expressions derived at NLO in all three versions of ChPT we were able to
examine the size of systematic uncertainties which currently plague the
accurate determination of the form factors by means of lattice QCD, namely:
the effects of quenched approximation, the impact of the chiral log terms on
the extrapolation toward the physical pion mass. The complete discussion with
details of calculations is presented in refs.~\citetwo{Qpaper}{PQpaper}.  
In the
case of partially quenched theory with $N_f=2$ degenerate dynamical flavours
we propose a simple chiral extrapolation strategy by which one can avoid the
effects of divergent quenched chiral logarithms~\cite{PQpaper}.

We also discuss the $\xi$ parameter which enters decisevely into the standard
CKM unitarity triangle analyses~\cite{doubleratio}.  In particular, we propose
to combine the $f_{B_s}/f_B$ and $f_K/f_\pi$ in double ratio in which the
large chiral logarithms cancel, thus reducing the otherwise huge uncertainties
in the chiral extrapolation of $f_{B_s}/f_B$. From the double ratio computed
on the lattice, and the experimentally determined $f_K/f_\pi$, one can predict
$f_{B_s}/f_B$, with a very small uncertainty.

\newabstract 
\label{abs:duerr}

\begin{center}
{\large\bf The Pion Mass in Finite Volume}\\[0.5cm]
{\bf Stephan D\"urr}$^1$ and Gilberto Colangelo$^2$\\[0.3cm]
$^1$DESY Zeuthen, 15738 Zeuthen, Germany\\
$^2$Institute for Theoretical Physics, University of Bern,
3012 Bern, Switzerland\\[0.6cm]
\end{center}

When studying QCD in a finite box $L^3\!\times\!T$ with toroidal boundary
conditions (as is standard in lattice calculations with $T\!\gg\!L$),
properties of the QCD Hamiltonian will reflect the finite spatial extent.
In particular, the relative shift
\begin{equation}
R_M(M_\pi,L)={M_\pi(L)-M_\pi\over M_\pi}
\qquad\mathrm{where}\quad M_\pi\!\equiv\!M_\pi(L\!=\!\infty)
\label{myeq1}
\end{equation}
is a systematic effect that one would like to have analytically predicted.

One way is due to L\"uscher who noted that the finite-size effect in
Euclidean space follows from the $\pi$-$\pi$ forward scattering amplitude in
Minkowski space \cite{Luscher:1985dn}
\begin{equation}
R_M(M_\pi,L)=-{3\over16\pi^2M_\pi^2 L}\;\int_{-\infty}^\infty\!dy\; 
F(\mathrm{i}y)\,e^{-\sqrt{M_\pi^2+y^2}\,L}+O(e^{-\overline{M}L})
\label{my:luscher}
\end{equation}
where the generic bound $\overline{M}\!\geq\!\sqrt{3/2}M$ can be specified to
$\overline{M}\!=\!\sqrt{2}M_\pi$.
To LO in the chiral expansion the amplitude $F$ is constant (in $y$), hence the
integral in (\ref{my:luscher}) can be done analytically \cite{Fukugita:1992hr}.
Using NLO and NNLO chiral input, the resulting expressions for $R_M$ become
more involved \citetwo{Colangelo:2002hy}{CD03}, 
necessitating in the latter case a
thorough discussion of the impact of the uncertainties of the NLO low-energy
constants involved.
The advantage is that one can explicitly watch the chiral convergence behavior
at fixed $M_\pi$ and $L$.

The other approach is to directly compute the pion mass in chiral perturbation
theory in a finite volume.
This has been done at 1-loop level in \cite{GaLeFSE1}, resulting in
\begin{equation}
R_M(M_\pi,L)={1\over2N_\mathrm{f}}\xi\,\tilde g_1(\lambda)+O(\xi^2)
\label{GL_FSE}
\end{equation}
with $\xi\!=\!M_\pi^2/(4\pi F_\pi)^2$ and $\lambda\!=\!M_\pi L$.
In \cite{CD03} we provide a representation of $\tilde g_1$ which is handy for
a numerical evaluation, and we include a table where the predictions of both
approaches are compared for a reasonable range of $M_\pi$ and $L$ values.


\newabstract 
\label{abs:villadoro}

\begin{center}
{\large\bf The Lattice Determination of {\boldmath $K\to\pi\pi$}\\[0.1cm] in 
the {\boldmath $I=0$} Channel}\\[0.5cm]
Giovanni Villadoro\\[0.3cm]
Dip. di Fisica, Univ. di Roma "La Sapienza" and INFN,\\
Sezione di Roma, P.le A.Moro 2, I-00185 Rome, Italy\\[0.3cm]
\end{center}

In this talk, a brief review was presented about recent developments in
understanding problems related to the extraction of $I=0$ $K\to\pi\pi$ matrix
elements from the lattice. In particular the talk focused on the interplay
between finite volume corrections, final state interaction and the quenching
approximation.  After a short introduction about historical milestones on the
subject, such as the Maiani-Testa no-go theorem and finite volume L\"uscher
formulae (see ref.~\cite{MTL}), the general strategy to extrapolate the
amplitude to the physical point, through the use of $\chi$PT to NLO, was
described \cite{spqr}.  Two recent procedures of implementing $\chi$PT in this
type of calculations and the relevant formulae can be found in
refs.~\citetwo{spqr}{LS}. 
Finite volume analysis has been discussed then in full,
quenched (q) and partially quenched (pq) QCD by using $\chi$PT.  In particular
it was discussed how, in qQCD, the lack of unitarity prevents the extraction
of matrix elements from correlators with multi-particle external states
\cite{LinQ}.  More recently, in ref.~\cite{LinPQ}, it has been shown, however,
that pqQCD can avoid this problem for particular choices of masses and
momenta. More in details, the $I=0$ $K\to\pi\pi$ matrix element can be
extracted from pq simulations provided that sea quarks be degenerate with the
valence ones and the c.o.m. energy be under threshold for production of
quenched flavors.  Outlook and consequences on actual numerical simulations
were presented in the conclusion.

\newabstract 
\label{abs:chiarappa}

\begin{center}
{\large\bf Investigation of {\boldmath $\epsilon$} Regime of ChPT:
Meson Correlation\\[0.1cm]  Functions}\\[0.5cm]
{{\bf T. Chiarappa}$^1$, W. Bietenholz$^2$, K. Jansen$^1$,
 K.-I. Nagai$^1$ and S. Shcheredin$^2$}\\[0.3cm]
$^1$NIC/Desy Zeuthen,\\
Platanenallee 6, D-15738 Zeuthen, Germany\\[0.3cm]
$^2$ Institut f\"{u}r Physik, Humboldt Universit\"{a}t zu Berlin,\\
Newtonstr. 15, D-12489 Berlin, Germany\\[0.3cm]
\end{center}

Surprisingly, physical information can be extracted
from unphysical setups, like from a finite volume via a finite size analysis. 
This framework is highly suitable for lattice simulations.\\
One analytic tool to study such a setup is 
the $\epsilon$-expansion of Chiral Perturbation Theory (ChPT) \cite{epsGL}, 
which allows to extract the characteristic Low Energy Constants (LEC) 
by confronting its prediction with numerical data.\\
In contrast to the infinite volume case, in a finite 
box ($V = L^{4}$) gauge field of fixed topological sectors 
play a prominent r\^{o}le \cite{LS}.\\
This turns out to be expecially important in the $\epsilon$-regime of ChPT, 
where the Compton wavelength of the pion does not fit at 
all into the box ($L m_{\pi} < 1$).\\
From a numerical point of view, the possibility to simulate 
chiral fermions on the lattice is one of the great achievement 
of the last years \cite{N-L-H}.\\
This new lattice world is very cost demanding, forcing us 
(so far) to restrict simulations to the quenched approximation.\\
Based on two recent analytical works \cite{mesons}, a numerical 
pilot study of the 2-point meson correlation functions in the 
$\epsilon$-regime of quenched chiral perturbation theory 
was presented \cite{Tsukuba}, using the overlap formalism 
with $\beta = 6.0$ and two lattices of physical volume 
$1.12$ fm and $0.93$ fm.\\
From the data relative to the axial correlation function, it was 
possible to extract the quenched bare pion decay constant 
to be $F_{\pi}^{b} = 86.7 \pm 4.0$ MeV.\\
The determination of other LEC is in progress.

\newabstract 
\label{abs:peris}
\begin{center}
{\large\bf {\boldmath $\epsilon'/\epsilon$} and the 
{\boldmath $\Delta\mathrm{I}=1/2$} 
Rule in the {\boldmath $1/N_c$} Expansion}\\[0.5cm]
S. Peris\\[0.3cm]
Grup de Fisica Teorica and IFAE\\ Universitat Autonoma de Barcelona\\
08193 Barcelona, Spain\\[0.3cm]
\end{center}

In this talk, after briefly discussing how to analytically compute
unfactorized contributions to weak matrix elements in the large-$N_c$
expansion \cite{one}, I reported on the new results for the matrix elements of
the $Q_6$ and $Q_4$ penguin operators obtained in \cite{two}. These results
include unfactorized contributions of
$\mathcal{O}(N_c^2\times\frac{n_f}{N_c})$ and exhibit analytic matching
between short-- and long--distance scale dependence in the
$\overline{\mathrm{MS}}$ scheme. It is found a numerically large and positive
contribution to the $\Delta I =1/2$ matrix element of $Q_6$ and hence to the
direct CP-violation parameter $\varepsilon'/\varepsilon$.

The implications of these results for a quenched theory were analyzed in
\cite{three} where the conclusion was reached that the pure quenching
artifact, $\alpha_{q}^{NS}$, the quenched octet coupling $\alpha_{q1}^{(8,1)}$
and its counterpart in the unquenched theory, $\alpha_{1}^{(8,1)}$\cite{ME},
obey the hierarchy $\alpha_{q}^{NS} \gg \alpha_{q1}^{(8,1)} \gg
\alpha_{1}^{(8,1)}$. Since $\alpha_{q}^{NS}$ is currently neglected in lattice
analyses of $\varepsilon'/\varepsilon$, the result just mentioned could help
explain the present discrepancy between lattice and experimental
results\cite{RBC}\cite{exp}.

Results were also obtained for the $\Delta I = 1/2$ rule in $K \rightarrow \pi
\pi$ amplitudes. Large ``eye--diagram'' contributions of
$\mathcal{O}(N_c^2\times\frac{1}{N_c})$ from the $Q_2$ operator were found
which are related to the unfactorized contributions from $Q_4$ mentioned
above. The results lead to an enhancement of the $\Delta I =1/2$ effective
coupling. A common origin for all the violation of factorization that we find
was discussed in terms of the relevant scales in the problem.

I am very thankful to M. Golterman, T. Hambye and E. de Rafael for a most
pleasant and fruitful collaboration.

\vspace{ -0.5cm}

\newabstract 
\label{abs:goeckeler}

\begin{center}
{\large\bf Nucleon Form Factors from the QCDSF Collaboration}\\[0.5cm]
M. G\"ockeler$^{1,2}$\\[0.3cm]
$^1$Institut f\"ur Theoretische Physik, Universit\"at Leipzig, \\
D-04109 Leipzig, Germany\\[0.3cm]
$^2$Institut f\"ur Theoretische Physik, Universit\"at Regensburg, \\
D-93040 Regensburg, Germany\\[0.3cm]
\end{center}

We have computed the electromagnetic form factors of the nucleon in
quenched lattice QCD, using non-perturbatively improved Wilson 
fermions~\cite{fofa}. The analysis focusses on the isovector case as
in this sector the quark-line disconnected contributions,
which we have not evaluated, cancel. We want to compare the pion-mass 
dependence of the Monte Carlo results with predictions of chiral
effective field theory. Since the pions in the simulations are relatively
heavy ($m_\pi > 500 \, \mbox{MeV}$) quenching effects are expected to
be small, and one can use ordinary chiral perturbation theory rather
than quenched chiral perturbation theory. More specifically, we consider 
formulae obtained within the small scale expansion~\citetwo{chpt}{chiralmag}, 
which has explicit nucleon and $\Delta$ degrees of freedom. 
As the momenta in our simulations are too large for a comparison of the 
form factors themselves, we fit the $Q^2$ dependence of our form factors 
with a dipole ansatz and compare only the resulting radii as well as the 
anomalous magnetic moment. While a reasonable connection between the
Monte Carlo data 
and the physical value is achieved for the Pauli radius and the anomalous
magnetic moment, the Dirac radius is more problematic~\cite{fofa}. 
It seems that the leading one-loop calculation in
the small scale expansion is not accurate
enough to describe the quark-mass dependence of the isovector Dirac
radius up to the masses used in present simulations. It remains to be 
seen whether additional terms in the expansion can solve this problem.
On the other hand, the simulations will progress towards smaller quark 
masses so that eventually contact with chiral perturbation theory should
be established.

\newabstract 
\label{abs:beane2}
\begin{center}
{\large\bf N and NN Properties in PQQCD}\\[0.5cm]
Silas R.~Beane$^{1,2}$\\[0.3cm]
$^1$Department of Physics, University of New Hampshire, 
Durham, NH 03824\\[0.3cm]
$^2$Jefferson Laboratory, 12000 Jefferson Avenue, 
Newport News, VA 23606\\[0.3cm]
\end{center}

\noindent Presently, unquenched lattice simulations with the physical
values of the light-quark masses are prohibitively time-consuming,
even on the fastest machines.  Relatively recently it was realized
that partially-quenched (PQ) simulations, in which the sea quarks are
more massive than the valence quarks, provide a rigorous method to
determine QCD observables and are much less time-consuming than their
QCD counterparts. Technology has been developed to describe
partially-quenched QCD (PQQCD) with PQ$\chi$-PT (Ref.~\cite{Pqqcd1}
and references therein). It is hoped that future lattice simulations
can be performed with sufficiently small quark masses where the chiral
expansion is convergent, and can be used to extrapolate down to the
quark masses of nature. Recently, meson and baryon properties have
been studied extensively in
PQ$\chi$-PT~\citesix{CSn}{BSn}{BSpv}{CSn2}{Leinweber:2002qb}{Arndt1}.
Two-nucleon systems have also been explored in
PQQCD~\citetwo{BSnn}{ABS}.  An important further motivation for
PQQCD is the additional ``knobs'' that are provided by exploring the
``plane'' of sea and valence quark masses and which increase the
consistency checks available to lattice simulations.

\newabstract 
\label{abs:d.arndt}

\begin{center}
{\large\bf Partially Quenched Chiral Perturbation Theory
 for the\\[0.1cm] Baryon and Meson Charge Radii}\\[0.5cm]
{\bf Daniel Arndt} and Brian C.\ Tiburzi\\[0.3cm]
Department of Physics, Box 351560,\\
University of Washington, Seattle, WA 98195-1560, USA\\[0.3cm]
\end{center}

The study of electromagnetic form factors 
of the hadrons
at low momentum transfer
provides important insight into the non-perturbative structure of QCD.
A number of
new lattice QCD calculations of these form factors
have recently appeared~\cite{lattice}.
These calculations use the quenched approximation
of QCD (QQCD);
partially quenched calculations (PQQCD)
are expected in the near future.
Since currently and foreseeably lattice
calculations cannot be performed with the physical masses of the 
light quarks,
it is crucial to know how to properly extrapolate the results
of these calculations
from the quark masses used on the lattice to those in nature.
To aid in this extrapolation,
low-energy effective theories%
---quenched and partially quenched chiral perturbation theory
(Q$\chi$PT, PQ$\chi$PT)---%
have been developed~\cite{QandPQ}.

We calculate~\cite{we} the electric charge radii 
of the $SU(3)$ meson and baryon octets
in Q$\chi$PT and PQ$\chi$PT.
We work in the isospin limit, 
up to next-to-leading order in the chiral expansion,
and to leading order in the heavy baryon expansion.
We find that,
while the expansions about the chiral limit for 
Q$\chi$PT and PQ$\chi$PT
are formally similar,
$<r^2>\sim\alpha+\beta\log m_Q+\dots$,
the QQCD result exhibits quenched oddities:
for $\Sigma^-$ and $\Xi^-$, for example, 
$\beta=0$ and these radii are
independent of the quark masses $m_Q$!
Such behavior, also found for the mesons, 
illustrates how QQCD has no
known connection to QCD.
PQQCD, however, is smoothly
connected to QCD
and the low-energy constants of PQ$\chi$PT
are found to be the same as those of $\chi$PT.
Hence, our PQ$\chi$PT result
will enable proper extrapolation of PQQCD lattice simulations
to QCD.

\newabstract 
\label{abs:lewis1}

\begin{center}
{\large\bf Strange Matrix Elements in the Nucleon}\\[0.5cm]
{\bf Randy Lewis}$^1$, Walter Wilcox$^2$ and R. M. Woloshyn$^3$\\[0.3cm]
$^1$Department of Physics, University of Regina, Regina, SK, S4S 0A2, Canada
\\[0.2cm]
$^2$Department of Physics, Baylor University, Waco, TX, 76798-7316, USA
\\[0.2cm]
$^3$TRIUMF, 4004 Wesbrook Mall, Vancouver, BC, V6T 2A3, Canada
\\[0.2cm]
\end{center}

The strange-quark current matrix elements of the 
nucleon probe non-valence degrees of freedom.
This physics can be studied within quenched lattice QCD by
invoking a stochastic method for the strange quark propagator\cite{stoch}.
Pioneering studies of the strangeness electromagnetic form factors demonstrated
that a signal can be difficult to extract from lattice
simulations\cite{previous}, so our recent work\cite{LWW} includes a study with
many more configurations than previously used, as well as two different
choices for the number of noises in the stochastic propagator.
A clear signal is seen for the scalar matrix element, but the electric and
magnetic matrix elements can only be bounded by an uncertainty consistent with
zero.

Ref.~\cite{LWW} discusses the chiral extrapolations of these strangeness
matrix
elements through quenched SU(3) chiral perturbation theory (ChPT) calculations.
Some pieces of the calculations can also be found in Ref.~\cite{otherChPT}.
It is noted that the scalar, electric and magnetic cases share the same
low energy constants (LEC's), so the established scalar signal from lattice
data constrains the electric and magnetic extrapolations.
There could be advantages to using SU(2) ChPT instead, and a calculation
exists\cite{CS}, but in SU(2) ChPT the scalar LEC's differ from
the electromagnetic LEC's, so it seems that the application
of SU(2) ChPT to electromagnetic matrix elements must wait for a clearer
lattice QCD signal.

\newabstract 
\label{abs:procura}

\begin{center}
{\large\bf Nucleon Mass and Nucleon Axial Charge from\\[0.1cm] 
Lattice QCD}\\[0.4cm]
{\bf M. Procura}$^{1,2}$, T.R. Hemmert$^1$ and W. Weise$^{1,2}$\\[0.3cm]
$^1$Physik-Department, Theoretische Physik, TU M{\"u}nchen,\\
D-85747 Garching, Germany\\[0.3cm]
$^2$ECT*, Villa Tambosi, I-38050 Villazzano (Trento), Italy.\\[0.3cm]
\end{center}

Lattice QCD on one side and Chiral Perturbation Theory ($\chi$PT) on the
other, are progressively developing as important tools to deal with the
non-perturbative aspects of hadron structure. At present, however, there is a
gap between the relatively large quark masses accessible in fully dynamical
lattice simulations of nucleon properties and the small quark masses relevant
for comparison with physical observables. In our work we explore the
consequences of a direct application of $\chi$PT in the range of presently
accessible quark masses on the lattice and the feasibility of a systematic
approach based on chiral effective Lagrangian for the extrapolation of nucleon
properties from lattice QCD.\\
In \cite{PHW} we focus on the quark mass dependence of the nucleon mass $M_N$
and present an extrapolation of this observable from two-flavor lattice QCD
results in the framework of relativistic baryon $\chi$PT up to order $p^4$ in
infrared regularization scheme. Already at leading one-loop order we obtain a
good chiral extrapolation function and the next-to-leading one-loop
corrections turn out to be reasonably small. From the $p^4$ extrapolation we
find $M_N= (0.91 \pm 0.07)\,{\rm{GeV}}$ together with the pion-nucleon sigma
term $\sigma_N= (53 \pm 8)\,{\rm{MeV}} $, at the physical value of the pion
mass. Including the empirical $M_N$ as a constraint gives the nucleon mass in
the chiral limit $M_0 \approx 0.89 \,{\rm{GeV}}$, and $\sigma_N \approx 47
\,{\rm{MeV}}$.\\
We study also the quark mass expansion of the axial-vector coupling constant
$g_A$ of the nucleon \cite{HPW}. We compare two versions of non-relativistic
chiral effective field theory: Heavy Baryon $\chi$PT and an extension of that
(Small Scale Expansion) which incorporates in a systematic framework explicit
$\Delta$(1232) resonance degrees of freedom. It turns out that, in order to
approach the physical value of $g_A$ in a leading one-loop calculation, the
inclusion of the explicit $\Delta$(1232) degrees of freedom is crucial. With
information on important higher order couplings constrained from analyses of
the $\pi N\rightarrow \pi\pi N$ reaction, a chiral extrapolation function
$g_A(m_\pi)$ is obtained, which works well from the chiral limit across the
physical point into the region of present lattice data.

\newabstract 
\label{abs:golterman}

\begin{center}
{\large\bf ``What should be addressed the next five years"}\\[0.5cm]
Maarten Golterman \\[0.2cm]
{\it Dept. of Physics and Astronomy, SFSU, 
San Francisco, CA 94044, USA} \\
[0.3cm]
\end{center}

I collected and discussed a number of methodological
issues I believe to be 
important for extracting hadronic physics 
from Lattice QCD, restricting myself to topics involving
the use of effective field theory.
Since almost all Lattice QCD computations are ``unphysical"
(wrong quark masses, unphysical momenta, small volume, {\it etc.}),
EFT techniques such as ChPT are essential
in extracting physics from Lattice QCD.  An important new application 
of EFT is its use in parameterizing in a systematic way $O(a)$ and
$O(a^2)$ corrections \citetwo{shoresh}{aubin}.  

(Partially) quenched Lattice QCD with the number
of sea quarks $N_f\ne 3$ leads to {\it uncontrolled} errors and
unphysical effects.
These can{\it not} be estimated/corrected for in ChPT, 
since the (partially) quenched
low-energy constants can be different, and run
differently.  Since many Lattice QCD results at present are 
still obtained at $N_f=0,2$, it's useful pursuing quantitative
{\it analytic} estimates of such differences; 
an example for the penguin $Q_6$
is in \cite{mgsp}.   Because of the importance of $N_f=3$ 
computations, one should consider using staggered fermions
for sea quarks -- even though there are issues
to be understood ({\it e.g.}
fractional powers of the fermion
determinant and ``mismatched" zero modes). 

An important issue is the convergence of ChPT, both
on the lattice (does it converge for typical lattice meson
masses? -- see {\it e.g.} \cite{rbckentuckyadelaide} for examples of
poor convergence) and in the real world (does it converge at $M_K$?).
It does not help to stick LECs computed on the lattice into 
real-world ChPT if that does not converge (getting
$M_K$ right on the lattice, and extrapolating in
$m_{u,d}$ \cite{savage} may help).  
I emphasized that LECs are interesting
and easier than physical quantities ($g_{8,27}$ are easier
than $\varepsilon'/\varepsilon$!), and phenomenological or
model estimates are often available.  Comparison with
analytic approaches \cite{peris}
should be interesting (example:
$\Pi_{LR}(Q^2)$, for $\Pi_{EM}(Q^2)$ see \cite{blum}), 
because they
have typically more difficulty outside the chiral limit, while
the lattice has trouble getting to the chiral limit.

\end{document}